\documentclass[journal,comsoc]{IEEEtran}
\usepackage{algorithm,algorithmicx,amsbsy,amsmath,amssymb,epsfig,bbm,mathrsfs,multirow,amsthm,cite}
\usepackage{xcolor}

\usepackage{subcaption}
\usepackage[labelformat=simple]{subcaption}
\renewcommand\thesubfigure{(\alph{subfigure})}
\usepackage{multirow}
\PassOptionsToPackage{hyphens}{url}\usepackage{hyperref}
\usepackage{eurosym}
\hypersetup{
	colorlinks=true,
	linkcolor=blue,
	filecolor=blue, 
	urlcolor=black,citecolor=blue,}

\DeclareMathAlphabet\mathbfcal{OMS}{cmsy}{b}{n}
\usepackage{array}

\begin{document}
	
	\title{Enabling and Emerging Technologies for Social Distancing: A Comprehensive Survey and Open Problems}
	
	\author{Cong T. Nguyen$^1$, Yuris Mulya Saputra$^1$, Nguyen Van Huynh$^1$, Ngoc-Tan Nguyen$^1$, Tran Viet Khoa$^1$, Bui~Minh~Tuan$^1$, Diep~N.~Nguyen$^1$, Dinh Thai Hoang$^1$, Thang X. Vu$^2$, Eryk Dutkiewicz$^1$, Symeon Chatzinotas$^2$, and Bj\"{o}rn Ottersten$^2$\\
		$^1$ School of Electrical and Data Engineering, University of Technology Sydney, Australia  \\
		$^2$ Interdisciplinary Centre for Security, Reliability and Trust, University of Luxembourg, Luxembourg	
		
	\thanks{\textbf{IEEE Copyright Notice}: This paper was accepted and published by IEEE Access. The published versions of this article are: ``A Comprehensive Survey of Enabling and Emerging Technologies for Social Distancing—Part I: Fundamentals and Enabling Technologies'' DOI: 10.1109/ACCESS.2020.3018140 and ``A Comprehensive Survey of Enabling and Emerging Technologies for Social Distancing—Part I: Fundamentals and Enabling Technologies'' DOI: 10.1109/ACCESS.2020.3018140.} 
		
		\vspace{-5mm}}

	\maketitle
	
	\begin{abstract}
		Social distancing plays a pivotal role in preventing the spread of viral diseases illnesses such as COVID-19. By minimizing the close physical contact among people, we can reduce the chances of catching the virus and spreading it across the community. This paper aims to provide a comprehensive survey on how emerging technologies, e.g., wireless and networking, artificial intelligence (AI) can enable, encourage, and even enforce social distancing practice. To that end, we first provide a comprehensive background of social distancing including basic concepts, measurements, models, and propose various practical social distancing scenarios. We then discuss enabling wireless technologies which are especially effective and can be widely adopted in practice to keep distance, encourage, and enforce social distancing in general. After that, other emerging and related technologies such as machine learning, computer vision, thermal, ultrasound, etc., are introduced. These technologies open many new solutions and directions to deal with problems in social distancing, e.g., symptom prediction, detection and monitoring quarantined people, and contact tracing. Finally, we provide important open issues and challenges (e.g., privacy-preserving, scheduling, and incentive mechanisms) in implementing social distancing in practice. As an example, instead of reacting with ad-hoc responses to COVID-19-like pandemics in the future, smart infrastructures (e.g., next-generation wireless systems like 6G, smart home/building, smart city, intelligent transportation systems) should incorporate a \emph{pandemic mode} in its standard architecture/design.
	\end{abstract}
	
	\begin{IEEEkeywords}
		Social distancing, pandemic, COVID-19, technologies, wireless, networking, positioning systems, AI, machine learning, data analytics, localization, privacy-preserving, scheduling, and incentive mechanism.
	\end{IEEEkeywords}
	
	\section{Introduction } 
	\label{sec:introduction}
	
	COVID-19 has completely changed the world's view on pandemics with dire consequences to global health and economy. Within only four months (from January to April 2020), 210 countries and territories around the world have reported more than three million infected people including more than two hundred thousand deaths~\cite{covid1}. Besides the global health crisis, COVID-19 has also been causing massive economic losses (e.g., a possible 25\% unemployment rate in the U.S.~\cite{covid2}, one million people lost their jobs in Canada during March 2020~\cite{covid3}, 1.4 million jobs lost in Australia~\cite{covid4}, and a projected global 3\% GDP loss~\cite{covid5}), resulting in a global recession as predicted by many experts~\cite{covid5,covid6,covid7}. In such context, there is an urgent need for solutions to contain the disease spread, thereby reducing its negative impacts and buying more time for pharmaceutical solution development.
	
\begin{figure}[h]
		\includegraphics[width=.5\textwidth]{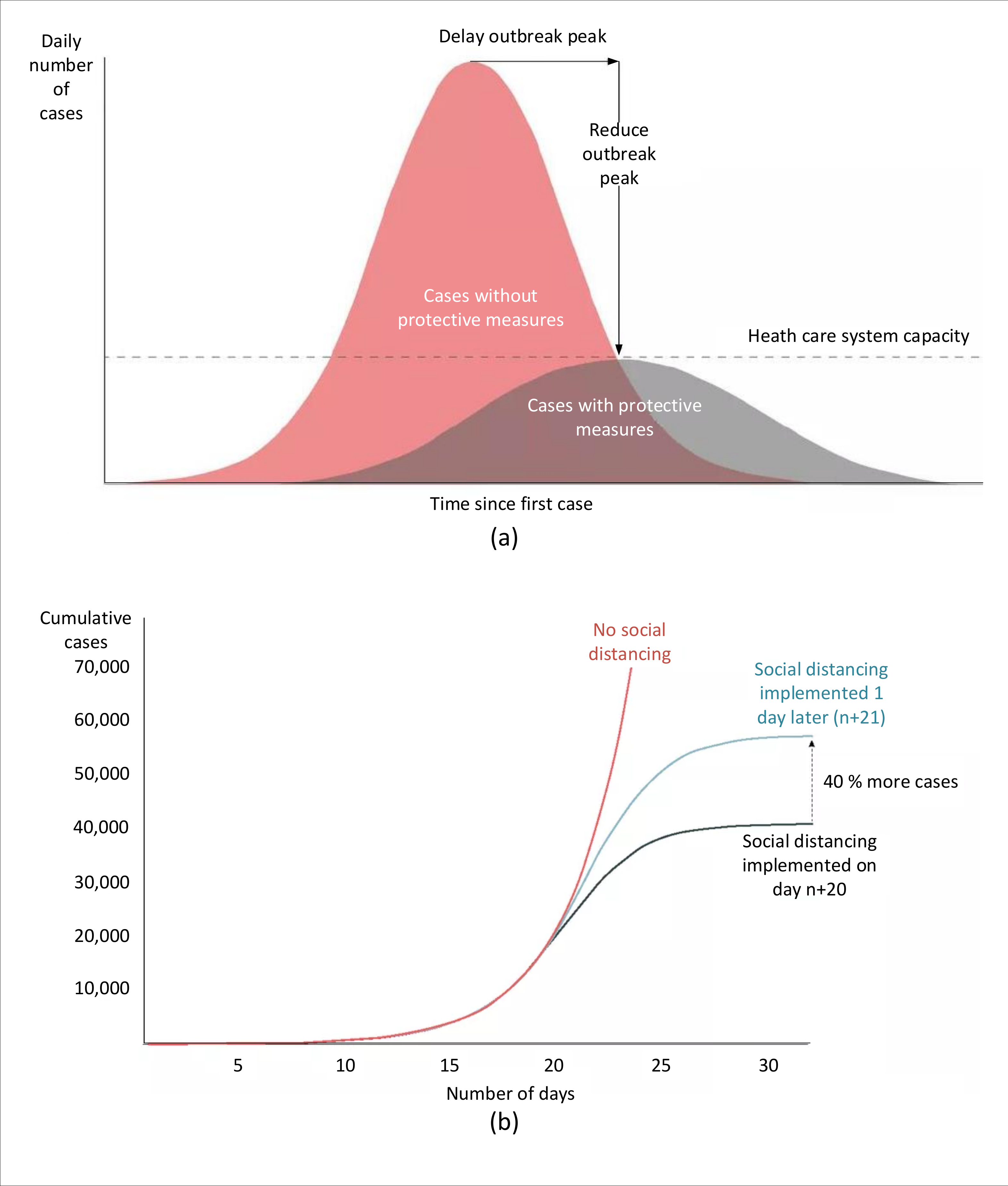}
		\centering
		\caption{Effects of social distancing~\cite{covid8}.}
		\label{Fig:effect}
	\end{figure}

	In the presence of contagious diseases such as SARS, H1N1, and COVID-19, social distancing is an effective non-pharmaceutical approach to limit the disease transmission~\cite{SDdef1,SDiso1,SDMath2}. Social distancing refers to measures that minimize the disease spread by reducing the frequency and closeness of human physical contacts, such as closing public places (e.g., schools and workplaces), avoiding mass gatherings, and keeping a sufficient distance amongst people~\cite{SDdef1,SDdef2}. By reducing the probability that the disease can be transmitted from an infected person to a healthy one, social distancing can significantly reduce the disease's spread and severity. If implemented properly at the early stages of a pandemic, social distancing measures can play a key role in reducing the infection rate and delay the disease's peak, thereby reducing the burden on the healthcare systems and lowering death rates~\cite{SDdef1,SDiso1,SDMath2}. Fig.~\ref{Fig:effect} illustrates the effects of social distancing measures on the daily number of cases~\cite{covid8}. As can be observed in Fig.~\ref{Fig:effect}(a), social distancing can reduce the peak number of infected cases~\cite{SDiso1} to ensure that the number of patients does not exceed the public healthcare capacity. Moreover, social distancing also delays the outbreak peak~\cite{SDiso1} so that there is more time to implement countermeasures. Furthermore, social distancing can reduce the final number of infected cases~\cite{SDiso1}, and the earlier social distancing is implemented, the stronger the effects will be as illustrated in Fig.~\ref{Fig:effect}(b)~\cite{covid8}.
	
	During the ongoing COVID-19 pandemic, many governments have implemented various social distancing measures such as travel restrictions, border control, closing public places, and warning their citizens to keep a 1.5-2 meters distance from each other when they have to go outside~\cite{intro1,intro2,intro3}. Nevertheless, such aggressive and large-scale measures are not easy to implement, e.g., not all public spaces can be closed, and people still have to go outside for food, healthcare, or essential work. In such context, technologies play a key role in facilitating social distancing measures. For example, wireless positioning systems can effectively help people to keep a safe distance by measuring the distances among people and alerting them when they are too close to each other. Moreover, other technologies such as Artificial Intelligence (AI) technologies can be used to facilitate or even enforce social distancing.

	In this article, we present a comprehensive survey on enabling and emerging technologies for social distancing. The main aims are to provide a comprehensive background on social distancing as well as effective technologies that can be used to facilitate the social distancing practice. In particular, we first present basic concepts of social distancing together with its measurements, models, effectiveness, and practical scenarios. After that, we review enabling wireless technologies which are especially effective in monitoring and keeping distance amongst people. Then, we discuss various emerging technologies, e.g., AI, thermal, computer vision, ultrasound, and visible light, which have been introduced recently in order to address many new issues related to social distancing, e.g., contact tracing, quarantined people detection and monitoring, and symptom prediction. Finally, some important open issues and challenges (e.g., privacy-preserving, scheduling, and incentive mechanisms) of implementing technologies for social distancing will be discussed. Furthermore, potential solutions together with future research directions are also highlighted and addressed. 
	\begin{figure*}[h]
		\centerline{\includegraphics[width=.99\linewidth]{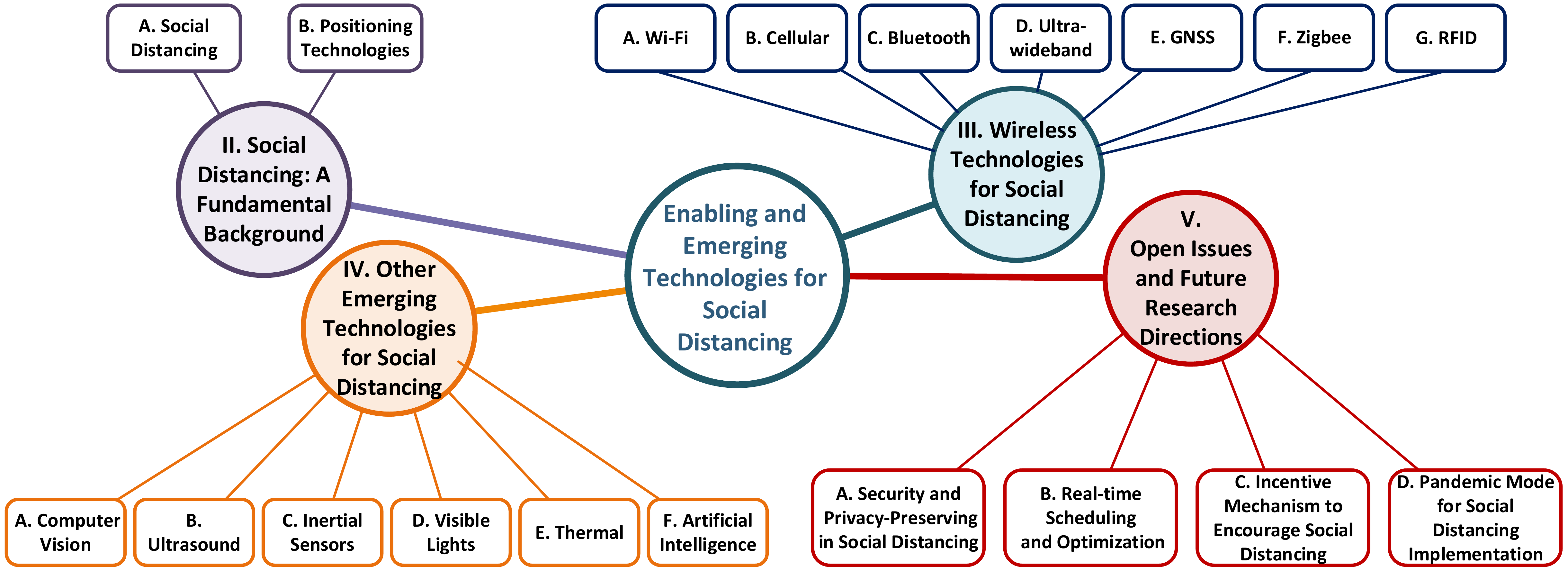}}
		\caption{The organization of this survey.}
		\label{fig:intro}
	\end{figure*}
	
	Although there are few surveys related to localization and positioning systems, e.g.,~\cite{intro4,intro6,intro7,intro10}, to the best of our knowledge, this is the first survey in the literature discussing technologies for social distancing. It is worth noting that, due to the increasingly complex development of many types of viruses as well as the rapid growth of social interaction and globalization, the concept of social distancing is not as simple as physical distancing. In fact, it also includes many non-pharmaceutical interventions or measures taken to prevent the spread of contagious diseases, such as monitoring, detection, and warning people (as we identify and propose in Table~\ref{tab:Scenarios}). Thanks to the significant development of emerging technologies, e.g., future wireless systems, AI, and data analytics, many new solutions have been introduced recently which can create favorable conditions for practicing social distancing.

	As illustrated in Fig.~\ref{fig:intro}, the rest of this paper is organized as follows. We first provide a brief overview of social distancing and distance measurement methods in Section~\ref{background}. Then, Section~\ref{wireless} and Section~\ref{other} discuss enabling wireless technologies and other emerging technologies for social distancing, respectively. After that, we discuss open issues and future research directions of technology-enabled social distancing in Section~\ref{open}, and conclusions are given in Section~\ref{conclusion}.

	\section{Social Distancing: A Fundamental Background}
	\label{background}
	\subsection{Social Distancing}
	\subsubsection{Definition and Classifications}
	Social distancing refers to the non-pharmaceutical measures to reduce the frequency of physical contacts and the contact distances between people during an infectious disease outbreak~\cite{SDdef3}. Social distancing methods can be classified into public and individual measures. Public measures include closing or reducing access to educational institutions and workplaces, canceling mass gatherings, travel restrictions, border control, and quarantining buildings. Individual measures consist of isolation, quarantine, and encouragement to keep physical distances between people~\cite{SDdef2}. Although these measures can cause some negative impacts on the economy and individual freedom, they play a crucial role in reducing the severity of a pandemic~\cite{SDdef3}.
	\subsubsection{Measurements and Models}
	The evaluation of social distancing measures is often based on several standardized approaches. One of the main criteria for social distancing measures selection is the basic reproduction number $R_o$ which represents on average how many people a case (i.e., an infectious person) will infect during its entire infectious period~\cite{SDMath1}. For example, $R_o<1$ indicates that every case will infect fewer than 1 person, and thus the disease is declining in the considered population. Since the value of $R_o$ represents how quickly the disease is spreading, $R_o$ has been one of the most important indicators for social distancing measures selection~\cite{SDMath2,SDiso1}. Mathematically, $R_o$ can be determined by
	\begin{equation}
	\label{R}
	R_o=\int_{0}^{\infty}b(a)F(a)da,
	\end{equation}
	where $b(a)$ is the average number of new cases an infectious person will infect per unit of time during the infectious period $a$, and $F(a)$ is the probability that the individual will remain infectious during the period $a$~\cite{SDMath1}.
	
	Beside showing the transmissibility of a disease, $R_o$ also gives some intuitive ideas on how to limit the disease spread. As observed from~\eqref{R}, $R_o$ can be reduced in different ways, i.e., to decrease $b(a)$ or $F(a)$. To reduce $b(a)$, there are several approaches such as to lower the number of contacts the infected individuals make per unit of time (e.g., avoid mass gatherings and public places closures) or to reduce the probability that a contact will infect a new person (e.g., by wearing masks). To reduce $F(a)$, the infected person needs to be cured or completely avoid contacts with the non-infected (e.g., isolation and quarantine).

	\subsubsection{Effectiveness}
	To evaluate the effectiveness of social distancing, a common approach is to measure the attack rate which is the percentage of infected people in a susceptible population (where no one is immune at the beginning of the disease) at the time of measurement~\cite{SDworkplace1}. The attack rate reflects the severity of a disease at a given time, and thus it has different values during the disease outbreak. Among these values, the peak attack rate is often considered and compared to the current healthcare capacity (e.g., intensive care unit capacity) to see the current system's ability to handle the peak number of patients. After the outbreak is over, data is often collected to determine the final attack rate which is the total number of infected cases over the entire course of the outbreak divided by the total population. 
	
	Social distancing measures are proven to be effective when implemented properly~\cite{SDworkplace1,SDworkplace2,SDworkplace3,SDschool1,SDschool2,SDiso1,SDquaran1}. Different types of social distancing measures may have diverse levels of effectiveness on the disease spread. In~\cite{SDworkplace1}, the effect of social distancing measures at workplaces is evaluated by an agent-based simulation approach. In particular, six different workplace strategies that reduce the number of workdays are simulated. The results show that, for a seasonal influenza ($R_o=1.4$), reducing the number of workdays can effectively reduce the final attack rate (e.g., up to 82\% if three consecutive workdays are reduced). Nevertheless, in a pandemic-level influenza ($R_o=2.0$), reducing the number of workdays has a significantly weaker impact, i.e., $3\%$ (one extra day off) to $21\%$ decrease (three extra consecutive days off). Several other studies present similar results. In~\cite{SDworkplace2}, it is shown that workplace social distancing can reduce the final attack rate by up to $39\%$ in a $R_o=1.4$ setting. Similarly,~\cite{SDworkplace3} shows that different types of measures can reduce the attack rate from 11\% to 20\% depending on the frequency of contacts among the employees.
	
	For school closure measures, studies also show positive effects. In~\cite{SDschool1}, a modeling technique is employed to examine the effects of four different social distancing measures under three varying $R_o$ settings. Among different types of measures, the school closure measure is shown to be able to reduce the final attack rate by $20\%$, $10\%$, and $5\%$, and the peak attack rate by $77\%$, $47\%$, and $32\%$ in the cases where $R_o<1.9$, $2.0 \leq R_0 \leq 2.4$, and $R_o>2.5$, respectively. Similarly, it is shown in~\cite{SDschool2} that prolonged school closure in a pandemic context can reduce the final attack rate by up to $17\%$ and the peak attack rate by up to $45\%$. 
	
	Another common social distancing measure is the isolation of the infected cases and cases with similar symptoms. In~\cite{SDiso1}, large-scale epidemic simulations are performed to evaluate different strategies for influenza pandemic mitigation. Among the simulated strategies, the results show that the proper implementation (such that an isolated individual reduces 90\% of its contact rate) of isolation can reduce the final attack rate by 7\% in a $R_o=2$ setting. Similarly, it is shown in~\cite{SDschool1} that isolation can reduce the final attack rate by $27\%$, $7\%$, and $5\%$, and the peak attack rate by $89\%$, $72\%$, and $53\%$ in the cases where $R_o<1.9$, $2.0 \leq R_0 \leq 2.4$, and $R_o>2.5$, respectively. 
	
	For household quarantines, studies have shown that this measure can be effective if the compliance level is sufficient. In~\cite{SDiso1}, the effects of voluntary quarantine of household for a duration of 14 days are examined. Simulations are carried out with the assumption that 50\% of households will comply, which leads to a 75\% reduction of external contact rates, while the internal contact rate will increase by 100\%. The results show that this measure can reduce the final attack rate by up to 6\% and the peak attack rate by up to 40\%. Similarly, in~\cite{SDquaran1}, simulations are performed to examine the impacts of different measures. For household quarantines, the result shows that this measure can reduce the final attack rate by 31\% and the peak attack rate by 68\% with $R_o=1.8$ and a compliance rate of 50\%.
	
	Apart from the abovementioned measures, the effectiveness of the other social distancing measures either received limited attention or was often considered in combination with another approach. In~\cite{SDiso1}, the effectiveness of travel restrictions and border control measures are examined. However, the results only show that different levels of travel restrictions (from 90\% to 99.9\%) can delay the peak attack rate by up to six weeks, while how travel restrictions affect the attack rate is not examined. Another type of measure that does not receive much attention is community contact reduction measures (e.g., avoid crowds and mass gatherings cancellation). In~\cite{SDschool1}, it is shown that this type of measure can reduce the final attack rate by $17\%$, $14\%$, and $10\%$, and the peak attack rate by $72\%$, $49\%$, and $38\%$ in the cases where $R_o<1.9$, $2.0 \leq R_o \leq 2.4$, and $R_o>2.5$, respectively.
	
	When combined together, social distancing measures are proven to be even more effective~\cite{SDschool1,SDiso1,SDcomb1}. It is shown in~\cite{SDschool1} that when all four measures, i.e., school closure, isolation, workplace nonattendance, and community contact reduction, are in effect, they can drastically reduce the attack rates in all the considered $R_o$ settings. In particular, the final attack rate can decrease from 65\% to only 3\% and the peak attack rate from 474 cases per 10 thousand to only five cases, in the highest $R_o$ setting. Similarly,~\cite{SDiso1} examines the effects when household quarantines, workplace closures, border control, and travel restrictions are combined. The results show that the final and peak attack rates are three times and six times, respectively, lower than when no policy is implemented. Moreover, the peak attack rate can be delayed by nearly three months in a $R_o=1.7$ setting. In~\cite{SDcomb1}, it is also shown that when four types of measures (i.e., school closure, household quarantines, workplace nonattendance, and community contact reduction) are in effect, the final attack rate can be reduced 3-4 times depending on $R_o$.
	
	There are several studies focusing on the negative impacts of social distancing. In~\cite{SDcost1}, simulations are performed to evaluate the benefit and cost of different social distancing strategies. In this study, simulations are carried out without and with social distancing under different caution levels settings. Simulation results are evaluated based on the benefits of the reduced infection rate and the economic cost of reducing contacts. The main finding of this work is that a favorable result can only be obtained by implementing social distancing measures with a high caution level. Since the economic cost is also considered, it is shown that implementing social distancing with an insufficient caution level gives worse results than that of the case without social distancing. In~\cite{SDcost2}, a game theoretical approach based on the classic SIR model is proposed to evaluate the benefits and costs of social distancing measures. Interestingly, the results show that in the case where $R_o<1$, the equilibrium behaviors include no social distancing measures. Moreover, social distancing measures are shown to achieve the highest economic benefit when $R_o\approx2$. 
	
\begin{figure}[h]
		\includegraphics[width=.5\textwidth]{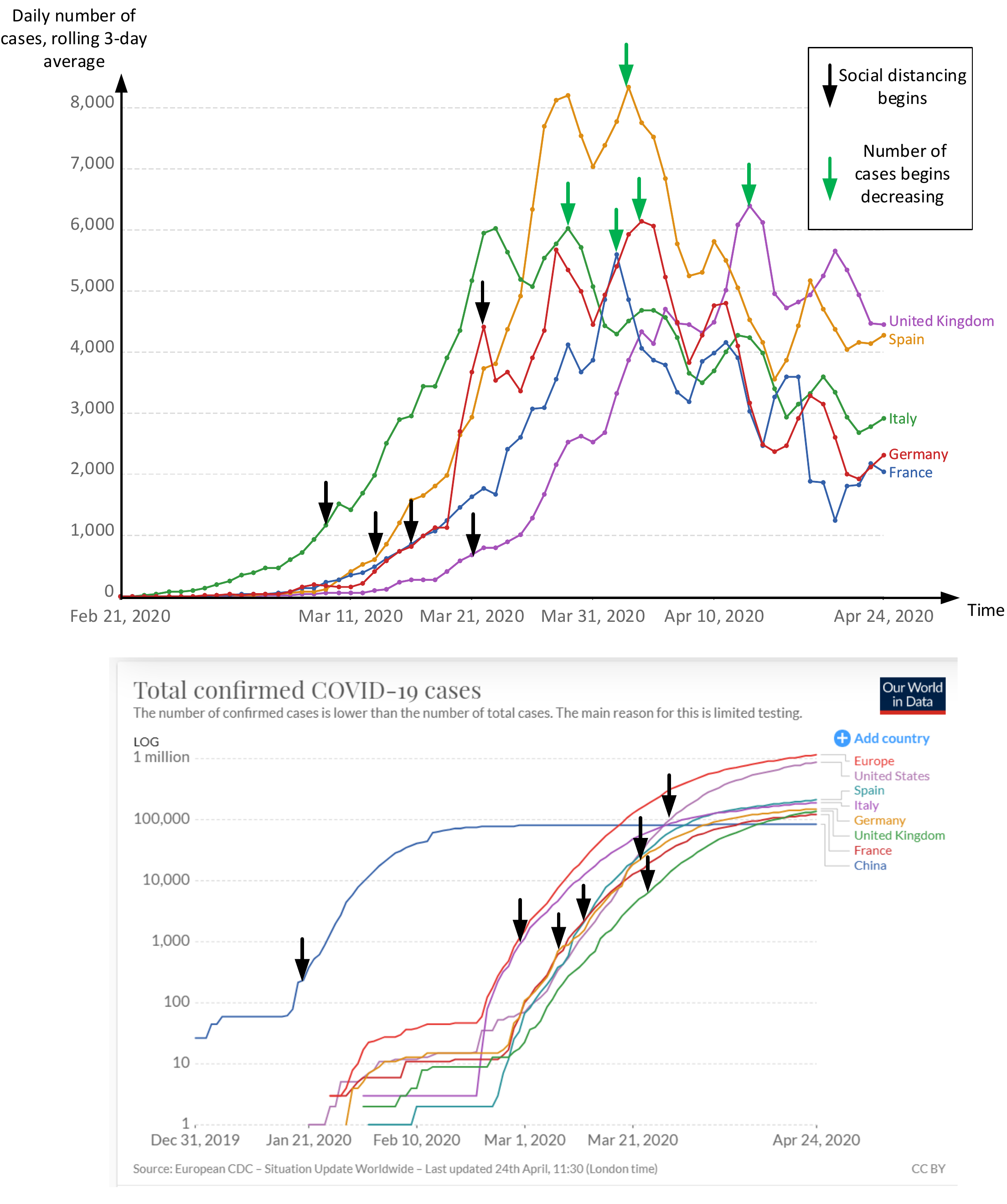}
		\centering
		\caption{Real-world effects of social distancing~\cite{SDreal}.}
		\label{Fig:real}
	\end{figure}

	In the current COVID-19 pandemic, the World Health Organization (WHO) estimates that the value of $R_o$ would be in the range of 2-2.5~\cite{SDesti1}. As can be seen from the abovementioned studies, social distancing measures can play a vital role in mitigating this pandemic with such $R_o$ values. For example, Fig.~\ref{Fig:real} illustrates the rolling 3-day average of daily new confirmed COVID-19 cases in several countries~\cite{SDreal}. Generally, after a country began implementing social distancing (e.g., lockdown at different levels) for 13-23 days, the daily number of new cases begins to drop. As can be also seen from the second graph, the curves representing the total number of cases become less steep after social distancing is implemented (i.e., flattening the curve).
	
	Despite its significant potential, it can be observed that social distancing is very effective only when applied properly. Nevertheless, it is not easy to implement because of many reasons such as the negative economic impacts, personal freedom violation, and difficulties in changing people's behaviors. Thus, technologies can play a key role in facilitating social distancing, which will be discussed in the next sections.
	
	\subsubsection{Practical Scenarios}
	The practical social distancing scenarios identified/proposed in this survey are categorized and illustrated in Fig.~\ref{Fig:scen}. More specific scenarios are summarized in Table~\ref{tab:Scenarios}. The scenarios can be briefly classified as follows:
	\begin{figure*}[h]
		\includegraphics[width=\textwidth]{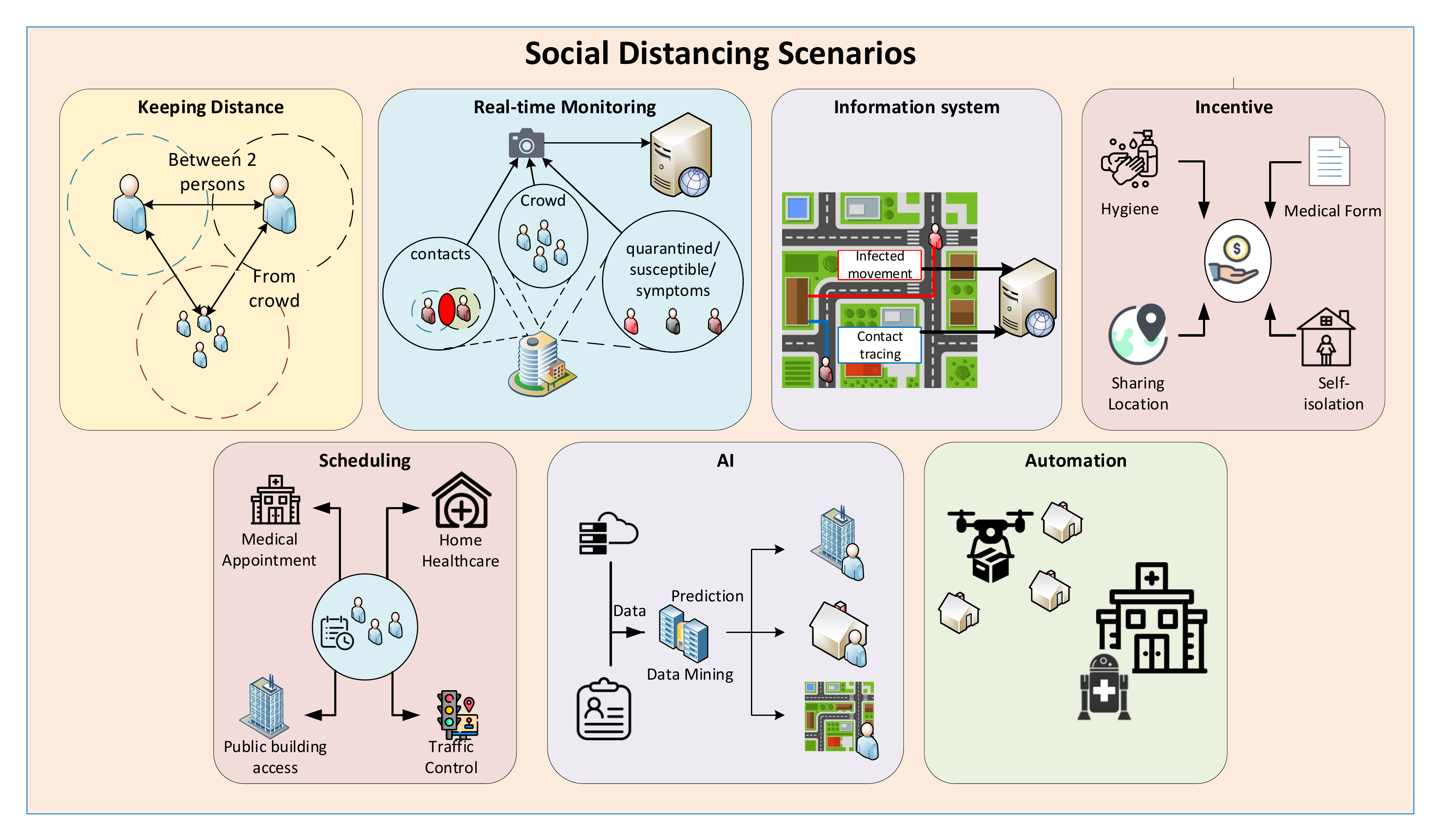}
		\centering
		\caption{Practical social distancing scenarios.}
		\label{Fig:scen}
	\end{figure*}

\begin{figure*}[]
		\includegraphics[width=.9\textwidth]{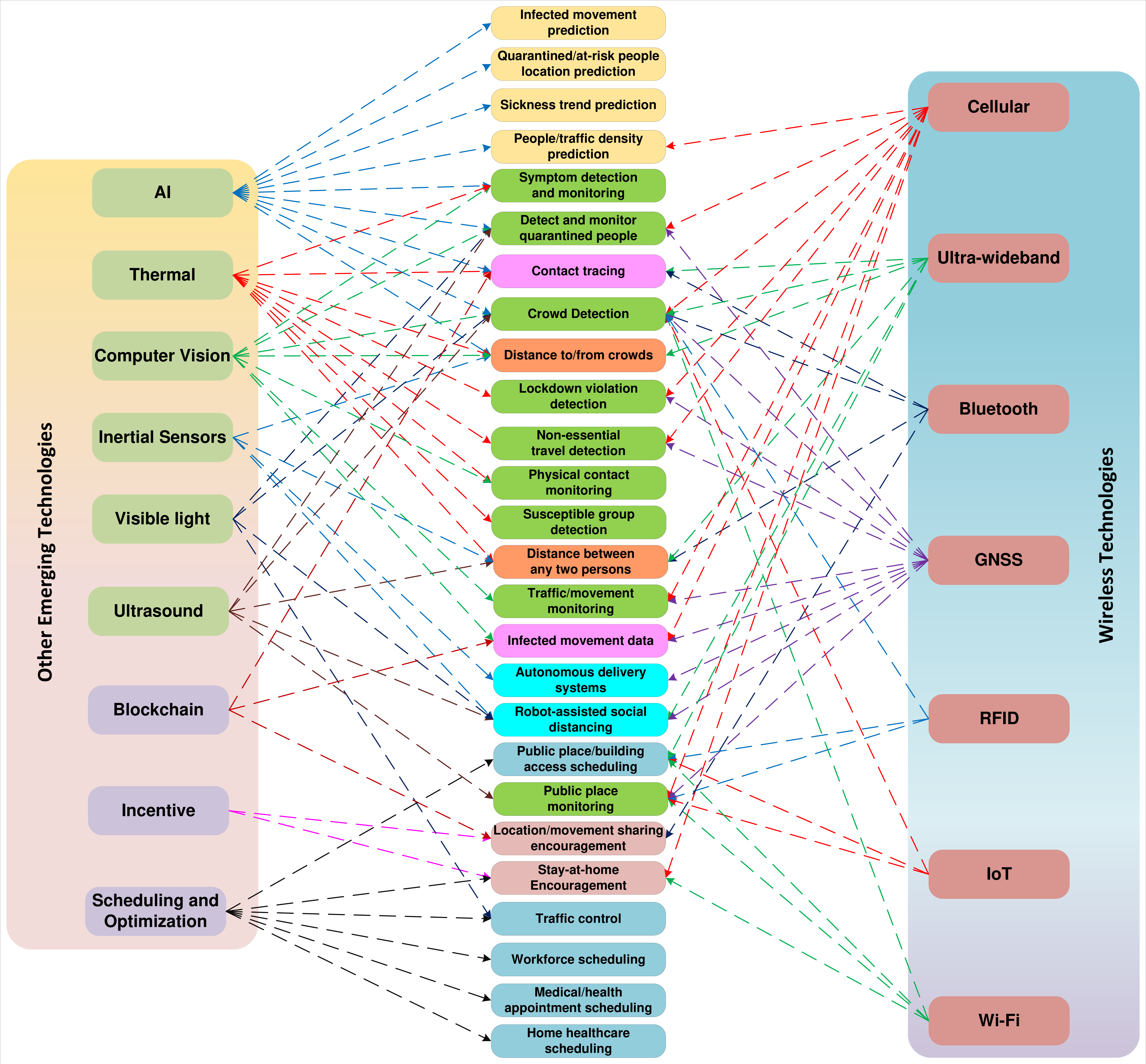}
		\centering
		\caption{Application of technologies to different social distancing scenarios.}
		\label{Fig:scen2}
	\end{figure*}

	\begin{itemize}
		\item \textit{Keeping distance}: In these scenarios, various positioning and AI technologies can assist in keeping sufficient distance (e.g., 1.5m apart) between people. Based on that, when a person gets too close to another or a crowd, the person can be alerted (e.g., by smartphones).
		\item \textit{Real-time monitoring}: Many wireless and related technologies can be utilized to monitor people and public places in real-time (without compromising citizens' privacy). The purposes of such monitoring are to gather meaningful data (e.g., numbers of people inside buildings, contacts, symptoms, crowds, and social distancing measures violations) to facilitate social distancing. Based on these data, appropriate measures can be carried out (e.g., limit access to buildings when there are too many people inside, avoid crowds, and alert/penalize violations).
		\item \textit{Information system}: Technologies such as Bluetooth, Ultra-wideband, Global Navigation Satellite Systems (GNSS), and thermal can be employed to collect the trajectory data of the infected individuals and the contacts that these individuals made. Based on this information, susceptible people who were at the same place or had contacts with the infected ones can take cautious actions (e.g., self-isolation, and test for the disease).
		\item\textit{Incentive}: Social distancing has negative impacts on personal freedom and the economy. Therefore, incentive mechanisms are needed to encourage people to comply with social distancing measures (e.g., incentivize people to share their movement data and self-isolate). Optimization techniques and technologies such as Bluetooth, Wi-Fi, and cellular together with economics tools like game theory, auctioning, and contract theory can facilitate those incentive mechanisms.
		\item \textit{Scheduling}: Various scheduling techniques can be employed to increase the efficiency of workforce and home healthcare service scheduling, thereby decreasing the number of employees at workplaces and patients at hospitals. Moreover, scheduling techniques can also be applied for traffic control to reduce the number of vehicles and pedestrians on the street. Furthermore, technologies such as Wi-Fi, Radio frequency identification (RFID), and Zigbee can be applied for building access scheduling.
		\item \textit{Automation}: In the social distancing context, autonomous vehicles such as medical robots and unmanned aerial vehicle (UAV) can be utilized to reduce the need for human presence in essential tasks, e.g., medical procedures and delivery services. Technologies such as ultra-wideband, GPS, ultrasound, and inertial sensors can be leveraged for the positioning and navigation of these autonomous vehicles.
		\item \textit{Modeling and Prediction}: AI technologies can be employed for pandemic data mining. The results can help to predict the future trends and movement of the infected and susceptible individuals. Moreover, AI-based classification algorithms can be leveraged to detect disease symptoms in public places.
	\end{itemize}
	The applications of technologies to specific social distancing scenarios are illustrated in Fig.~\ref{Fig:scen2}.

	\begin{table*}[!h]
		\caption{Practical Social Distancing Scenarios}
		\begin{tabular}{|>{\raggedright\arraybackslash}m{1.5cm}|>{\raggedright\arraybackslash}m{3cm}|>{\raggedright\arraybackslash}m{5cm}|>{\raggedright\arraybackslash}m{3.3cm}|>{\raggedright\arraybackslash}m{3cm}|}
			\hline
			
			& \multicolumn{1}{|>{\centering\arraybackslash}m{3cm}|}{\multirow{1}{*}{\textbf{Scenarios}}} & \multicolumn{1}{|>{\centering\arraybackslash}m{5cm}|}{\multirow{1}{*}{\textbf{Description}}} & \multicolumn{1}{|>{\centering\arraybackslash}m{3.3cm}|}{\multirow{1}{*}{\textbf{Technologies}}}   & \multicolumn{1}{|>{\centering\arraybackslash}m{3cm}|}{\multirow{1}{*}{\textbf{References}}} \\ \hline \hline
			\multirow{2}{*}{\begin{tabular}[c]{@{}l@{}}Keeping \\ Distance\end{tabular}}& Distance between any two people & Detect and monitor the distance between any two people &Bluetooth, Ultrasound, Thermal, Inertial, Ultra-wideband&\cite{Wang2013Bluetooth,Activebat_1,Firefly_1,DecaWave,INS2}\\ \cline{2-5}
			 & Distance to/from crowds                                                            & Alert when approaching a crowd                                                                                                                                                                                              &     AI, Thermal, Inertial, Ultra-wideband, Computer Vision   &\cite{BeSpoon, Jia:2016,hybrid:smartphone,Optotrak,INS2}                \\ \hline
			\multirow{9}{*}{\begin{tabular}[c]{@{}l@{}}Real-time \\ Monitoring\end{tabular}}    & Public place monitoring & Monitor and gauge the number of people inside/at a public place&                    Wi-Fi, RFID, Zigbee, Cellular, GNSS, Ultrasound& \cite{Choi2017,Bartoletti2017}       \\ \cline{2-5} 
			& Physical contact  monitoring& Monitor physical contacts, e.g., handshakes, hugs, between people&       Computer Vision, Thermal    & \cite{Firefly_1,Optotrak}             \\ \cline{2-5} 
			&Symptom detection and monitoring& Detect and monitor sickness symptoms, e.g., body temperature, coughs &Computer Vision, Thermal, AI	& \cite{vision:cough}  \\ \cline{2-5} 
			& Susceptible group detection& Monitor highly susceptible groups &         Thermal      	&\cite{Thermal_temp,Thermal_temp2}           \\ \cline{2-5} 
			&Detect and monitor quarantined people & Detect and monitor quarantine people (e.g., for complying/violating the isolation/quarantine requirement)& AI, Cellular, Ultrasound, Visible Lights, Computer Vision 
		  & \cite{Brighente:2019,vision:deepface}              \\ \cline{2-5} 
			& Crowd detection & Detect crowds/gatherings in public places   & Wi-Fi, Bluetooth, RFID, Zigbee, AI, Ultra-wideband, Cellular, GNSS, Computer Vision, Visible Light &  \cite{Luo2019dynamic,Agiwal2016,DecaWave,Wang2015RSSI,Luoh2013Zigbee,Xiao2018One}\\ \cline{2-5} 
			&Non-essential  travel detection & Using location information to determine if the trip is essential (e.g., medical facilities and gasoline stations)  or not (e.g., restaurants and cinemas) &                 Cellular, GNSS, Thermal   & \cite{travel,quantify,Thermal_Trac}         \\ \cline{2-5} 
			& Traffic/movement monitoring                    &  Detect the vehicles on the street when isolation measures are in effect                                                                                                    &                    GNSS, Vision, Cellular  &  \cite{travel,quantify,Koivisto2017}     \\ \cline{2-5} 
			&  Lockdown violation detection  &  Detect violations of public place's closure or lockdown.  &        GNSS, Cellular, Thermal        	&  \cite{travel,quantify,Koivisto2017,Thermal_camera} \\ \hline
			\multirow{2}{*}{\begin{tabular}[c]{@{}l@{}}Information \\ System\end{tabular}}    & Infected movement data  & Track the infected people's movement to notify susceptible people who were at the same places &  Cellular, Blockchain, GNSS, Computer Vision & \cite{hybrid:position10,hybrid:indoor1,hybrid:indoor4,hybrid:indoor6}                 \\ \cline{2-5} 
			& Contact tracing                                                                    &  Trace the contacts that an infected individual made      &  Bluetooth, Blockchain Ultra-wideband, Thermal, AI &  \cite{News3,DecaWave,Firefly_1,Jia:2016} \\ \hline
			\multirow{2}{*}{Incentive}	  & Location/movement sharing encouragement  & Encourage people to share their movement data        &  Bluetooth, Blockchain, incentive mechanism  &   \cite{News3,PACT,Tian:2017}           \\ \cline{2-5} 
			& Stay-at-home encouragement                                                                & Incentivize people to stay home                                                                                                                                                               &  Wi-Fi, Cellular, incentive mechanism   &  \cite{Tan2018,Poularakis2016,Rahman:2017,Wang:2019,opti3} \\ \hline
			\multirow{5}{*}{Scheduling}                                                       &  Workforce scheduling                  &  Limit the number of people at the workplaces                                                                                                                   &               Scheduling     &  \cite{sched1,sched2,sched4,sched3}    \\ \cline{2-5} 
			& Medical/health appointment scheduling                 &  Schedule medical appointments to reduce the number of patients &Scheduling & \cite{sched7,sched8,sched9,sched10,sched11}  \\ \cline{2-5} 
			&  Home healthcare scheduling               &  Optimize home healthcare services to reduce the number of patients at the hospitals                                                                                               &              Scheduling  & \cite{sched12,sched13,sched14,sched15,sched16}            \\ \cline{2-5} 
			&  Public place/building access scheduling & Control the number of people inside public buildings                                                                                                                              &              Ultra-wideband, Wi-Fi, RFID, Zigbee   &   \cite{Xiao2018One,Chang2017FitLoc,Luoh2013Zigbee,Bartoletti2017}  \\ \cline{2-5} 
			& Traffic control                                                                    &  Regulate and reduce vehicles and pedestrians density                                                                                                                            &                Visible Light, Scheduling & \cite{sched17,sched18,Jin2017}          \\ \hline
			\multirow{2}{*}{Automation}                                                       &Robot-assisted social distancing &  Improve positioning and navigation of  robots, especially medical robots inside  hospitals  &  Ultra-wideband, GNSS, Visible Lights, Inertial, Ultrasound,    &  \cite{INS9,INS11,GNSS_CHinaserv,Medical_robot,Guan2020} \\ \cline{2-5} 
			&  Autonomous delivery systems (e.g., UAVs, ...) &  Reduce the number of people going outside (food, merchandise, etc., delivery)  &      GNSS, Inertial & \cite{GNSS_Drone,INS12,INS13}          \\ \hline
			\multirow{5}{*}{\begin{tabular}[c]{@{}l@{}}Modeling\\and\\ Prediction\end{tabular}} &  Infected movement prediction  &  Predict infected people's movement &                 AI   	& \cite{Cho:2016}  \\ \cline{2-5} 
			&  Quarantined/at-risk people location prediction  & Predict quarantined and at-risk people's current location to enforce them stay at isolation/protection facility &            AI  & \cite{Brighente:2019,Yin:2020}    \\ \cline{2-5} 
			& People/traffic density prediction  & Predict people density and traffic density &             Cellular, AI  & \cite{Polese:2019,Alawe2018,Wang2019,Sciancalepore2017,Bega2019,Sun2017}   
			\\ \cline{2-5} 
			&  Sickness trend prediction   &  Predict sickness trends in specific areas     &   AI & \cite{Hossain:2020}     \\ \hline
		\end{tabular}
		\label{tab:Scenarios}
	\end{table*}
	
	\subsection{Positioning Technologies}
	Since the main principle of social distancing is to increase the distances of human contacts, approaches to determine the positions and measure the distance between people can play a vital role in facilitating social distancing measures. Using ubiquitous technologies, such as Wi-fi, cellular, and GNSS, positioning (localization) systems are crucial to many practical social distancing scenarios such as distance keeping, public places monitoring, contact tracing, and automation.
	\subsubsection {Overview of Positioning Systems}

	Fig.~\ref{Fig:process} illustrates the general process and several popular methods of a positioning system~\cite{Posi1}. Generally, a positioning system aims to continuously track the position of an object in real-time~\cite{intro10}. To achieve this goal, firstly, signals are transmitted from the target to the receiving nodes (e.g., sensors). From the received signals, useful properties such as arrival time, signal direction, and signal strength (depending on the measurement methods) are extracted in the signal measurement phase. Based on these features, the position of the target can be calculated using various methods in the position calculation phase~\cite{Posi1}. Several effective signal measurements and position calculation methods are presented in the rest of this section.

	\subsubsection{Signal Measurements}
	Typical signal measurement methods can be classified based on the extracted property of the received signal. Among them, time-based methods use the arrival time of the signal to determine the distance between the receiving nodes and the target~\cite{Posi1}. Time-based methods can be further classified as follows:
	\begin{itemize}
		\item \textit{Time-of-Arrival} (TOA)~\cite{Posi4}: This method determines the distance $D$ between the receiving node and the target based on the time it takes for the signal to travel from the target to the node, i.e.,
		\begin{equation}
		\label{TOA}
		D=ct,
		\end{equation}
		where $c$ is the speed of the signal transmission and $t$ is the time for the signal to reach the receiving node. 
		\item \textit{Time Difference-of-Arrival} (TDOA)~\cite{Posi4}: This method uses two kinds of signal with different speeds and calculates $D$ based on the difference between them, i.e.,
		\begin{equation}
		\label{TDOA}
		\dfrac{D}{c_1}-\dfrac{D}{c_2}=t_1-t_2,
		\end{equation}
		where $c_1$, $c_2$, $t_1$, and $t_2$ are the speeds and arrival time of the two signals, respectively.
		\item\textit{ Round Trip Time }(RTT)~\cite{Posi1}: The RTT method measures the duration in which the signal travels to the targets and comes back, i.e.,
		\begin{equation}
		\label{RTT}
		D=\dfrac{t_{RT}-\Delta t}{2},
		\end{equation}
		where $t_{RT}$ is the time of the whole round trip, and $\Delta t$ is the predetermined delay between when the target receives the signal and when the target starts sending back.
	\end{itemize}
	A common disadvantage of the TOA and TDOA methods is that they require synchronized clocks at the node and the target to determine $t, t_1$ and $t_2$. That may be costly to be implemented as it requires frequent calibrations to maintain the accuracy. Although the RTT method does not require clock synchronization, it needs to acquire the delay $\Delta t$ which cannot be predicted in many circumstances~\cite{Posi2}. Consequently, extra efforts are needed to determine $\Delta t$.
	
	Unlike the time-based methods, the \textit{Angle-of-Arrival} (AOA) method determines $D$ by measuring the angle of the incoming signals by using directional antennas or array of antennas. The measured angles can then be used in the \textit{triangulation} method to geometrically determine the target position. However, the main disadvantage of this method is that it requires extra directional antennas which are costly to implement~\cite{Posi1}.
	
	The \textit{Received Signal Strength Indicaton} (RSSI) method measures the attenuation of the signals to determine the distance. Typically, the relationship between the RSSI and distance can be formulated as follows~\cite{Mazuelas2019Robust}:
	\begin{equation}
	P_R = \alpha - 10 n \log_{10}(d) + X,
	\end{equation}
	where $P_R$ is the RSSI value at the receiver (e.g., access point), $d$ represents the distance from the user device to the access point, $X$ is a random variable (caused by the shadowing effect) which follows the Gaussian distribution with zero mean. $\alpha$ is a constant value which can be known in advance and depends on fading, antennas gain, and emitted power of the user device. $n$ is the path loss exponent which depends on the channel environment between each user device and the access point. Thus, based on the RSSI level of the received signals, the access point can estimate the position of the user device in indoor environments.
	
	\subsubsection{Position Calculation}
\begin{figure*}[!t]
		\includegraphics[width=\textwidth]{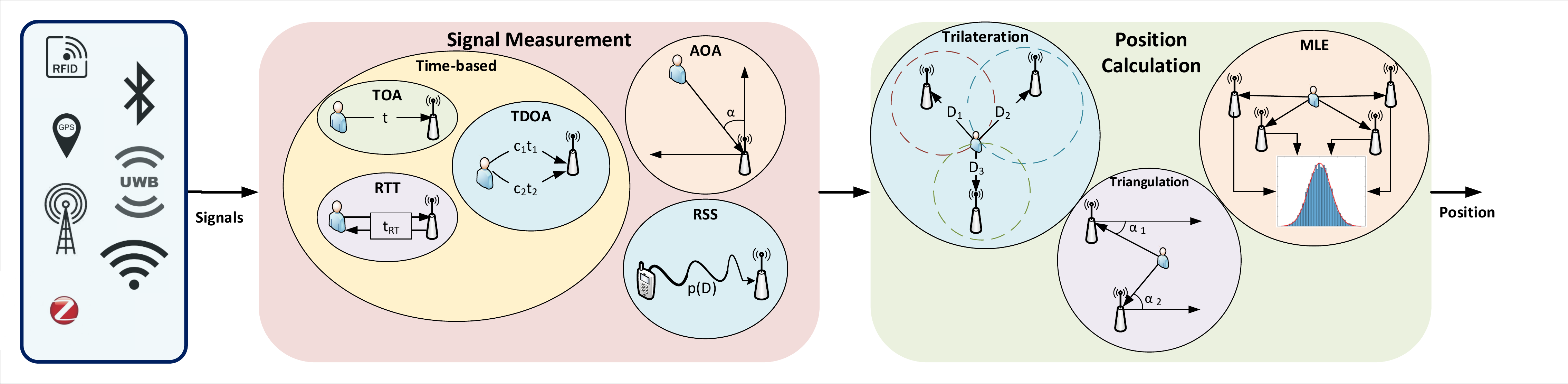}
		\centering
		\caption{General principle of positioning systems.}
		\label{Fig:process}
	\end{figure*}

	Based on the measured signal properties, different methods are employed to calculate the target's position. Among them, \textit{Trilateration} is a common method which uses three reference nodes and the distances between them to the target to calculate the position~\cite{Posi1}, as illustrated in Fig.~\ref{Fig:process}. More specifically, using the coordinates $(x_1,y_1),(x_2,y_2), (x_3,y_3)$ of the reference nodes and the corresponding measured distances $D_1, D_2$, and $D_3$, the coordinate $(x,y)$ of the target can be determined by
	\begin{equation}
	\label{trilateration}
	\begin{cases}
	\sqrt{(x_1-x)^2+(y_1-y)^2}=D_1,\\
	\sqrt{(x_2-x)^2+(y_2-y)^2}=D_2,\\
	\sqrt{(x_3-x)^2+(y_3-y)^2}=D_3.
	\end{cases} 
	\end{equation}
	
	Instead of using distances, the	\textit{Triangulation} method uses the angles of the signal (from the AOA method) to determine the target's position. As illustrated in Fig.~\ref{Fig:process}, if the coordinates of two reference nodes and the corresponding measured angles $\alpha_1, \alpha_2$ are known, the target's position can be geometrically determined~\cite{Posi1}.
	
	To address the uncertainty in measurements, the \textit{Maximum Likelihood Estimation} (MLE) method is often employed. This method utilizes the signal measurements from a number of reference nodes (usually three or more) and applies some statistical approaches such as the minimum variance estimation method~\cite{Posi3} to calculate the target's position while minimizing the impact of noises in the environment~\cite{Posi1}.
	\section{Wireless Technologies for Social Distancing}
	\label{wireless}
	To enable social distancing, many wireless technologies can be adopted such as Wi-Fi, Cellular, Bluetooth, Ultra-wideband, GNSS, Zigbee, and RFID. In this section, we first briefly provide the fundamentals of these technologies and then explain how they can enable, encourage, and enforce people to practice social distancing. After that, we discuss the potential applications, advantages, limitations, and feasibility of these technologies.
	
	\subsection{Wi-Fi}
	
	Due to the fact that Wi-Fi technology is widely deployed in indoor environments, this technology can be considered to be a promising solution to practice social distancing inside multi-story buildings, airports, alleys, parking garages, and underground locations where GPS and other satellite technologies may not be available or provide low accuracy~\cite{Yang2015WIFi}. In a Wi-Fi system, a wireless transmitter, known as a wireless access point (AP), is required to transmit radio signals to communicate with user devices in its coverage area. Currently, Wi-Fi enabled wireless devices are working on the IEEE 802.11 standards. Wi-Fi 6 (based on 802.11ax technology) is the latest version of Wi-Fi standards which provides high-throughput and reliable communications~\cite{WiFi6}. We discuss few example scenarios of social distancing that can be enabled by Wi-Fi as follows.
	
	\subsubsection{Crowd Detection}
	One potential application of Wi-Fi technology in social distancing is positioning~\cite{Mazuelas2019Robust}-\cite{Lim2013Radio}. Based on the location of users, the authority can detect crowds inside a building and force them to maintain a safe distance. This is an essential factor to practice social distancing during a pandemic outbreak in indoor public places such as train stations and airports. There are two main reasons making Wi-Fi technology possible in social distancing. First, due to the convenience of hardware facilities, we can quickly deploy Wi-Fi systems for user positioning with very low cost and efforts~\cite{Wen2019Survey}. Second, with recent advances in Wi-Fi-based indoor positioning, Wi-Fi can provide reliable and precise location services to enable social distancing. The most common and easiest way for indoor positioning is to calculate the user's location based on the RSSI of the received signals from the user device~\cite{Mazuelas2019Robust},~\cite{Naggar2019Indoor}. However, the accuracy of this solution much depends on the propagation model. Thus, in~\cite{Mazuelas2019Robust}, the authors present a new method to dynamically estimate the channel model from the user device to the access point. The key idea of this solution is continuously determining the RSSI values in real-time to obtain the estimated channel model that is close to the real channel model. Once the propagation is estimated, the distance between the access point and the user device can be accurately determined. After that, the user's location will be derived by using the \textit{trilateration} mechanism.
	
	Differently, the authors in~\cite{Naggar2019Indoor} propose to adopt the inertial navigation system (INS) to significantly increase the accuracy of conventional RSSI-based methods. The key idea of this solution is using a Kalman filter to combine and fill the signal database with the INS data. As such, the authors can obtain the average distance error as small as 0.6m. The above RSSI-based solutions can be easily adopted to detect crowds in indoor environments. Then, the local authorities can take appropriate actions to disperse the crowds or suggest other people to not go to the place. For example, if there are too many people in a supermarket, the authorities can notify and recommend new coming customers to go to other supermarkets or come in another time so that they can avoid crowds.
	
	\subsubsection{Crowd Detection in Dynamic Environments}
	Although the RSSI-based solution can detect the user's location with sufficient accuracy, it may not be effective in dynamic and complicated indoor environments such as airports or train stations~\cite{Luo2019dynamic},~\cite{Guo2019Robust},~\cite{Alshami2017Adaptive}. This is due to the effects of non-line of sight (NLOS) on the wireless signals between the user's device and the access point, especially in dynamic and complicated environments in which the wireless signals are greatly scattered by obstacle shadows or people (e.g., running and walking)~\cite{Luo2019dynamic}. Another RSSI-based indoor localization technique is the fingerprinting approach (or radio map) that locates devices based on a previously built database. In particular, this database contains the signal fingerprints corresponding to several access points in a specific area. Nevertheless, collecting fingerprint data is time-consuming and laborious~\cite{Wang2019An}, especially in large areas such as airports or train stations. In addition, it is infeasible to directly apply the pre-obtained fingerprint database to new areas for localization~\cite{Chang2017FitLoc}. The main reason is that the adjustment process to apply the fingerprint database of an area to another is time-consuming and usually requires human intervention.
	
	To address these problems, several solutions~\cite{Chang2017FitLoc,Luo2019dynamic, Guo2019Robust,Wang2019An,Alshami2017Adaptive} are proposed to enable indoor localization in dynamic and complicated areas such as airports and train stations. With these solutions, the authorities can detect crowds and force people to leave to enable social distancing during pandemic outbreaks. Specifically, in~\cite{Luo2019dynamic}, the authors show that when the environment changes, e.g., the presence of people in the line of sight between the user device and the access point, the performance of conventional RSSI-based localization techniques is greatly decreased. Thus, the authors propose an adaptive signal model fingerprinting algorithm to adapt to the dynamic of the environment by detecting users' positions and updating the database simultaneously. In~\cite{Chang2017FitLoc}, the authors propose a new localization technique to locate multiple users in different areas by performing a fine-grained localization. In addition, the authors introduce a transfer mechanism to adjust the fingerprint database over multiple areas to minimize human intervention.
	
	An interesting design is proposed in~\cite{Adib2013See} to locate and track people by using Wi-Fi technology, namely Wi-Vi (stands for Wi-Fi Vision). This technology allows the authorities to track people in indoor environments and detect potential crowds, so that they can take appropriate actions to enable social distancing, e.g., inform people not to go to potentially crowded places. In particular, Wi-Vi uses an MIMO interference nulling to remove reflections from static objects and only focuses on moving objects, e.g., a user. Moreover, the authors propose to consider the movement of a user as an antenna array and then track the user by observing its RF beams. If there are many people having the same direction, e.g., going to the same place, the authorities can notify them to avoid forming crowds. Thus, Wi-Vi can be considered as a promising technology to enable social distancing.
	
	However, to efficiently detect crowds, Wi-Fi-based localization systems may require several transceivers attached to each access point to obtain high accuracy. Another problem is to differentiate between human and machine terminals. To address this problem, fingerprint databases can be used to detect machine terminals which are usually placed at known locations. Nevertheless, this solution may not be feasible if we consider autonomous robots in the environments, and thus can be a potential research direction.
	
	\subsubsection{Public Place Monitoring and Access Scheduling}
	Another way to apply Wi-Fi technology in social distancing is by controlling the number of people inside a building, e.g., supermarket, shopping mall, and university. Specifically, with various Wi-Fi access points implemented inside the building, the number of people currently inside the building can be estimated based on the number of connections from user devices to the access points. Based on this information, several actions can be made, such as forcing people to queue before entering the building to maintain a safe number of people inside the facilities at the same time. Another application is notifying people who want to go to the building. Specifically, based on the number of people inside the building, the authority can encourage/force them to stay home or come at a different time if the place is too crowded. However, the accuracy of this approach depends on many factors such as the number of smart devices one person possesses, how many devices can be connected to a network simultaneously, and whether the user connects to the access point as many people completely rely on their cellular connections.
	
	\subsubsection{Stay-at-home Encouragement}
	Wi-Fi technology can also be used to encourage people to stay at home by detecting the frequency of moving outside their houses for a particular time, e.g., a day. Specifically, when user devices move far away from the access point inside their houses, the connection between them will be weak or lost. Based on this information, the access points can estimate the frequency of moving out of their house and then notify the users to encourage them to stay at home as much as possible.
	
\textit{Summary:} Wi-Fi technology is a prominent solution to quickly and effectively enable, encourage, and force people to practice social distancing. With the current advances of Wi-Fi, the accuracy of localization systems can be significantly improved, resulting in effective and precise applications for social distancing. However, Wi-Fi-based technology is mainly used for indoor environments as this technology requires several access points for localization which may not be feasible for outdoor environments. For outdoor environments, other wireless technologies, e.g., Bluetooth, GPS, and cellular technologies, can be considered.

	\subsection{Cellular }
\begin{figure*}[!]
		\includegraphics[width=.7\textwidth]{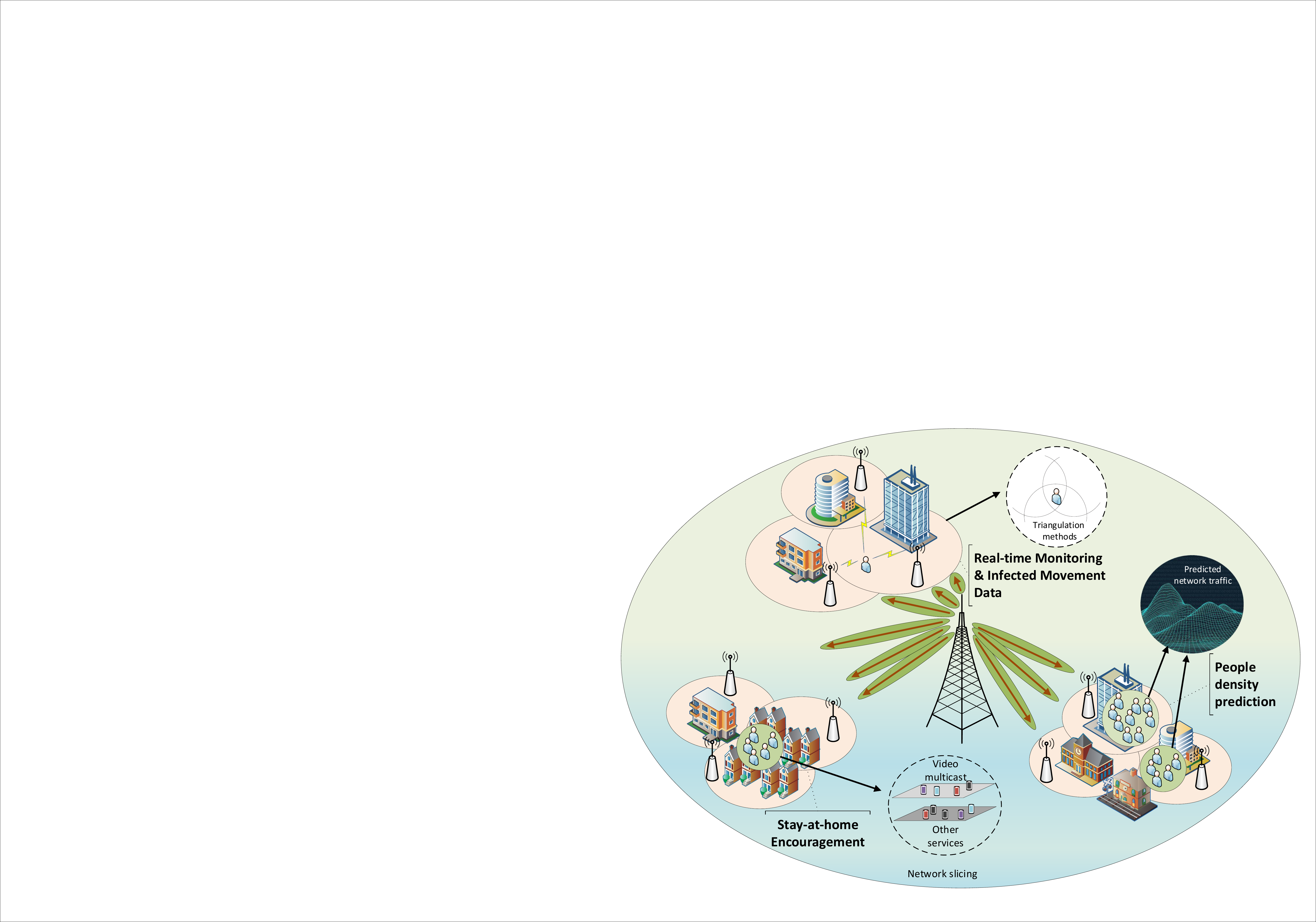}
		\centering
		\caption{Cellular communications systems to support social distancing.}
		\label{Fig:Cellular}
	\end{figure*}

	Over the past four decades, cellular networks have seen tremendous growth throughout four generations and become the primary way of digital communications. 
	The fifth generation (5G) of cellular networks is coming around 2020 with the first standard. According to the Cisco mobile traffic forecast, there will be more than 13 billion mobile devices connected to the Internet by 2023~\cite{Cisco2018}. That positions the cellular technology at the center to enable social distancing in many circumstances including real-time monitoring, people density prediction and encouraging stay-at-home by enabling 5G live broadcasting, as illustrated in Fig.~\ref{Fig:Cellular}.
	
	\subsubsection{Real-time Monitoring}
	Individual tracking and mobility pattern monitoring are potential approaches using cellular technology to practice social distancing as shown in Fig.~\ref{Fig:Cellular}(a). According to the 3GPP standard, the current cellular networks, i.e., LTE and LTE-A, are employing various localization methods such as Assisted-GNSS (A-GNSS), Enhanced Cell-ID (E-CID), and Observed TDoA (O-TDoA) as specified in the Release 9; Uplink-TDoA (U-TDoA) included in the Release 11; and with the aids of other technologies like Wi-Fi, Bluetooth, and Terrestrial Beacon System (TBS) as stated in the Release 13~\cite{Rosado2018,Laoudias2018}.
	Cellphone location data collected by the current cellular network is normally used for network operations and managers~\cite{Laoudias2018} such as network planning and optimization to enhance the Quality of Service (QoS) rather than user applications due to privacy and network resource concerns. However, in the context of social distancing, user tracking based on data of user movement history can be very effective, e.g.,  for quarantined people detection, and infected people tracing. The authorities can check whether infected people are violating quarantine requirements or not. In cases they do not follow the requirements, the authorities can send warning messages or even perform some aggressive measures, e.g., fines and arrests, to force them to self-isolate. 
	
	Moreover, when a user has been exposed to the virus, the user's mobility history can be extracted to investigate the spread of the virus. In these cases, the cellular technology can outperform other wireless technologies in term of availability and popularity. For example, localization services relying on wireless technologies such as GPS always need to be run in the foreground application (i.e., the availability), while this service is a part of cellular network operations. In addition, Ultra-wideband and Zigbee technologies require additional hardware~\cite{BeSpoon,Chu2011High} (i.e., the popularity).
	Incoming 5G networks with the presence of key technologies such as mm-Wave communications, D2D communications, and Ultra-dense networks (UDNs)~\cite{Agiwal2016} are capable of performing a high precision localization. Two positioning schemes exploiting the mm-Wave communications are proposed in~\cite{Palacios2019} based on the validation of triangulation measurements and angle of differences of arrival (ADoA). The simulation results show that the triangulate-validate and ADoA methods can obtain a sub-meter accuracy level with a probability of 85\% and 70\%, respectively in a $18m \times 16m$ indoor area.  
	The authors in~\cite{Koivisto2017} propose a positioning scheme in UDNs using a cascaded Extended Kalman filter (EKF) structure to fuse the DoA and ToA estimations from the reference nodes. The proposed scheme can localize a moving target at speed 50 km/h with a sub-meter level accuracy in an outdoor environment. It can be used for tracking vehicles and monitoring the traffic density.
	
	Recently, some governments have required telecom companies to share cellphone location data to implement social distancing to deal with COVID-19. For instance, Taiwan deployed an ``electronic fence" exploiting the cellular-based triangulation methods to ensure that the quarantined cases stay in their homes~\cite{Taiwan2020}. The local officials call them twice a day to ensure they do not leave their phones at home and visit them within 15 minutes after their phones are turned off or they move away from their home. The Moscow government is also said to be planning to use SIM card data for tracking foreigners and residents who have close contacts with foreigners when the border closure order is lifted~\cite{Moscow2020}. 
	However, individual tracking using cellular technology has risen concerns about privacy~\cite{Privacy1,TheVerge}. Instead, group/crowd detecting and monitoring based on shared location data which is anonymous and aggregated from carriers become the key approach utilized by several governments such as Italy, Germany, Austria, the UK, Korea, and Australia~\cite{European,UK,Korea,Australia}. This approach is intended to alleviate privacy concerns compared with individual-level tracking (i.e., it satisfies the EU privacy rules~\cite{TheVerge}). The metadata can be used to obtain the mobility patterns, thus the governments can monitor whether people are complying with the lock-down rules or not. It can be also employed to model the spread of the virus to aid the governments to analyze and evaluate the effectiveness of ongoing quarantine measures during the outbreak.

	\subsubsection{People Density Prediction}
	In addition to the real-time crowd monitoring and modeling the spread of the virus, the movement history data can be utilized to predict the network traffic thanks to the large-scale location data provided by carriers and the recent advances of machine learning. There are various works on network traffic prediction proposed in~\cite{Alawe2018,Wang2019,Sciancalepore2017,Bega2019,Sun2017} using the history of users' movements. Furthermore, the number of users in a specific area can be also estimated from the network traffic of that area as illustrated in Fig.~\ref{Fig:Cellular}(b). Thus, the authorities can predict the crowd gathering in public places (e.g., shopping malls, airports, and train stations) relying on the corresponding forecasted network traffic. Then, appropriate actions can be performed by the authorities to prevent crowd gathering in these places. For example, if the predicted number of people entering a shopping mall exceeds a threshold, the authorities can notify customers to avoid coming to this place at this time or recommend them to go to other shopping malls having lower densities. In addition, this method can be also applied in residential areas to study how often people stay home as well as predict when they go out or the places they come to. This can provide significant data input for network traffic forecasts in public places. In addition, if they regularly go to non-essential places, the authorities can warn or force them to stay at home as much as possible.

	\subsubsection{Stay-at-home Encouragement}
	To implement social distancing, many people must do their daily activities remotely from their home such as working, studying, and entertainment. Therefore, some video conference applications used to work from home or study online have witnessed an explosion of downloads. For example, the Zoom application has achieved an increase by 1,270\% from 22 Feb to 22 Mar in 2020~\cite{Zoom} and the number of newly registered users of Microsoft Teams has also risen 775\% monthly in Italy after the full lock-down was started~\cite{MicrosoftTeam}. 
	As a result, 5G live broadcasting technology can be used to encourage people to stay at home while minimizing the impact on their work, or study (Fig.~\ref{Fig:Cellular}(c)). Especially, this is probably applicable to cases where landline Internet is not available. There are many works to enhance the quality of video multicast/broadcast applications by utilizing the advances of 5G networks~\cite{Tan2018,Poularakis2016,Montalban2018,Lv2017,Zhang2017}. Video multicast/broadcast services are defined as an ultra high definition slice in an MIMO system~\cite{Tan2018}. To improve the spectral efficiency for video multicast/broadcast in the proposed system, the authors introduce a hybrid digital-analog scheme to tackle channel condition and antenna heterogeneity. Another possible solution that can significantly improve qualities for video multicasting/broadcasting is data caching. A novel caching paradigm proposed in~\cite{Poularakis2016} is applied for multicast services in heterogeneous networks. With the awareness of multicast files, the proposed caching policy can select files efficiently for the caches. Studies in~\cite{Montalban2018,Lv2017} propose using NOMA techniques to support multicast/broadcast by increasing the spectrum efficiency in multi-user environments. Finally, the authors in~\cite{Zhang2017} propose a video multicast orchestration scheme for 5G UDNs which can help to improve the spectrum efficiency.
	
	\subsubsection{Infected Movement Data}
	Due to the omnipresence of mobile phones and the near world-wide coverage of cellular signals, cellular technology can be an effective tool to track the movement of people. Unlike in the quarantined people detection scenarios where these people may deliberately leave their phones at home, people do not have any reason to do so in the infected movement data scenario. Therefore, cellular can be an effective technology in this scenario. The authors of~\cite{hybrid:position10} summarize the methods to trace human position in outdoor environments using base stations and indoor environments using access points. However, the positioning accuracy for outdoor environments still needs to be improved because a small error by using the cellular network technology can cause a big error in the distance measurement.
	
	\textit{Summary:} Cellular technology can be considered to be one of most important approaches to assist social distancing. It can be deployed on a large scale due to its convenience and ubiquitousness compared to other wireless technologies. It can be used to track quarantined or infected individuals. Furthermore, it can provide a unique solution to not only monitor crowds in real-time, but also allow the local authorities to predict the forming of crowds in public areas (e.g., airports, train stations, and shopping malls) based on the forecasted network traffic. The low latency feature of 5G networks in data processing using edge/fog computing enables quick responses of the authorities (e.g., send notifications instantly), for example, to prevent close contact. However, the use of subscriber's location data for social distancing measures is subject to great privacy concerns from citizens (to be discussed more details in Section~\ref{open}). 
	\subsection{Bluetooth}
	\begin{figure*}[!]
		\centering
		\includegraphics[scale=0.53]{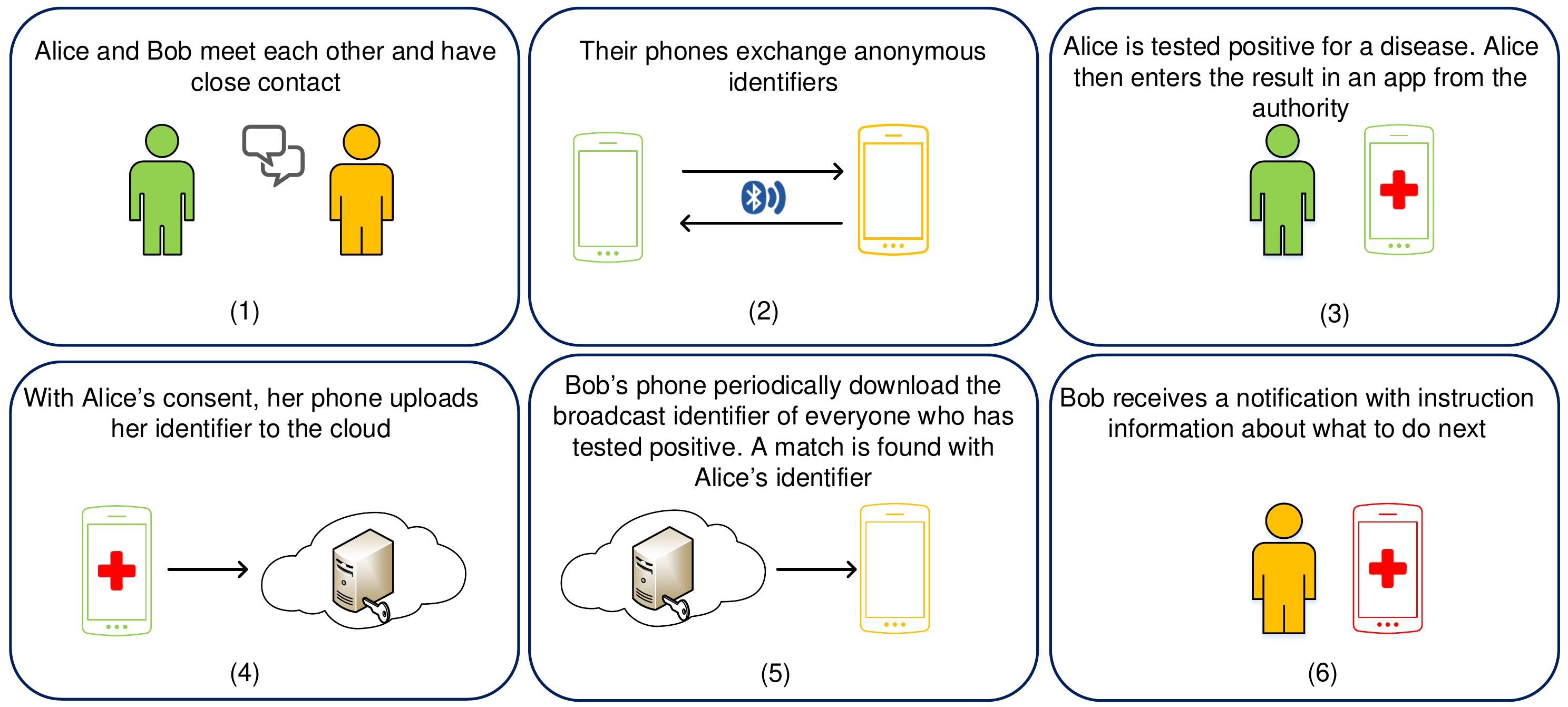}
		\caption{Contact tracing application based on Bluetooth technology~\cite{News3}.}
		\label{Fig.contactTracing}
	\end{figure*}
	
	With the explosive growth of Bluetooth-enabled devices, Bluetooth technology is another solution for social distancing in both indoor and outdoor environments. In particular, Bluetooth is a wireless technology used for short-range wireless communications in the range from 2.4 to 2.485 GHz~\cite{Bluetooth},~\cite{Todtenberg2019Survey}. Bluetooth devices can automatically detect and connect to other devices nearby, forming a kind of ad-hoc called piconet~\cite{Todtenberg2019Survey}. Recently, Bluetooth Low Energy (BLE) has been introduced as an extended version of the classic Bluetooth to reduce the energy usage of devices and improve the communication performance~\cite{Todtenberg2019Survey}. Given the above, the BLE localization technology possesses several advantages compared with those of the Wi-Fi localization. First, the BLE signals have a higher sample rate than that of the Wi-Fi signals (i.e., 0.25 Hz $\sim$ 2 Hz)~\cite{Zhuang2016Smartphone}. Second, the BLE technology consumes less power than that of the Wi-Fi technology, and thus it can be implemented widely in handheld devices. Third, the BLE signals can be obtained from most smart devices, while Wi-Fi signals can be obtained from only access points. Finally, BLE beacons are usually powered by battery, and thereby they are more flexible and easier to deploy than Wi-Fi. It is worth noting that Bluetooth is mainly used for infrastructureless adhoc communications in contrast to other technologies.
	
	\subsubsection{Contact Tracing}
	One application of Bluetooth in social distancing is contact tracing~\cite{News1},~\cite{News2} as illustrated in Fig.~\ref{Fig.contactTracing}. The key idea is using Bluetooth to detect other users in close proximity with their information (e.g., identifier) stored in a person's Bluetooth device, e.g., a mobile phone. When there is an infected case, the authorities can ask people to share these records as a part of a contact tracing investigation. Thereby, the authorities can detect people who may have close contact with the infected one and notify them promptly to prevent the spreading of diseases. Several attempts to use Bluetooth in contact tracing have been reported. Apple and Google have recently introduced a mobile application (running on both iPhone and Android devices) that can detect other smartphones nearby using Bluetooth technology~\cite{News3}. If a person is tested positive for a disease, he/she will enter the result in the app to inform others about that. Then, people who may have close contact with the positive case will be notified and instructed about what to do next. Note that a Wi-Fi or cellular connection would be also required to enable the app. Similar apps have been recently launched in Singapore~\cite{TraceTogether}, Europe~\cite{Pan}, and India~\cite{DROR}.
	
	\subsubsection{Crowd Detection}
	Bluetooth technology can be used to detect crowds in indoor environments to practice social distancing with the latest advances in Bluetooth localization techniques \cite{Wang2015RSSI,Wang2013Bluetooth}. In particular, based on signals received from users' Bluetooth devices, a central controller can calculate the positions of users and detect/predict crowds in indoor environments. If a crowd is detected, the local manager can force people to leave to practice social distancing. In addition, they can advise people who want to go to the place to come at a different time if the place is too crowded at the moment. In~\cite{Wang2015RSSI}, the authors point out that with the development of Bluetooth Low Energy, Bluetooth-based indoor localization can be considered as a practical method to locate Bluetooth devices in indoor environments due to its low battery cost and high communication performance. The authors then propose indoor localization schemes that collect RSSI measurements to detect the user's location by using the triangulation mechanism. 
	
	In~\cite{Faragher2015Location}, the authors show that the BLE technology is strongly affected by the fast fading and interference, resulting in a low accuracy when detecting the user's device. To improve the accuracy of the BLE positioning, the authors run several experimental tests to choose the optimal parameters to set up BLE localization systems. The authors demonstrate that the BLE-based indoor localization can achieve a better performance than that of Wi-Fi localization systems. The authors of~\cite{Naghdi2020Detecting} point out that the accuracy of BLE-based localization is strongly affected by advertising channels, human movements, and human obstacles. To address these problems, they propose a dynamic AI model that can detect human obstacles by using three BLE advertising channels. Then, the RSSI values will be compensated accordingly.
	
	In~\cite{Pei2017Evaluation} and~\cite{Pei2012The}, the authors show Wi-Fi-based and Bluetooth based localization systems can be strongly affected by the interference from other wireless devices operating at 2.4 GHz bands. To mitigate the interference, Wi-Fi devices can use 802.11b and 802.11g/n standards which deploy direct-sequence spread spectrum and orthogonal frequency-division multiplexing signaling methods. Similarly, Bluetooth devices can avoid interference from other wireless devices, e.g., Wi-Fi enabled devices, by using the spread-spectrum frequency hopping technique to randomly use one of 79 different frequencies in Bluetooth bands. As such, the interference from other devices is significantly reduced, thereby improving the accuracy of localization systems.
	
	\subsubsection{Distance Between Two People}
\begin{figure}[!]
		\centering
		\includegraphics[scale=0.33]{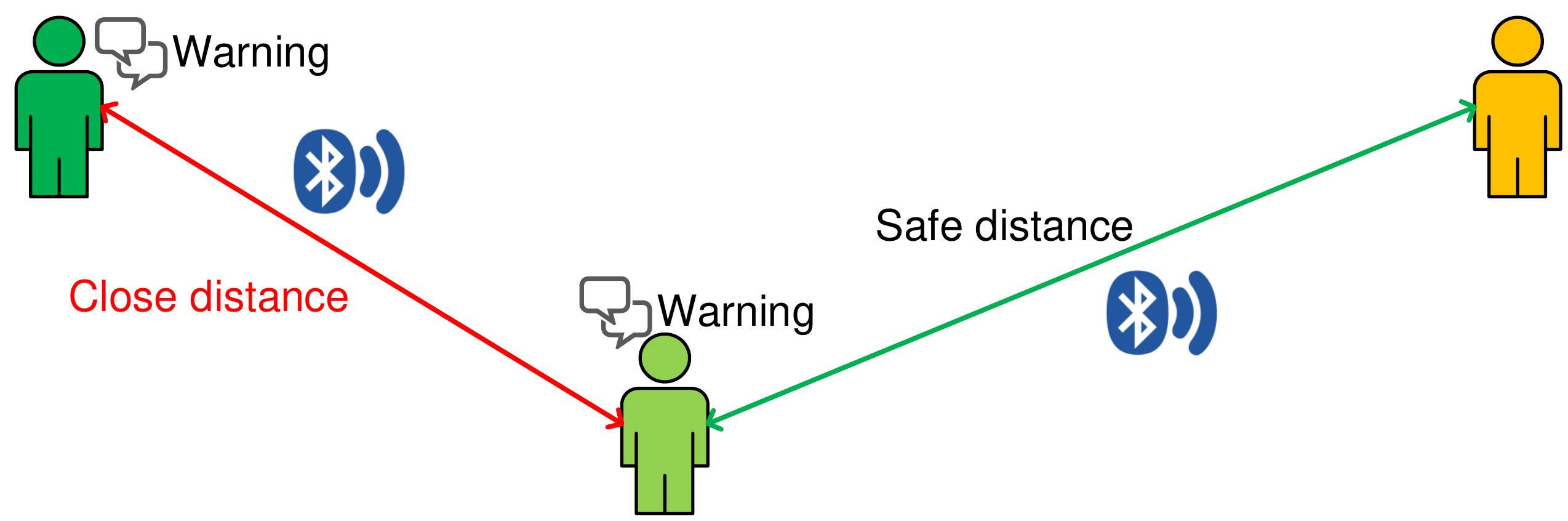}
		\caption{Distance between any two persons based on Bluetooth technology.}
		\label{Fig:bluetoothdistance}
	\end{figure}

	Bluetooth can also be used to determine the distance between two persons by using their Bluetooth-enabled devices, e.g., smartphone or smartwatch, as depicted in Fig.~\ref{Fig:bluetoothdistance}. Specifically, similar to the Wi-Fi technology, based on RSSI levels, a device can calculate the distances between it and other nearby devices~\cite{Wang2013Bluetooth}. It is worth noting that Bluetooth technology can allow a device to connect to multiple devices at the same time~\cite{Bluetooth}. Thus, the device can simultaneously detect distances to multiple devices in its coverage. If the distance is less than a given threshold, e.g. 1.5 meters~\cite{intro1}, the devices can warn and/or encourage users to practice social distancing.
	
	\textit{Summary:} Bluetooth technology is a very promising solution to enable social distancing. However, the privacy of users needs to be taken into account as the applications require users to share information with the authorities and third parties. This can be a research direction to ensure privacy and encourage people to share their information to prevent the spreading of diseases. In addition, several drawbacks of Bluetooth technology in social distancing which need to be considered such as the accuracy of localization techniques when the users' devices are located inside the pockets or bags and their devices always need to turn on the Bluetooth mode. Furthermore, combining Bluetooth and other technologies (e.g., Wi-Fi~\cite{Tejada2012Combine}) to improve the localization accuracy is also an open research direction.

	\subsection{Ultra-wideband }
	
	Ultra-WideBand (UWB) technology has been deemed to be a promising candidate for precise Indoor Positioning Systems (IPSs) that can sustain an accuracy at the centimeter level in the ranges from short to medium. This is thanks to its unique characteristics (e.g., high time-domain resolution, immunity of multipath, low-cost implementation, low power consumption, and good penetration)~\cite{Sahinoglu2008}. Due to the wide bandwidth nature of UWB signals (at least 500 MHz as specified by FCC~\cite{FCC}), the impulse radio (IR) UWB technology has the capability of generating a series of very short duration Gaussian pulses in time-domain which enables its advantages compared with other RF technology. Pulse position modulation with \textit{time hopping} (TH-PPM) is the most popular modulation scheme exploited in the impulse radio based UWB~\cite{Oppermann2004}. This pulse can directly propagate in the radio channel without requiring additional carrier modulation. The baseband-like architecture of the IR-UWB facilitates extremely simple and low-power transmitters.
	Thus, the advantages of the IR-UWB technology can greatly support social distancing, even better than other wireless technologies (e.g., higher accuracy in indoor positioning applications) or provide exclusive solutions (e.g., device-free tracking/counting) for some scenarios, as discussed below. 
	
	\subsubsection{Real-time Monitoring}
	In this section, we review some social distancing scenarios using Ultra-wideband technology for real-time monitoring such as crowd detection (e.g., tracking users' location), public place monitoring and access scheduling (e.g., counting the number of people in a specific area).
	
	\paragraph{Crowd detection} 
	One of the major solutions for crowd detection is tracking locations of people in public areas. There are many commercial products exploiting the IR-UWB technology for real-time localization in both daily life and factories such as DecaWave~\cite{DecaWave}, BeSpoon~\cite{BeSpoon}, Zebra~\cite{Zebra}, Ubisense~\cite{Ubisense}. DecaWave and BeSpoon claim their products based on ranging measurements can offer an accuracy under 10 cm~\cite{DecaWave,BeSpoon}. Furthermore, Ubisense and Zebra provide industrial products which can obtain a high accuracy even in cluttered, indoor factory environments~\cite{Zebra},~\cite{Ubisense}. All of them support real-time positioning for multiple mobile tags by using the triangulation techniques based on the absolute locations of reference nodes or anchors (e.g., UWB transceivers). Especially, the Dimension4 sensor invented by Ubisense can be integrated with a built-in GPS module for outdoor tracking purposes. Experiments conducted to evaluate holistically the performance of three commercial products (i.e., DecaWave, BeSpoon, and Ubisense) under indoor industrial environment setting (with the presence of NLOS) can be found in~\cite{Ruiz2017}. The availability of commercial UWB-based localization systems enables real-time people tracking in public places by localizing their UWB-supported phones, or personal belongings equipped with tags (e.g., keys and shoes). Thus, the authorities can detect the crowd to notify them and other people in the area, disperse the crowd or even predict and prevent the forming of the crowd by using AI/Machine learning algorithms based on the previously collected data. 
	
	
	Recently, device-free localization (or passive positioning) techniques have witnessed significant interest. This is thanks to the capability to tackle inherent problems of aforementioned communication-based localization approaches: (i) privacy issues (e.g., tracking targets do not need to communicate with an access point/network coordinator, thus it can protect private information of the target), and (ii) physical obstacles (e.g., LOS communications have significant influence by obstacles)~\cite{Shit2019}. The high time-domain resolution feature of the IR-UWB technology enables the device-free localization methods relying on the changes of very short pulses properties between two transceivers because of absorption, scattering, diffraction, reflection, and refraction~\cite{Gulmezoglu2015,Ledergerber2020}. In particular, the authors in~\cite{Gulmezoglu2015} use monostatic radar modules (i.e., P410 platform) equipped with one transmitter and one receiver for multi-target tracking based on Gaussian mixture probability hypothesis density (GM-PHD) filters. Information (including raw signal, bandpass signal, motion filtered signal, and detection list) extracted from the reflected signals is used to estimate the locations of targets with an accuracy at the decimeter level. To improve the accuracy, a multi-static is deployed in~\cite{Ledergerber2020} to track a person in real-time by determining the difference between the newly channel impulse response with the presence of a new object with that of the previous one without the object. The location of the object can be found with the mean error of only 3 cm by applying a leading edge detection algorithm on the difference between the two measurements. However, the limitation of this work is that it can track only one target at a time. Motivated by the above works, we can easily deploy device-free localization techniques for crowd detection in public areas without revealing any personal information and hardware requirements on target objects. Thereby, the authorities can locate the exact locations of crowds and have appropriate actions to disperse crowds or force them to practice social distancing.
	
	\begin{figure}[!]
		\centering
		\includegraphics[scale=0.4]{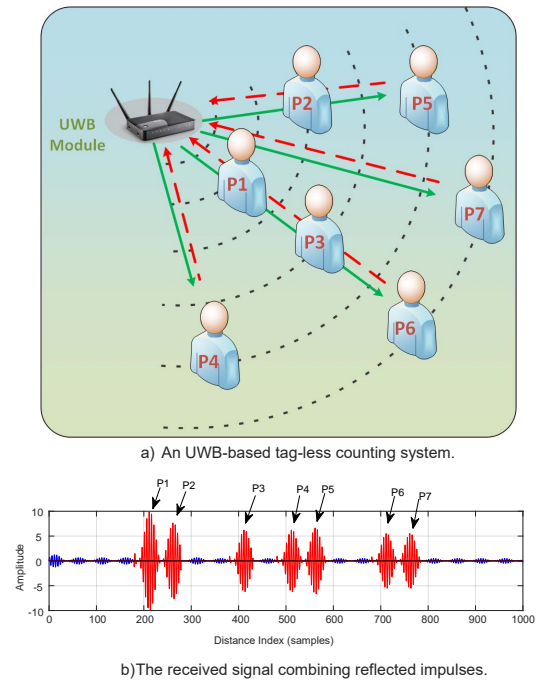}
		\caption{Tag-less counting technique using the UWB technology.}
		\label{Fig: Counting_UWB}
	\end{figure}
	
	\paragraph{Public place monitoring and access scheduling} 
	A simple solution for public place monitoring is referred to as device (or tag)-free counting techniques~\cite{Choi2017,Bartoletti2017}. Specifically, the authors in~\cite{Choi2017} propose an advanced people counting algorithm using the revelation of the received signal pattern according to the number of people illustrated in Fig.~\ref{Fig: Counting_UWB}(a). This method enables people counting even with the presence of dense multipath signals in the environment which is not able to be performed by counting techniques based on detecting single signals corresponding to individual persons. For example, other counting approaches using Wi-Fi and Zigbee rely on the number of connections from users to an access point (i.e., Wi-Fi) or a network coordinator (i.e., Zigbee). Major clusters are picked up to find main pulses having maximum amplitude. A joint probability density function derived from these main pulses is utilized to derive maximum likelihood (ML) equation. Then, the estimated number of people is determined to be the figure having the maximum likelihood as shown in Fig.~\ref{Fig: Counting_UWB}(b). Similarly, the solution in~\cite{Bartoletti2017} also provides a counting approach without positioning targets by using the \textit{crowd-centric} method based on energy detection. Without requiring hardware deployment like Wi-Fi and Zigbee, the approaches proposed in~\cite{Choi2017,Bartoletti2017} can provide a low-cost and high-privacy solution to detect the number of people in public areas. Further actions can be conducted by the local manager to maintain social distancing such as scheduling people to enter the place based on the counting information or giving advice to other people who are planning to go to the crowded place to come at a different time.
	
	\subsubsection{Keeping Distance and Contact Tracing}

	Similar to Bluetooth, the IR-UWB technology can also be applied to maintain the distances between people as well as close physical contact tracing by using ranging methods with high precision in both indoor and outdoor environments~\cite{DecaWave},~\cite{AppleU1}. While DecaWave provides a ranging measurement using sensors and tags~\cite{DecaWave}, Apple has already brought this feature to their phones (e.g., Iphone 11 series) for their primitive location-based services (e.g., finding objects and improving AirDrop)~\cite{AppleU1}. These approaches use time-based ranging techniques like ToF, TDoA or combined ToF and AoA to measure the distances to nearby sensors, tags, or phones. However, these products can be employed to detect close proximity between users in public places. Thanks to the IR-UWB technology, they can frequently broadcast pilot messages containing some information (e.g., their specific IDs, timestamp, etc.) to nearby devices for ranging measurements with extremely low energy consumption. Then, surrounding devices can utilize the information of the received messages to estimate the distance from the source device and warn the users if they are too close to each other (e.g., less than 1.5m in a pandemic situation~\cite{intro1}). In addition, these devices can also store other information like who had close contacts with them along with the distances and duration periods. This information is very important because it can be used to trace close contacts in the future (e.g., investigate the spread of the virus in a pandemic) with minimal privacy violation.

	In order to help people to avoid crowds, especially vulnerable or at-risk groups, in indoor environments such as shopping malls, hospitals, and office buildings, BeSpoon introduces a commercial product that allows moving targets to self-localize their positions very accurately (i.e., less than 10 cm over 600 m in LOS environments) in a short time by using the IR-UWB technology~\cite{BeSpoon}. This product provides both evaluation kits and an ultra-compact UWB module which can be easily integrated into off-the-shelf products (e.g., shopping trolleys or baskets) for localization and navigation purposes. A SnapLoc platform proposed in~\cite{Grobwindhager2019} allows an unlimited number of tags to self estimate their locations at position update rates up to 2.3 kHz. It uses the TDoA technique based on all simultaneous responded information from reference nodes integrated into one single channel impulse response. By combining with the positioning service (i.e., to provide locations of other people in a specific area), a navigation application exploiting commercial products like BeSpoon can be developed to assist people (e.g., customers) for self-detecting their current locations as well as crowds' locations along the way, thereby assisting them to plan their moves and navigate to stay away from crowds.
	
	\textit{Summary:} With the aforementioned potential applications, the IR-UWB systems can be considered to be an outstanding solution to handle social distancing in both indoor and outdoor environments. The IR-UWB based localization systems discussed in~\cite{DecaWave,BeSpoon,Zebra,Ubisense} can be employed for detecting and monitoring crowds in public places with a low-cost deployment. Although this technology can also be used to monitor the positions of self-isolated people to check whether they may violate the quarantine requirements or not, it is less attractive than other RF technologies like Wi-Fi or cellular which do not require to install additional hardware for tracking purposes. In addition, UWB-enabled phones like iPhone 11 series can assist users in practicing social distancing without localization and navigation services. However, this solution only works with a modern iPhone equipped with a UWB chip. Last but not least, the device-free technology presented in~\cite{Gulmezoglu2015,Ledergerber2020,Choi2017,Bartoletti2017} is a great advantage of the IR-UWB technology compared with other wireless technologies for the crowd detecting and monitoring in public places with acceptable accuracy at the decimeter level~\cite{Gulmezoglu2015}.

	\subsection{Global Navigation Satellite Systems (GNSS)}
The GNSS has been being the most widely used for positioning purposes in the outdoor environment nowadays. 
GNSS satellites orbit the Earth and continuously broadcast navigation messages. When a receiver receives the navigation messages from the satellites, it calculates the distances from its location to the satellites based on the transmitting time in the messages. Basically, to calculate the current location of a user, it requires at least three different navigation messages from three different satellites (based on the \textit{Trilateration} mechanism in Section~\ref{background}). However, in practice, to achieve a high accuracy in calculating the location of a user, at least four different messages from four satellites are required (the fourth one is to address the time synchronization problem at the receiver)~\cite{GNSSprinciples}. Currently, some GNSS systems (e.g., Galileo~\cite{Galileo}) can achieve an accuracy of less than 1m. As a result, GNSS systems can be considered to be a very promising solution to enable social distancing practice. 

\subsubsection{Real-time Monitoring}
Due to the outstanding features of GNSS technology in locating people, especially in outdoor environments, this technology is very useful for tracking people to practice social distancing. Specifically, most smartphones are currently equipped with GPS devices which can be used to track locations of mobile users when needed. In the context of a pandemic outbreak, e.g., COVID-19, many suspects, for example, returning from an infected area, will be required to be self-isolated. Thus, to monitor these people, the authorities can ask them to wear GPS-based positioning devices to make sure that they do not leave their residences during the quarantine~\cite{travel,quantify}. The main advantage of using GNSS technology compared to Wi-Fi or Infrared-based solutions for people tracking is that this technology allows to monitor people anywhere and anytime globally, and thus even the suspects move from one city to another city, the authorities still can track and monitor them. However, one of the major disadvantages of this technology is that it depends on the satellite signals. Thus, in some areas with weak or high interference signals (e.g., inside a building or in crowded areas), the location accuracy is very low~\cite{Multipath,GPS_indoor,GPS_indoor1}. To overcome this limitation, pseudolites have been proposed. Pseudolites are ground-based transceivers that can act as an alternative for satellites to transmit GNSS signals. These pseudolites can be installed in the areas where the satellite signals are weak to enhance the positioning accuracy of the GPS. Nevertheless, pseudolites have not been widely deployed because of their high price and strict time synchronization requirement~\cite{pseudo}.
\subsubsection{Automation}
Another useful application of GNSS to practice social distancing is automation. It comes from the fact that GNSS is especially important for navigation in autonomous systems, such as robots, UAVs, and self-driving cars. Thus, in a pandemic outbreak when people are required to stay at home, GNSS-based autonomous services play a key role to minimize physical contact between people. For example, customers can shop online and receive their items with drone delivery services. Such kind of services has been introduced recently by some large retail corporations such as Amazon and DHL. Similarly, robotaxi services have been introduced recently in some countries to deal with COVID-19 outbreak~\cite{GNSS_Drone,GNSS_CHinaserv}. It can be clearly seen that these GNSS-based autonomous services can contribute a significant part in implementing social distancing in practice by minimizing the required human presence for delivery and transportation. 

\subsubsection{Keeping Distance and Crowd Detection}
In~\cite{GNSS_app}, the authors introduce a GNSS service which can be used to determine the locations of users, thereby can warn them if they violate the social distancing requirements. In particular, in this service, mobile users are required to install a mobile application which can track the location of the users based on GPS technology. Then, the users' locations will be updated constantly to the service provider. Thus, based on the users' locations, the service provider can determine whether the user violates the social distancing requirements or not. For example, if there are more than two users locating too close to each other (e.g., less than two meters), the service provider can send warning messages to remind the users. Furthermore, in the cases where a user goes to restricted areas, e.g., isolated areas, they will receive warning messages to be aware of using protection measures. 
\subsubsection{Infected Movement Data}
In the infected movement data scenario, GNSS can be a very effective technology because of its world-wide coverage and positioning accuracy is not the main concern. For the outdoor environment, using GNSS alone can be sufficient for tracking the location of infected people. With the omnipresence of smartphones with built-in GPS feature, the movement path of the infected people can be easily determined. However, the main concern in this scenario is that people have to turn on GPS service on their smartphones, which necessitate mechanisms to incentivize people to share their movement information. This issue will be further discussed in Section~\ref{open}.

\textit{Summary:} Although this GNSS-based service has many advantages in practicing social distancing, e.g., tracking users, keeping distance, and group monitoring, it has some shortcomings which limit its applications in practice. Specifically, this service requires tracking locations of users based on GPS in a real-time manner, which may cause some extra-implementation costs and privacy issues for users. Furthermore, in terms of determining the distance between two people, the accuracy of GNSS services is not high in general, especially for distances less than two meters. Thus, some recent advanced GNSS technologies like~\cite{Differential_gps1,RTK_2,RTK_3,pseudo} can be used to improve the accuracy of the GPS. However, these technologies are still quite expensive and have not been widely deployed for public services, and thus more research in this direction should be further explored. For the privacy issues, they will be discussed in Section V with several solutions such as location information protection and personal identity protection.
	
	\subsection{Zigbee}
	Zigbee is also a potential technology that can help to enable
	social distancing. In particular, Zigbee is a standard-based wireless communication technology for low-cost and low-power wireless networks such as wireless sensor networks. A Zigbee system consists of a central hub, e.g., network coordinator, and Zigbee-enabled devices. Zigbee-enabled devices can communicate with each other at the range of up to 65 feet (20 meters) with an unlimited number of hops. Compared with Wi-Fi and Bluetooth technologies, Zigbee is designed to be cheaper and simpler, making it possible for low-cost and low-power communications for smart devices~\cite{Luoh2013Zigbee},~\cite{Zigbee}. Moreover, Zigbee can operate at several frequencies, such as 2.4 GHz, 868 MHz, and 915 MHz. Given the above, Zigbee is ideal for constructing mesh networks with long battery life and reliable communications~\cite{Zigbee}. As a result, Zigbee can be considered as a promising candidate in several applications that enable social distancing during a pandemic outbreak.
	
	\subsubsection{Crowd Detection}
	One promising application of Zigbee is detecting and tracking users' location in indoor environments. The key idea is that based on the RSSI level of the received signals from the user's Zigbee-enabled device, the Zigbee control hub can determine the location of the user. Several research works report that Zigbee localization systems can achieve high accuracy with low-power and low-cost devices~\cite{Luoh2013Zigbee}. Based on the location of users, the central hub can detect crowds, i.e., many users in the same area, and notify the local manager to ask people to practice social distancing during a pandemic outbreak. With the state-of-the-art mechanisms in the literature, the accuracy of Zigbee localization systems is significantly improved, making it feasible for social distancing. In~\cite{Chu2011High}, the authors propose a novel framework to enhance the localization accuracy of Zigbee devices by considering the effect of ``drift phenomenon'' when users move from a place to another place in indoor environments. The authors then demonstrate that the proposed framework can increase the accuracy by up to 60\% compared with conventional solutions.
	
	Differently, in~\cite{Fang2012An}, the authors introduce an ensemble mechanism to further improve the localization accuracy. In particular, instead of using the RSSI level, the proposed solution combines the gradient-based search, the linear least square approximation, and multidimensional scaling methods together with spatial dependent weights of the environment to approximate the target's location. In~\cite{Niu2015ZIL}, the authors propose an energy-efficient indoor localization system that can obtain Wi-Fi fingerprints by using ZigBee interference signatures. The key idea of this work is using ZigBee interfaces to detect Wi-Fi access points which can significantly save energy compared with using Wi-Fi interfaces. Furthermore, a K-nearest neighbor method with the Manhattan distance is introduced to increase the accuracy of the localization system. The experimental results show that the proposed solution can save 68\% of energy compared with the method using Wi-Fi interfaces. The accuracy is also improved by 87\% compared to state-of-the-art WiFi fingerprint-based approaches.
	
	\subsubsection{Public Place Monitoring and Access Scheduling}
	In a Zigbee system, there is a central hub, known as the network coordinator, to control other connected devices in the network. Thus, Zigbee can be used to control the number of people in indoor environments. Specifically, when a person equipped with a Zigbee-enabled device (e.g., ID card or member card) enters the place, the device will connect to the Zigbee central hub. As such, the central hub is able to calculate the total number of people inside the place at a given time. Based on this information, the local manager can ask people to queue before entering the place if it is too crowded.
	
	\textit{Summary:} Zigbee technology can play an important role in enabling social distancing during pandemic outbreaks. However, Zigbee is a new technology and has not been widely adopted in our daily life, and thus limiting its practical applications. Nevertheless, with the support from leading companies such as Amazon, Google, Apple, and Texas Instruments~\cite{Zigbee}, the number of Zigbee-enable devices is expected to explosively increase in the near future. Furthermore, combining Zigbee with other technologies (e.g., Wi-Fi~\cite{Niu2015ZIL}) is also a promising research direction to improve the performance of localization systems in terms of the accuracy and robustness.
	
	\subsection{RFID}
	RFID plays a key role in real-time object localizing and tracking~\cite{Xiao2018One}. An RFID localization system usually consists of three main components: (i) RFID readers, (ii) RFID tags, and (iii) a data processing system~\cite{Wang2019Active}. Typically, RFID tags can be categorized into two types: (i) active tags and (ii) passive tags. A passive RFID tag can operate without requiring any power source, and it is powered by the electromagnetic field generated by the RFID reader. In contrast, an active RFID tag has its own power source, e.g., a battery, and continuously broadcasts its own signals. Active RFID tags are usually used in localization systems. Thus, RFID technology can be considered as a potential technology to practice social distancing.
	
	\subsubsection{Crowd Detection}
	One potential application of RFID technology is locating users in the indoor environment based on recent RFID-based localization solutions~\cite{Xiao2018One}-\cite{Yang2013Efficient}. To that end, each user is equipped with an RFID tag, e.g., the staff ID or member cards. Based on the backscattered signals from the RFID tag, the RFID reader can determine the location of the user. If there are too many people in the same area, the system can notify the authorities to take appropriate actions, e.g., force people to leave the area to practice social distancing. Several recent mechanisms in the literature can be adopted to make this application possible during pandemic outbreaks. In~\cite{Huang2015Real}, the authors propose an RFID-based localization system for indoor environments with high localization granularity and accuracy. The key idea of this solution is reducing the RSSI shifts, localization error, and computational complexity by using Heron-bilateration estimation and Kalman-filter drift removal. In~\cite{Megalou2019Fingerprint}, the authors propose to use a moving robot to enhance the accuracy of a real-time RFID-based localization system. In particular, the robot is able to perform Simultaneous Localization and Mapping (SLAM), and thus it can continuously interrogate all RFID tags in its area. Then, based on passive RFID tags at known locations, we can estimate the location of target tags by properly manipulating the measured backscattered power. Alternatively, in~\cite{Xiao2018One}, the authors propose to equipped two RFID tags at the target instead of only one as in conventional solutions to improve the accuracy of localization techniques. Adding one more RFID tag possesses several advantages: (i) easy to implement and adjust the RFID reader's antenna, (ii) enabling fine-grained calculation, and (iii) enabling accurate calibration. The experimental results then show that equipping two tags at the user can greatly increase the localization accuracy of the system.
	
	However, the RFID technology has several limitations due to the fact that both the receiver and the RF source are in the RFID reader. Specifically, the modulated signals backscattered from the RFID tag are strongly affected by the round-trip path loss from the receiver and the RF source. In addition, the RFID system can also be affected by the doubly near-far problem~\cite{Huynh2018Survey}. To address these problems, a few recent works propose to use bistatic and ambient backscattered communication technologies (extended version of RFID) for localization~\cite{Jameel2019Application},~\cite{Vasisht2018Inbody}. The key idea is separating the RF source from the receiver. The RF source now can be a dedicated carrier emitter or an ambient RF source. The tag can then transmit data to the receiver by backscattering the RF signals generated by the RF source. Based on the received signals, the receiver can estimate the location of the tag. In~\cite{Vasisht2018Inbody}, the authors propose a localization system based on backscatter communications to locate patients in a hospital. In particular, each patient is equipped with a backscatter tag which can backscatter signals broadcast by an RF source. Then, the location of a patient can be detected by a localization algorithm, namely Remix, based on the backscattered signals from the backscatter tags. Remix consists of two processes. First, the algorithm approximates the distance from the tag to the receiver based on the backscattered signals. Second, the signal paths are modeled with linear splines. Then, an optimization problem is solved to find the effective distances corresponding to the paths that close to the actual paths from the tag to the receiver. As a result, Remix can accurately estimate the position of the backscatter tag by modeling the spline structure. Based on the users' locations, Remix can detect crowds in the hospital and advice the authorities to take appropriate actions to practice social distancing. Note that this solution can also be deployed to detect crowds in other places such as workplaces, schools, and supermarkets where backscatter tags can be easily attached to users/customers' cards, e.g., staff cards, student cards, and member cards.
	
	\subsubsection{Public Place Monitoring and Access Scheduling}
	Another application of RFID in social distancing is monitoring the number of people inside a place, e.g., a building or supermarket. In particular, an RFID reader will be deployed at the main gate of a place, and users are equipped with RFID tags (can be both active and passive tags). The users' tags can broadcast their ID (active) or send their ID upon receiving RF signals from the RFID reader (passive). When a user enters the place, the RFID reader can receive the user's ID and increase the counter. As such, the RFID reader can calculate the number of people inside the place. If there are too many people, the system can notify the local manager to force people to queue before entering the place to practice social distancing. This solution can be deployed in supermarkets or workplaces where the customer/staff usually have member/staff ID cards which can be equipped with RFID tags.
	
	\textit{Summary:} RFID technology is a potential solution to enable social distancing. However, unlike other wireless technologies, RFID technology has not been widely adopted in practice due to its complexity in implementation. Specifically, to be able to detect the location of people by using RFID technology, they need to be equipped with RFID tags. However, RFID tags are not readily available likes Wi-Fi access points or Bluetooth. Thus, applications of RFID technology for social distancing are still limited in practice.
	
	Table~\ref{Tab:sum} summarizes the technologies discussed in this Section. Technologies that have a wide communication range such as cellular and GNSS are effective solutions for the scenarios where it is necessary to track people's location over a large area (e.g., the infected movement data scenario). On the other hand, technologies with a shorter communication range (e.g., Wi-Fi, Bluetooth, Zigbee, and RFID) are more suitable for scenarios that involve indoor environments such as public place monitoring. Moreover, technologies that can achieve a high positioning accuracy (e.g., Ultra-wideband and Bluetooth) can be employed to keep a safe distance between any two people, except for GNSS since it requires a high cost to maintain a sufficient accuracy. Furthermore, most of these technologies are ready to be implemented and integrated with existing systems such as smartphones. However, user privacy is an open issue for most wireless technologies. Furthermore, other emerging wireless technologies such as LoRaWAN, Z-Wave, and NFC~\cite{newtech} have not been well investigated in the literature for positioning systems, and thus they could be potential research directions for social distancing in the future. 
	
	\begin{table*}
		\caption{Summary of Wireless Technologies Applications to Social Distancing}
		\label{Tab:sum}
		
		\begin{centering}
			\begin{tabular}{|>{\raggedright\arraybackslash}m{1.5cm}|>{\raggedright\arraybackslash}m{2cm}|>{\raggedright\arraybackslash}m{1.5cm}|>{\raggedright\arraybackslash}m{1.5cm}|>{\raggedright\arraybackslash}m{0.9cm}|>{\raggedright\arraybackslash}m{1.2cm}|>{\raggedright\arraybackslash}m{1.8cm}|>{\raggedright\arraybackslash}m{4.0cm}|}
				\hline 
				\multicolumn{1}{|>{\centering\arraybackslash}m{1.5cm}|}{\multirow{1}{*}{\textbf{Technology}}} & \multicolumn{1}{|>{\centering\arraybackslash}m{2cm}|}{\multirow{1}{*}{\textbf{Range}}}&
				\multicolumn{1}{>{\centering\arraybackslash}m{1.5cm}|}{\multirow{1}{*}{\textbf{Accuracy}}} & \multicolumn{1}{>{\centering\arraybackslash}m{1.5cm}|}{\multirow{1}{*}{\textbf{Cost}}} & \multicolumn{1}{>{\centering\arraybackslash}m{0.9cm}|}{\multirow{1}{*}{\textbf{Privacy}}}&
				\multicolumn{1}{>{\centering\arraybackslash}m{1.2cm}|}{\multirow{1}{*}{\textbf{Readiness}}} & \multicolumn{1}{>{\centering\arraybackslash}m{1.8cm}|}{\multirow{1}{*}{\textbf{Indoor/Outdoor}}} & \multicolumn{1}{>{\centering\arraybackslash}m{4.0cm}|}{\multirow{1}{*}{\textbf{Integrate with existing systems}}}\\
				\hline 
				\hline 
				Wi-Fi \cite{Yang2015WIFi,WiFi6,Mazuelas2019Robust,Luo2019dynamic,Chang2017FitLoc, SurveyIndoor}& 
				Typically up to 50m indoors and 100m outdoors \cite{RangeWifi} &   1m - 5m \cite{Naggar2019Indoor}   &       Low   &    Low     &     High      &       Indoor          &         Mobile phones, smart home, smart city~\cite{SurveyIndoor,Lan2019Modified}\\
				\hline

				Cellular~\cite{Wang2019,Alawe2018,Sun2017,Tan2018,Zhang2017}  & Short to Long & Less than 50cm \cite{Koivisto2017} & Low & Low to High & High & Both   & Smartphone, Smartwatch, Mobile phones~\cite{Rahman:2017}, Intelligent Transportation Systems~\cite{Lim:2017, Polese:2019, Challita:2019} \\ \hline

				Ultra-wideband \cite{DecaWave,Ubisense,Zebra,BeSpoon,Ledergerber2020,Choi2017,AppleU1, Grobwindhager2019} & Short to Medium & Less than 10cm \cite{DecaWave}, \cite{BeSpoon},~\cite{Ledergerber2020} & Low & High & Medium to High &  Both & Smartphone~\cite{AppleU1}, Smart industry~\cite{BeSpoon,Zebra,DecaWave}, Public transport management~\cite{Ubisense}, Commercial vehicles~\cite{Ubisense,DecaWave}                              \\ \hline

				Bluetooth \cite{Faragher2015Location,Naghdi2020Detecting, SurveyIndoor,Wang2013Bluetooth, News3,TraceTogether,DROR,Wang2015RSSI,Pei2012The,Pei2017Evaluation}     & Outdoor: 55m - 78m. Office: 15m - 19m. Home: 26m - 35m \cite{BluetoothRange} &  1m - 2m \cite{Naghdi2020Detecting}             &   Low   &   Low      &    High       &         Both        &     Mobile phones, smart home, smart city~\cite{SurveyIndoor,Tejada2012Combine, Wang2013Bluetooth}   \\\hline
				
				Zigbee \cite{Luoh2013Zigbee, Fang2012An,Niu2015ZIL, SurveyIndoor}        & Up to 300+ m (line of sight). Up to 75m - 100m indoor~\cite{Zigbee}  &3m - 5m \cite{Luoh2013Zigbee}                &   High   &    Low     &    Low       &       Indoor          &        Mobile phones, smart home, smart city~\cite{Bianchi2019RSSI, Chen2009RSSI,Subaashini2013Zigbee} \\
				\hline
				
				GNSS \cite{travel,quantify,Multipath,GPS_indoor,GPS_indoor1,GNSS_Drone,GNSS_CHinaserv,GNSS_app,Differential_gps1,RTK_2,RTK_3}
				& World-wide coverage &Outdoor less than 10cm \cite{GNSS_Smartphones}, indoor in m   &Low - High (depend on accuracy)  &Medium &Yes&Outdoor& Mobile phones~\cite{Cho:2016, Yin:2020, Liu:2017}, Intelligent Transportation Systems~\cite{Lim:2017,Cui:2018} \\
				\hline
				RFID \cite{Xiao2018One, Huang2015Real, Yang2013Efficient, SurveyIndoor}          & Active: 100m or more. Passive: $\sim$ 10m~\cite{RFIDRange} & Less than 1m \cite{Huang2015Real}           &  High    &    Low     &       Medium     &    Indoor             &       Mobile phones, smart home, smart city~\cite{Shahid2020Chipless,Ma2020The} \\
				\hline
				
			\end{tabular}
			\par\end{centering}
	\end{table*}
	
	\section{Other Emerging Technologies for Social Distancing}
	\label{other}
	In addition to the wireless technologies, other emerging technologies such as AI, computer vision, ultrasound, inertial sensors, visible lights, and thermal also can all contribute to facilitating social distancing. In this section, we provide a brief overview of each technology and discuss how it can be applied for different social distancing scenarios.

	\subsection{Computer Vision}
	\begin{figure*}[!]
		\begin{subfigure}{.3\textwidth}
			\centering
			\includegraphics[width=1\linewidth]{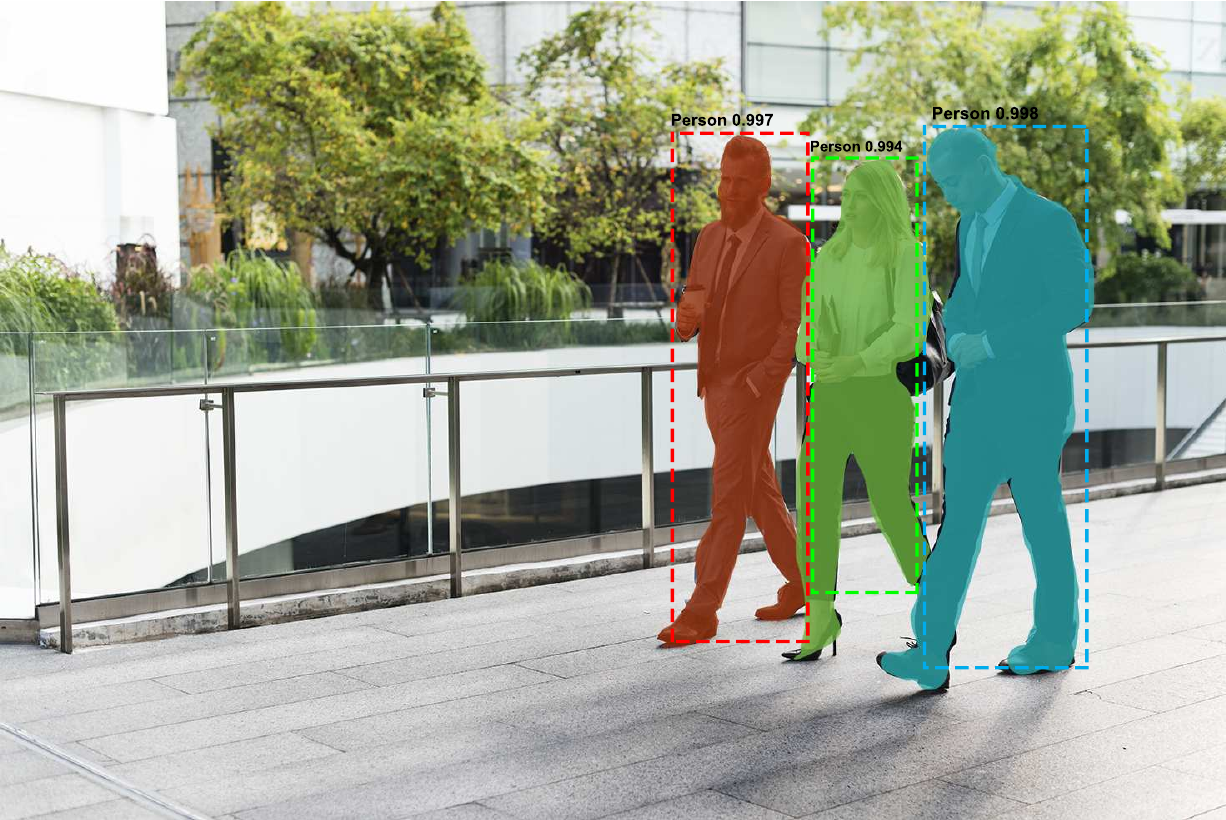}
			\caption{Human Detection}
			\label{fig:humandetection}
		\end{subfigure}
		\begin{subfigure}{.4\textwidth}
			\centering
			\begin{subfigure}{.49\textwidth}
				\renewcommand{\thesubfigure}{(\alph{subfigure}1)}	
				\centering
				\includegraphics[width=.8\linewidth]{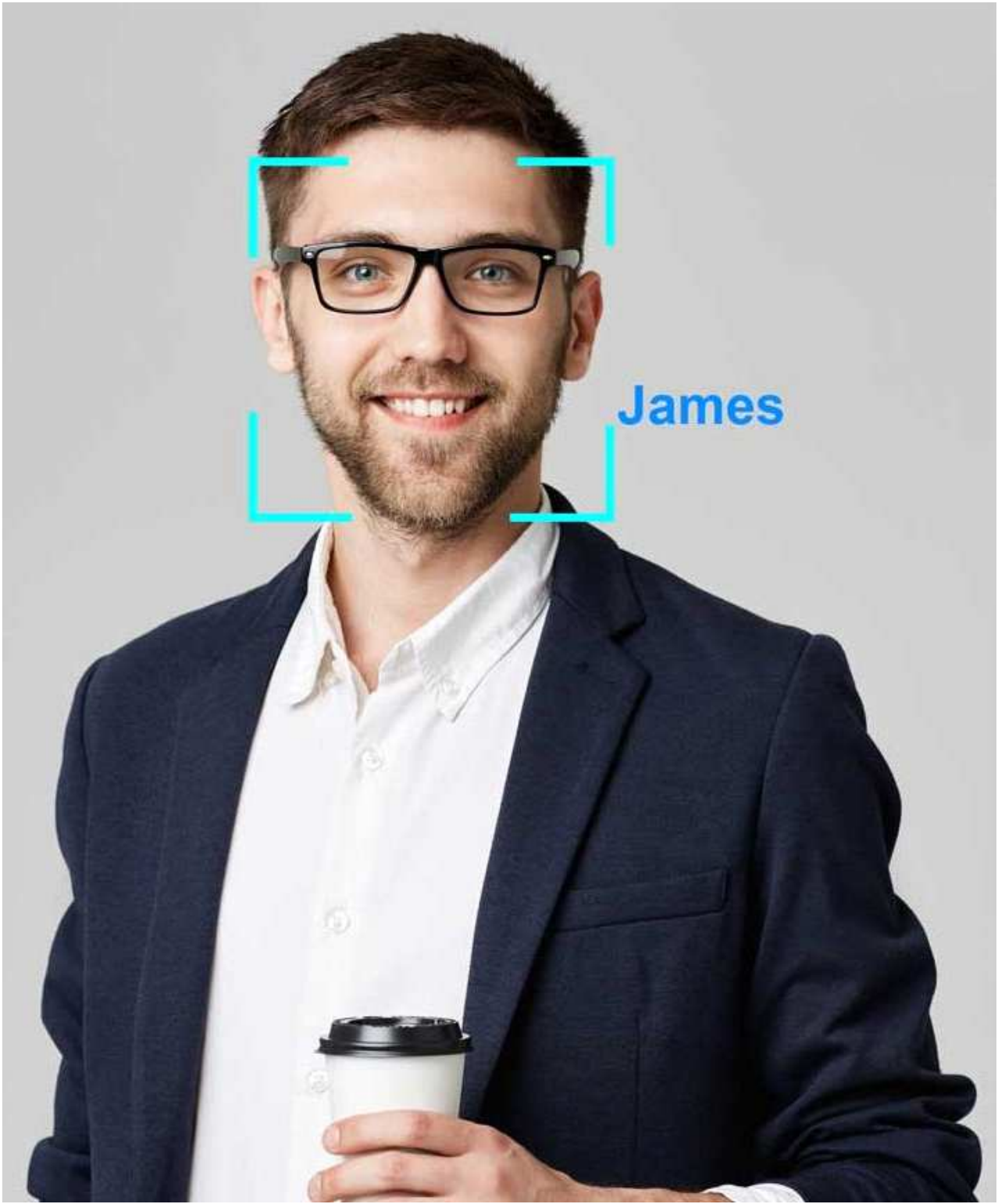}
				\caption{} 
				\label{fig:facerecognition1}
			\end{subfigure}
			\begin{subfigure}{.49\textwidth}
				\addtocounter{subfigure}{-1}	
				\renewcommand\thesubfigure{(\alph{subfigure}2)}	
				\centering
				\includegraphics[width=.8\linewidth]{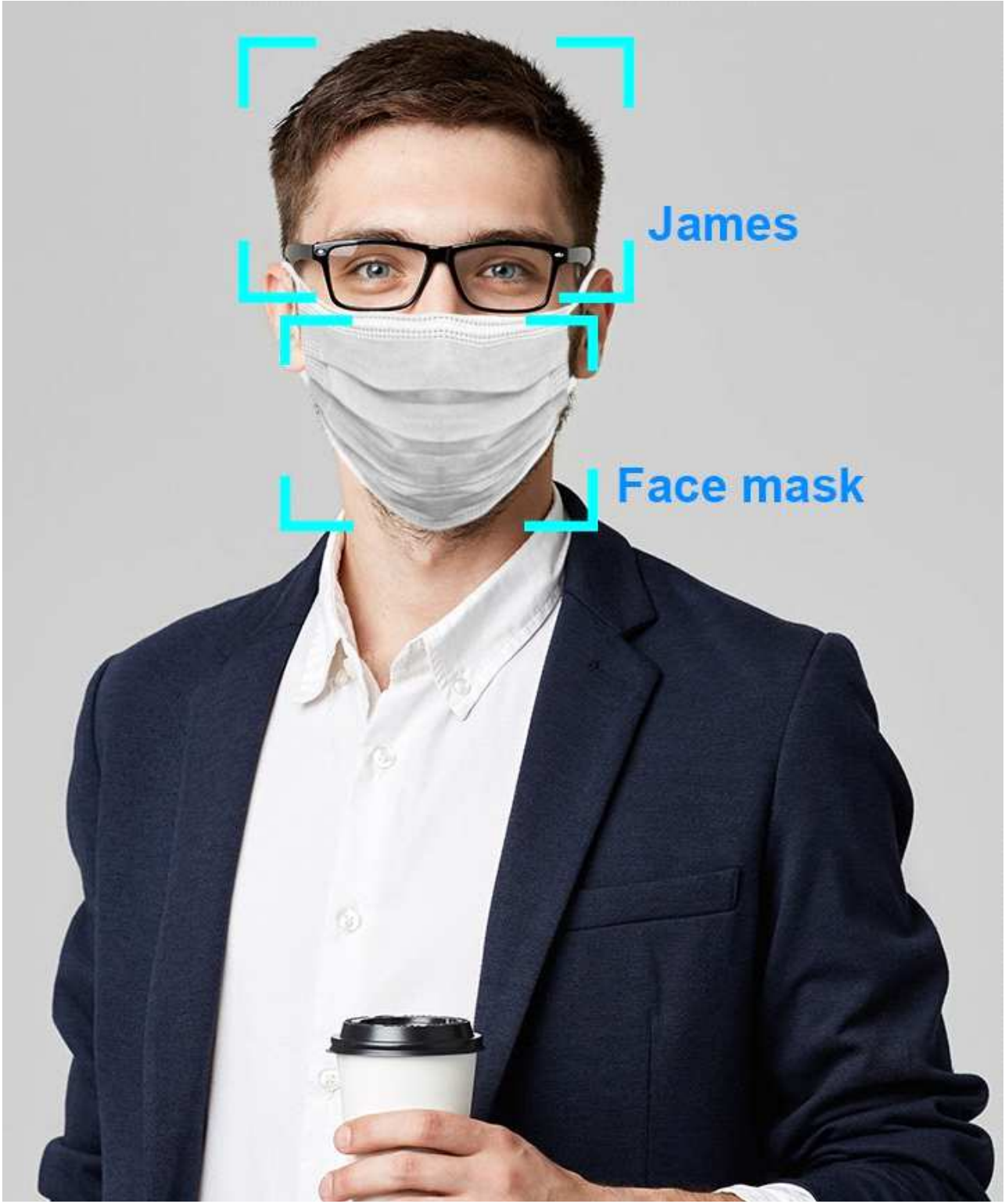}
				\caption{}
				\label{fig:facerecognition2} 	
			\end{subfigure}
			\addtocounter{subfigure}{-1}
			\caption{Face Recognition}
			\label{fig:facerecognition}
		\end{subfigure}
		\begin{subfigure}{.3\textwidth}
			\centering
			\includegraphics[width=.9\linewidth]{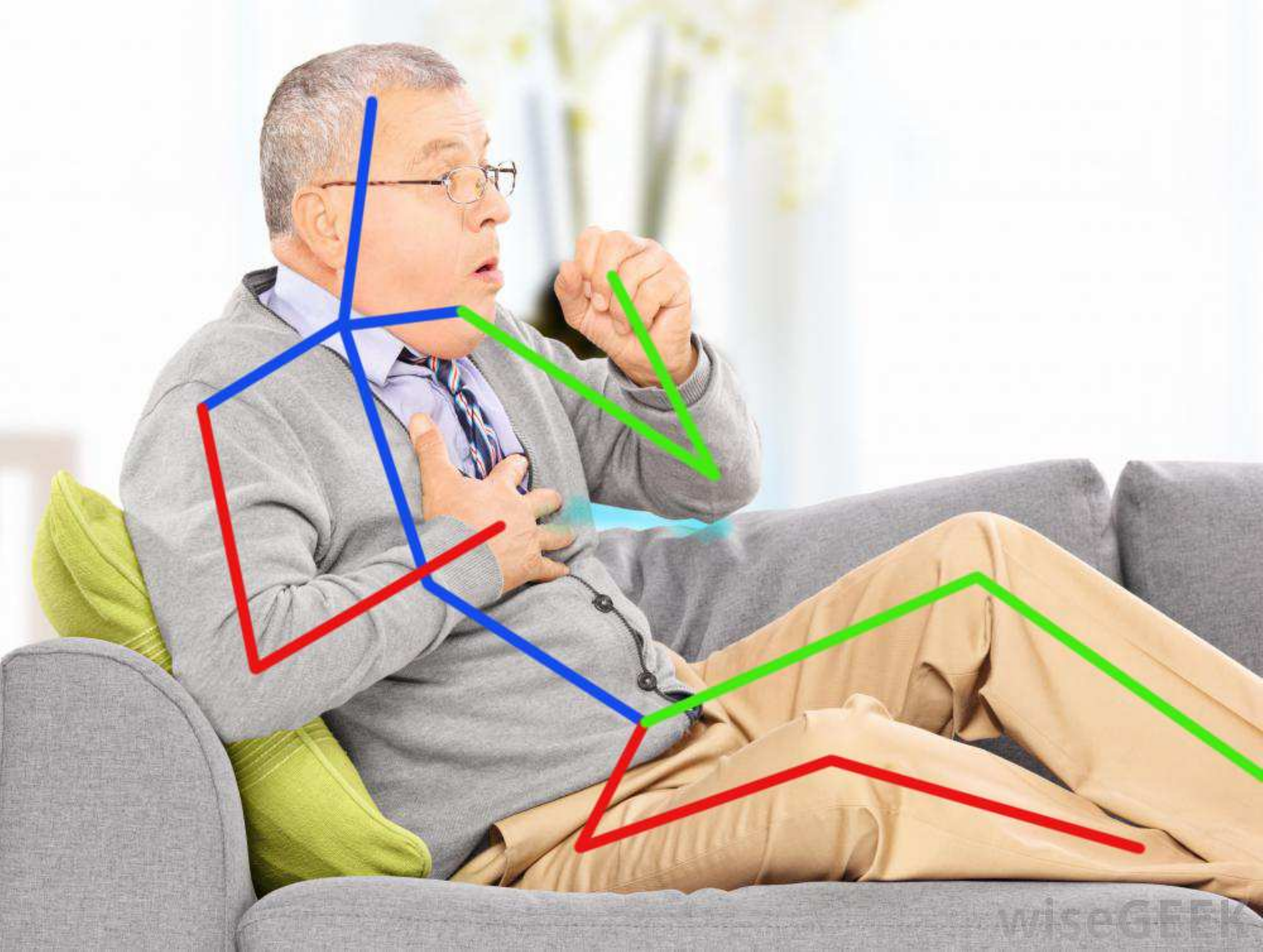}
			\caption{Pose Estimation}
			\label{fig:poseestimation}
		\end{subfigure}
		\caption{Computer vision technologies for social distancing (a) human detection to identify the number of people in the public place, (b) face recognition to identify (b1) the full face of isolated person, (b2) person with mask or person behind the mask and (c) pose estimation to detect one with coughing symptom.}
		\label{fig:computervision}
	\end{figure*}
	
	Computer vision technology trains computers to interpret and understand visual data such as digital images or videos. Thanks to recent breakthroughs in AI (e.g., in pattern recognition and deep learning), computer vision has enabled computers to accurately identify and classify objects~\cite{vision:computervision2}. Such capabilities can play an important role in enabling, encouraging, and enforcing social distancing. For example, computer vision can turn surveillance cameras into ``smart'' cameras which can not only monitor people but also can detect, recognize, and identify whether people comply with social distancing requirements or not. In this section, we discuss several social distancing scenarios where computer vision technology can be leveraged, including public place monitoring, and high-risk people (quarantined people and people with symptoms) monitoring and detection.

	\subsubsection{Public Place Monitoring}
	Despite government restrictions and recommendations about social gathering, some people still do not comply with, which can cause the virus infection to the community. In such context, human detection features in object detection~\cite{vision:liu2020deep}, a major sub-field of computer vision, can help to detect crowds in public areas through real-time images from surveillance cameras. An example scenario is described in Fig.~\ref{fig:humandetection}. If the number of people in an area does not meet the social distancing requirement (e.g., gathering above 10 people), the authorities can be notified to take appropriate actions.

	There are two main approaches to detect humans from images in object detection namely region-based and unified-based techniques. The former detects humans from images in two stages including the region proposal and the procession according to the regions~\cite{vision:Zhao2019deep}. Based on this approach, several frameworks including Fast-RCNN~\cite{vision:fast-rcnn} and Faster-RCNN~\cite{vision:faster_rcnn} are developed in combination with Convolution Neural Network (CNN)~\cite{vision:DCNN}. In~\cite{vision:mask_rcnn}, the authors improve the Faster-RCNN by proposing the Mask Regions with CNN features (Mask RCNN) method which masks the bounding box to detect the object with high accuracy while adding a minor overhead to the Faster-RCNN. Mask RCNN outperforms previous methods by simplifying the training process and improving the accuracy in detecting humans in the images for calculating the density of people in a particular area.
	
	Although the above region-based approach has high recognition accuracy~\cite{vision:mask_rcnn}, it has high complexity, which is unsuitable for devices with limited computational capacity. To address this, the unified approach is more appropriate to implement, which can reduce the computational complexity by detecting humans from images with only one step. This approach maps the pixels from image to the bounding box grid and class probabilities to detect humans or objects in real-time. Following this direction, the You Only Look Once (YOLO) method proposed in~\cite{vision:yolov3} can detect/predict objects (even small ones) in real-time with high accuracy. In addition, in~\cite{vision:ssd}, the authors propose the Single Shot Multibox Detector (SSD) framework which uses a convolution network on the image to calculate a feature map and then predict the bounding box. Through experimental results, they demonstrate that this method can detect objects faster and more accurately than those of both YOLO and Faster-RCNN. For public place monitoring, both YOLO~\cite{vision:yolov3} and SSD~\cite{vision:ssd} can be used to detect fast and accurately humans from real-time images or videos of surveillance cameras. After identifying people, we can use a real-time automatic counter to count and identify whether the number of gathering people complying with social distancing requirements or not.

	\subsubsection{Detecting and Monitoring Quarantined People}
	To prevent the spread of the virus from an infected person to others, the infected person or people who had physical contact with them must be isolated at the restricted areas or at home. For example, citizens who come back from highly infected countries/regions of COVID-19 are often requested to be quarantined or self-isolate for 14 days. Due to the lack of facilities, most countries require these people to self-isolate at home. In this case, the face recognition capability of computer vision can help to enforce this requirement by analyzing the images or videos from cameras to identify these people (i.e., to check whether they breach the self-isolation requirements or not). If these people are detected in public, the authorities can be notified to take appropriate actions.
	
	Unlike object detection, the dataset including the full face images of the isolated people needs to be built. The face recognition system firstly learns from this dataset and then analyzes the images from public surveillance cameras to identify their appearances as in Fig.~\ref{fig:facerecognition1}. The authors in~\cite{vision:deepface} propose a framework named DeepFace using Deep Neutral Network (DNN) which can detect with an accuracy of 97.35\% and 91.4\% in Labeled Faces in the Wild (LFW) and YouTube Faces (YTF) dataset, respectively. To improve the accuracy in detecting human from surveillance cameras, some advanced techniques can be implemented such as~\cite{vision:deepid},~\cite{vision:deepid2} and~\cite{vision:deepid3}.

	To prevent the spread of infectious diseases such as COVID-19, people are often required to wear masks in public places, which necessitates approaches to recognize or identify people with or without masks as illustrated in Fig.~\ref{fig:facerecognition2}. For example, the cameras in front of a public building can recognize and send warning messages (e.g., a beep sound) to remind the person who does not wear a mask when he/she intends to get into the building. This idea is introduced in~\cite{vision:masks} by using the CNN to detect people who do not wear the masks. However, this work is justs at the first step, which still requires much more efforts to demonstrate the effectiveness as well as improve the accuracy.

	\subsubsection{Symptoms Detection and Monitoring} 
	After a few days of being infected with the virus, the infected person may have some symptoms such as coughing or sneezing. To minimize spreading the virus to others, it would be very helpful if we can detect these symptoms from people in public and inform them or the authorities. The idea here is similar to that of using thermal imaging cameras at airports or train stations. Specifically, detecting human behaviors, motion, and pose in computer vision can play a pivotal role~\cite{vision:humanbehavior}. Pose estimation captures a person with different parts (as illustrated in Fig.~\ref{fig:poseestimation}) then detects human behaviors by studying the parts' movements and their correlation. For example, a coughing person in Fig.~\ref{fig:poseestimation} usually moves his hand near his head and his head would have a vibration.

	Recognition of human behaviors from surveillance cameras is a challenging problem because the same behaviors may have different implications, depending on the relationship with the context and other movements~\cite{vision:humanbehavior2}. The recent advances in AI/ML are instrumentals in correlating different movements/parts to interpret the associated behavior. In~\cite{vision:deeppose} the authors propose to use CNN~\cite{vision:DCNN} to enhance the accuracy of the model of the interaction between different body parts. In addition, the authors in~\cite{vision:2dpose} introduce several methods to detect body parts of multiple people in 2D images, and the authors in~\cite{vision:3dpose} propose methods to estimate 3D poses from matching of 2D pose estimation with a 3D pose library. These works can be further developed for future studies to detect people with symptoms of the disease such as coughing or sneezing in real-time. To improve the accuracy of the symptom detection in social distancing, computer vision-based behavior detection methods can be combined with other technologies, e.g., thermal imaging. 
	\subsubsection{Infected Movement Data}
	To prevent the spread of the virus, tracing the path of an infected person plays an important role to find out the people who were in the same place as the infected person. For this purpose, computer vision technology can not only detect the infected people by facial recognition but also contribute to the positioning process. In~\cite{hybrid:indoor1}, the movement of people is determined by analyzing the key point of transition frames captured from smartphone cameras. This method can draw the trajectory of movements and the location with an accuracy around two meters. In~\cite{hybrid:indoor4}, the authors propose to combine the human detection techniques of computer vision with digital map information to improve the accuracy. In this study, the user path from cameras is mapped to the digital map which has the GPS coordinates. This method can achieve a very high accuracy within two meters. In another approach, the authors in~\cite{hybrid:indoor6} propose to use both smartphones' cameras and inertial-sensor-based systems to accurately localize targets (with only 6.9 cm error). This approach uses the fusion of keypoints and squared planar markers to enhance the accuracy of cameras to compensate for the errors of inertial sensors.
	
	\subsubsection{Keeping Distance}
Computer vision can also be very helpful to support people in keeping distance to/from the crowds. In~\cite{hybrid:smartphone}, the authors develop an on-device machine-learning-based system leveraging radar sensors and cameras of a smartphone. When the radar sensor detects the surrounding moving objects, the smartphone camera can be utilized to capture its surrounding environment. Taking into account the recorded data, the smartphone can train the data using machine learning algorithms to determine the existence of nearby people and its distance from those people with respect to the social distancing requirements. We can also use a smartphone to estimate the distance between the mobile user and other people using radar sensors and cameras along with machine learning algorithms. 
	
	\textit{Summary:} Computer vision can be utilized in several social distancing scenarios, especially the ones that require people monitoring and detection. Particularly, computer vision is the only method that can differentiate between people and identify complex features such as masks and symptoms. To further improve the effectiveness of computer vision in the social distancing context, future research should focus on increasing the accuracy and reducing the complexity of computer vision methods, so that they can be integrated into existing systems such as surveillance cameras. 
	\subsection{Ultrasound}

	The ultrasound or ultrasonic positioning system (UPS) is usually used in the indoor environment with the accuracy of centimeters~\cite{Airborne_ultrasound}. The system includes ultrasonic beacons (UBs) as tags or nodes attached to users and transceivers. Beacon units broadcast periodically ultrasonic pulses and radio frequency (RF) messages simultaneously with their unique ID numbers. Based on these pulses and messages, the receiver's position can be determined by position calculation methods such as \textit{trilateration} or \textit{triangulation}~\cite{ultrasound_Principle}. In comparison with other RF-based ranging methods, the UPS does not require a line of sight between the transmitter and the receiver, and it also does not interfere with electromagnetic waves. However, since the propagation of the ultrasound wave is limited, most UPS applications for social distancing are only limited within the indoor environment.
	
	\subsubsection{Keeping Distance}
For this purpose, UPS can be used to position and notify people. One of the first well known UPS systems is Active Bat (AB)~\cite{Activebat_1} based on the time-of-flight of the ultrasonic pulse. Typically, an AB system consists of an ultrasonic receiver matrix located on the ceiling or wall, a transmitter attached to each target, and a centralized computation system to calculate the objects' positions. As presented in~\cite{Activebat_1}, by using a receiver matrix with 16 sensors, the AB system can achieve very high positioning accuracy, i.e., less than 14 centimeters. However, a limitation of this system is its high complexity, especially if a large number of ultrasonic sensors are deployed. 
	
\begin{figure*}[!]
		\centering{\includegraphics[width=.8\linewidth]{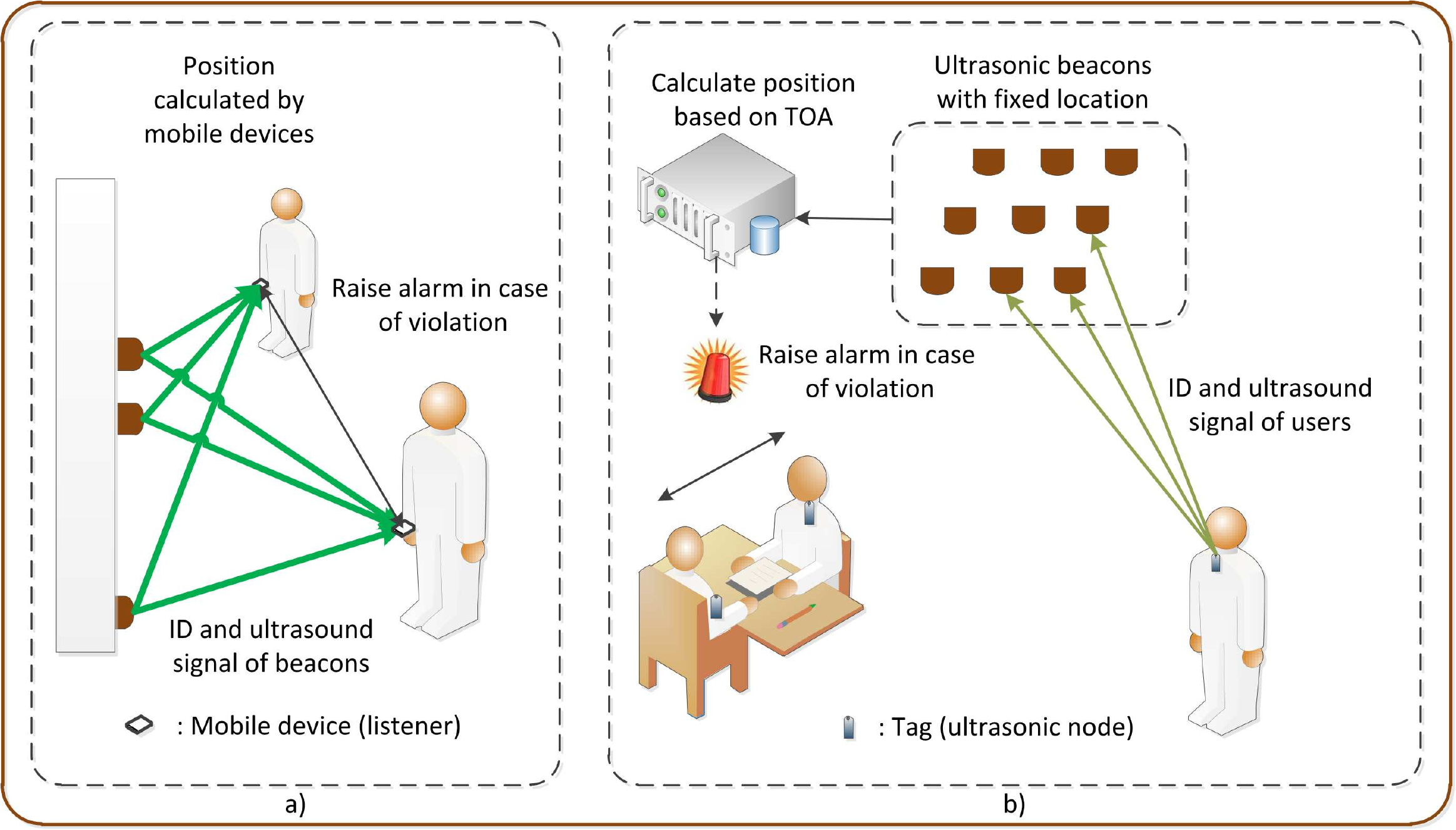}}
		\caption{Ultrasound application for keeping distance using a) Cricket system~\cite{Cricket}, and b) Active Bat system~\cite{Activebat_1}.}
		\label{fig:Ultra_SD}
	\end{figure*}

	Another limitation of the AB system is the privacy risk for users since the location of users under the AB system is calculated at the central server. To address that, the Criket (CK) system is proposed in~\cite{Cricket} wherein the position calculation is executed at the receivers. Specifically, a receiver in the CK system passively receives RF and ultrasound signals from UBs located on the wall or ceiling, and then the receiver calculates its position by itself based on UBs' ID and coordinates. Since the receivers do not transmit any signals, the privacy of users will not be compromised. Fig.~\ref{fig:Ultra_SD} demonstrates the two systems in the keeping distance application. 
	
	\subsubsection{Real-time Monitoring}
	In the context of social distancing, UPS can be an effective solution for real-time monitoring scenarios, especially gauging the number of people in public buildings. In particular, the main characteristic that makes UPS different from other positioning technologies is \textit{confinement}, i.e, the ultrasound signal is confined within the same room as the UBs~\cite{ultrasound_Principle}. Among the other positioning technologies, only infrared technology shares the same characteristic. Nevertheless, infrared signals prone to interference from sunlight and other thermal sources, and they also suffer from line-of-sight loss~\cite{ultrasound_Principle}. As a result, ultrasound is the most efficient technology for \textit{binary positioning}~\cite{ultrasound_Principle}, i.e., determine if the object is in the same room as the UBs or not. Thus, UPS can be particularly useful in the social distancing scenarios where the exact positions of people are not as necessary as the number of people inside a room (e.g., to limit the number of people). This technology is more efficient because it needs a few reference nodes (e.g., UBs) to determine the binary positions of people, which can significantly reduce implementation costs.

	\subsubsection{Automation} 
	Ultrasound can also be applied in the social distancing scenarios that utilize medical robots or UAV. Mobile robots, especially medical robots, can play a key role in reducing the physical contact rates between the healthcare staff (e.g., doctors and nurses) and the patients inside a hospital, thereby maintaining a suitable social distancing level. In such scenarios, UPS can help to improve the navigation of medical robots. In~\cite{Medical_robot}, a navigation system based on Wi-Fi and ultrasound is proposed for indoor robot navigation. To deal with the uncertainties which are very common in crowded places like hospitals, the system employs a Partially Observable Markov
	Decision Process, and a novel algorithm is also introduced to minimize the calibration efforts.

	In the social distancing context, besides the outdoor applications, UAVs can also be employed to reduce the necessity of human physical presence. For example, UAVs can be used to deliver goods inside a building or to manage warehouse inventory. However, most of the existing studies focus on UAV navigation for the outdoor environment, which often relies on GNSS for UAV positioning. Since GNSS's accuracy is low for the indoor environment, these methods cannot be applied directly for UAV navigation inside a building. To address that limitation, a navigation system is proposed in~\cite{UAV_Ul}, which utilizes ultrasound, inertial sensors, GNSS, and cameras to provide precise (less than 10cm) indoor navigation for multiple UAVs.

	\textit{Summary:} Ultrasound can be applied in several social distancing scenarios. In the keeping distance scenarios, UPS systems such as AB and CK can be applied directly to localize and notify people to keep a \textit{safe} distance. Moreover, due to its confinement characteristic, ultrasound is one of the most efficient technology for binary positioning, which is particularly useful for monitoring and gauging the number of people inside the same room. In the automation scenarios, ultrasound can facilitate UAVs and medical robots navigations, especially for the indoor environment.
	

	

	\subsection{Inertial Sensors}
	
\begin{figure}[!]
		\centering
		\includegraphics[width=.5\textwidth]{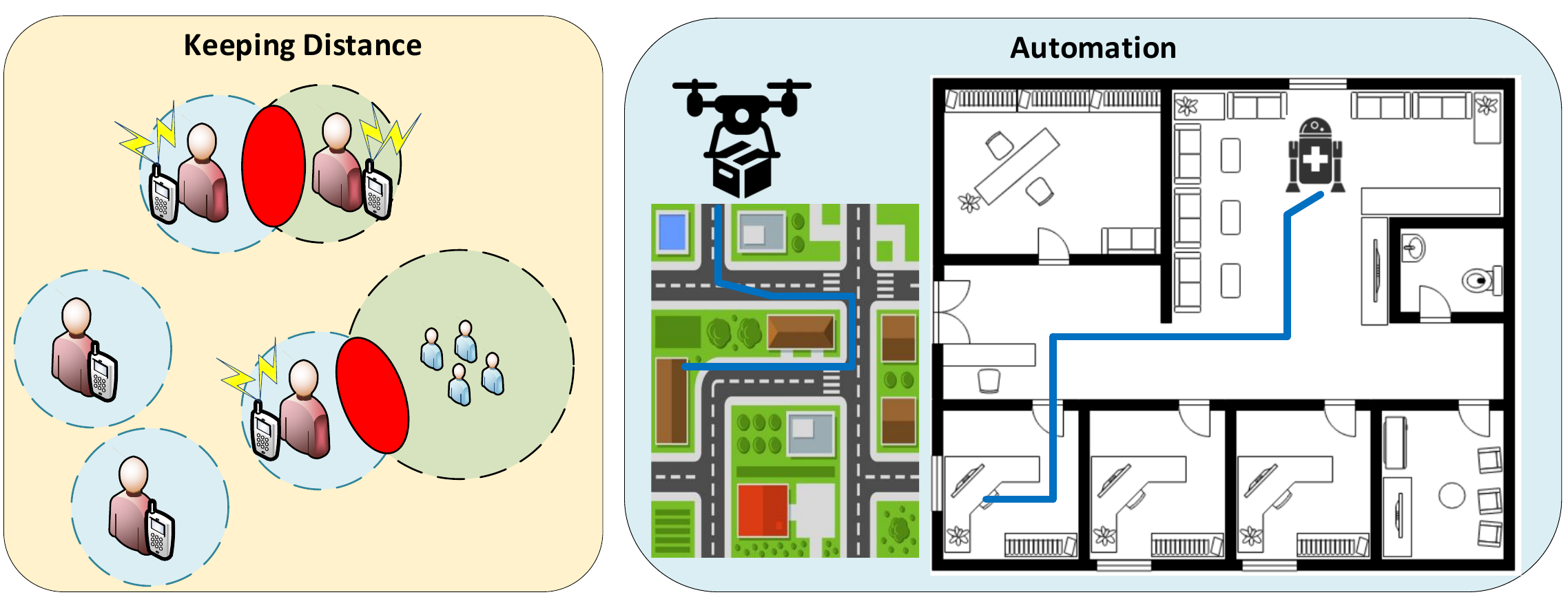}
		\caption{Inertial-sensors-based systems for several social distancing scenarios.}
		\label{fig:INS}
	\end{figure}

	In the context of social distancing, inertial-sensors-based systems can be applied in distance keeping and automation scenarios as illustrated in Fig.~\ref{fig:INS}. For example, positioning applications utilizing built-in inertial sensors can be developed for smartphones which can alert the users when they get close to each other or a crowd. Moreover, inertial sensors can be integrated into robots and vehicle positioning systems, which can facilitate autonomous delivery services and medical robot navigation. All of these scenarios can contribute to reducing the physical contact rate between people.
	
	Inertial sensors consist of two special types of sensors, namely gyroscopes and accelerometers, attached to an object to measure its rotation and acceleration. Based on the measured rotation and acceleration data, the orientation and position displacements of the object can be determined~\cite{INS1}. Because inertial sensors do not require any external reference system to function, they have been one of the most common sensors for dead reckoning, i.e., calculation of the current position is based on a previously determined position, navigation systems. Such navigation systems can provide accurate positioning within a short time frame. However, since the current position is determined based on the previously calculated positions, the errors accumulate over time, i.e., integration drift. Therefore, Inertial-Navigation-System (INS) is often used in combination with other positioning systems, e.g., GPS, to periodically reset the base position~\cite{INS1}.

	\subsubsection{Keeping Distance} Traditionally, INS has been widely used for aviation, marine, and land vehicle navigation. Recently, the ever-increase presence of smartphones has enabled many INS applications for pedestrian positioning and navigation, which can support social distancing scenarios. Moreover, INS is one of the few technologies that can enable accurate pedestrian positioning for the outdoor environment, especially when combined with other outdoor positioning technologies such as GPS. In~\cite{INS2}, a smartphone-based positioning system is proposed. The system makes use of a smartphone's built-in sensors, including gyroscopes, accelerometers, and magnetometers (sensors that measure magnetism), to calculate the smartphone's position. In particular, magnetometers are combined with gyroscopes to improve the accuracy of rotation measurements. This is done by correlating their measurements via a novel algorithm which uses four different thresholds to determine the weights of the gyroscope and magnetometers measurements in the correlation function. 
	
	In~\cite{INS3}, a novel indoor positioning system is developed using Wi-Fi and INS technologies. In this system, INS is utilized for the area where Wi-Fi coverage is limited, while Wi-Fi positioning is used to compensate INS's integration drift. Another positioning system using inertial sensors and Wi-Fi is presented in~\cite{INS4}, where Wi-Fi fingerprinting technique is used to improve the accuracy of the dead reckoning navigation. Because of the integration drift, a dead reckoning navigation system needs to frequently update its position by referencing to an external node. In the proposed system, a Wi-Fi fingerprinting map is set up in advance and the dead reckoning system can use the map to update its position. Moreover, in~\cite{hybrid:indoor7}, the authors propose using Kalman filter to combine the measurement data from Wi-Fi and INS, which can reduce the error to 1.53 meters.  
	
	Beside Wi-Fi, INS can be used in combination with other positioning technologies. In~\cite{INS5} and~\cite{INS6}, INS has been combined with the UWB technology for pedestrian positioning and tracking. Generally, INS helps to reduce UWB's high implementation cost and complexity, while INS's integration drift can be compensated. Particularly, INS is employed to compensate for the UWB's low dynamic range and proneness to external radio disturbances in~\cite{INS5}. To enable the combination, an information fusion technique using the extended Kalman filter is proposed to fuse the measurement data coming from both the INS and UWB sensors. The result shows that the hybrid system can achieve better performance than both the individual systems. In~\cite{INS6}, the information fusion problem between the INS and UWB is optimized to minimize the uncertainties in the measurements. As a result, the positioning accuracy can be significantly improved.

	\subsubsection{Automation}Beside pedestrian positioning, INS can also be applied for social distancing scenarios involving autonomous vehicles, e.g., medical robots and drone delivery. Generally, INS has been commonly used for medical robot applications, including surgeon assists, patient motion assists, and delivery robots. In this section, we will only focus on the medical and delivery robot applications for social distancing purposes. In~\cite{INS9}, a novel INS system is developed specifically for mobile robot navigation. In addition, an error model is proposed to increase the accuracy of the involved inertial measurements. A Kalman filter is also proposed to precisely estimate the velocity and orientation of the robot in the presence of noises. A novel data fusion algorithm, leveraging an adaptive Kalman filter is presented in~\cite{INS11} for indoor robot positioning based on an INS/UWB hybrid system. 
	
	Unlike INS for mobile robots that are mostly developed for the indoor environment, INS for UAV focuses on outdoor applications. Note that UAV navigation must also consider its altitude, which adds more complexity. The authors of~\cite{INS12} leverage inertial sensors and cameras to determine the UAV's position, velocity, and altitude. Particularly, the cameras attached to the UAV capture the images of the surrounding environment and send them to a control station. This station will then process the images to determine the UAV's pose in regards to the surroundings. The pose's data is then combined with the inertial sensors data via a Kalman filter to determine the UAV's position and velocity. Similarly, a system combining inertial and vision sensors is developed in~\cite{INS13} for UAV positioning and navigation. The system utilizes two observers which have inertial and vision sensors. The first observer calculates the orientation based on gyroscope and vision sensors, and the second observer determines the position and velocity based on data from the accelerometers and vision sensors. The experimental results show that the vision sensors measurements can be used to compensate for the inertial sensors errors, thereby achieving a high accuracy even with low-cost inertial sensors.
	
	\textit{Summary:} The omnipresence of smartphones with built-in inertial sensors has opened many opportunities for developing positioning systems based on INS. For the distance keeping scenarios, INS positioning systems, especially for pedestrians, can play a vital role as they are readily available. In the other scenarios such as medical robot navigation and UAV delivery, INS-based techniques can help to increase the efficiency (more accurate path, and lower traveling time) of the existing navigation systems.

	\subsection{Visible Light}
	
	The recent development in the light-emitting diodes (LEDs) technology has enabled the use of existing light infrastructures for communication and localization purposes due to attractive features of visible lights such as reliability, robustness, and security~\cite{Komine2004,Rajagopal2012,Pathak2015}. Visible light communication (VLC) systems usually comprise two major components, i.e., LED lights corresponding to transmitters to send necessary information (e.g., user data and positioning information) via visible lights and photodetectors (e.g., photodiodes) or imaging sensors (e.g., camera) playing the role of receivers~\cite{intro7}. Due to the ubiquitous presence of LED lights, VLC can be leveraged in many social distancing scenarios as discussed below.
	
	\begin{figure*}[!]
		\includegraphics[width=1\textwidth]{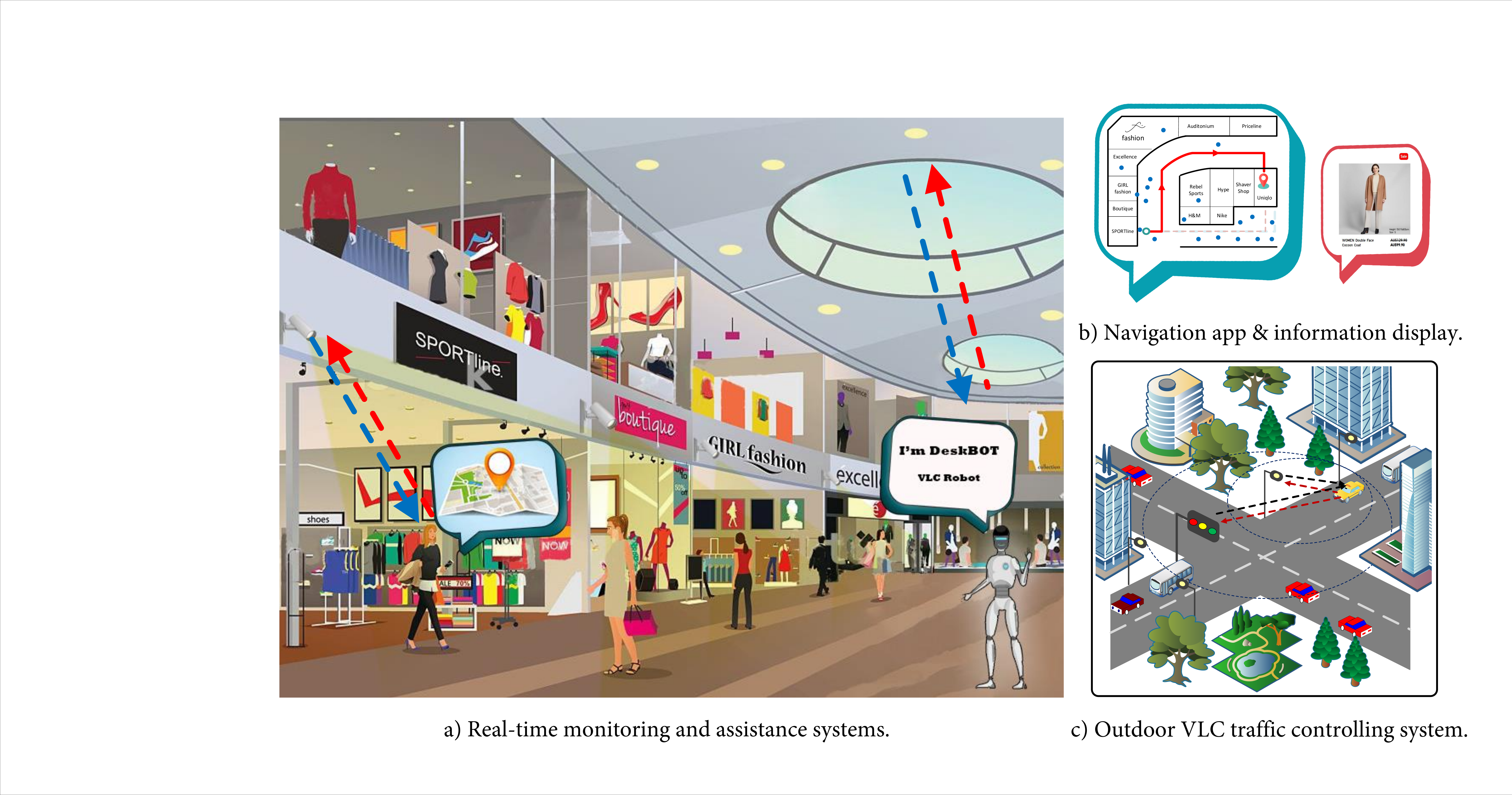}
		\centering
		\caption{Visible light communications supporting social distancing.}
		\label{Fig:VLC}
	\end{figure*} 
	
	\subsubsection{Real-time Monitoring}
	Communication systems using visible light (e.g., LED-based communications) can provide precise navigation and localization solutions in indoor environments. Utilizing this technology, some applications can be implemented to support social distancing such as tracking individuals who are being quarantined, detecting and monitoring crowds in public places as shown in Fig.~\ref{Fig:VLC}(a). 
	\paragraph{Photodiode-based VLC systems} 
	Due to many advantages such as low cost and easy to implement, the VLC receiver using photodiodes can be employed as a ``tag'' that is integrated into mobile targets such as trolleys/shopping carts, autonomous robots, etc. People attached with these tags can perform self-positioning based on the \textit{triangulation} method so that they can avoid crowded areas. Furthermore, the tags' locations can be collected by the authorities to monitor people in public areas. Based on this location data, further actions can be carried out such as warning people by varying the color temperature of the lights in the crowded areas. It is worth noting that this solution will not reveal any personal information of users (e.g., customers) because it only requires communications between VLC-based tags and light fixtures. However, most VLC systems only provide half-duplex communications due to the fact that LED lights operate as the role of transmitters. Therefore, they should be combined with other wireless technologies like Bluetooth~\cite{Qualcom2016,AtriusNavigator}, and Infrared~\cite{Chen2018} to enable an uplink communication with the server for location information exchange. Moreover, to improve the accuracy of positioning people in indoor environments using photodiodes, some advanced techniques can be used such as data fusion of AOA and RSS methods proposed in~\cite{Sahin2015} and the AOA method using a multi-LED element lighting fixture introduced in~\cite{Eroglu2015}. 
	One main disadvantage of the photodiode-based VLC systems is the need for hardware (i.e., the photodiode receiver) mounted on smart trolleys/shopping carts to receive light signals. Consequently, the system might fail to detect the locations of people who do not carry them. Nevertheless, pureLiFi company has recently invented a tiny optical front end which can be integrated into smartphones to take benefits of the photodiode receiver in high accuracy VLC-based localization services~\cite{pureLifi}. 
	
	\paragraph{Camera-based VLC systems} 
	The rapid development of smartphones has enabled VLC-based applications on handheld devices such as indoor localization and navigation applications (e.g., smart retail systems~\cite{Qualcom2016,AtriusNavigator,Philips}). These systems use front-facing cameras of mobile phones to receive visible light signals contained positioning information (e.g., the LED light's ID or location) from visible light beacons~\cite{Kuo2014}. The captured photos collected regularly by the front-facing camera are sent to a cloud/fog server for image processing to alleviate the computation on the phone. Then, the beacon's ID and coordinates can be extracted and sent back to the phone. After that, the AoA algorithm is implemented to estimate the location and orientation of the phone. 
	An attractive use case of the camera-based VLC systems~\cite{Qualcom2016,AtriusNavigator,Philips} is to assist users to quickly find specific products in shopping malls, or supermarkets. Thus, we can adopt this function to implement tracking and monitoring crowds in public places as well as assisting people to avoid crowds in a proactive manner. It is worth noting that this solution is more convenient than using photodiodes since it uses front-facing cameras of smartphones as the VLC receivers, thus everyone using smartphones can be tracked. 
	However, due to continuous photo shooting, these positioning applications are very energy-consuming, which is a major drawback of camera-based VLC systems when they are used for tracking people.

	\subsubsection{Automation}
	In public places, there is always a need for assistance in specific circumstances (e.g., information or physical supports for customers, older and disabled people). For instance, supporting staff in supermarkets can assist customers to find products or help older/disabled people to carry their goods. Similar assistance scenarios can be seen in hospitals, banks, and libraries. This results in an increase in close physical contacts between customers and assistants. Therefore, autonomous assistance systems using VLC technology can be employed to minimize the physical contacts as shown in Fig.~\ref{Fig:VLC}(a) and (b).
	
	\paragraph{Information assistance} 
	Beside the navigation purpose, the smart retail systems~\cite{Qualcom2016,AtriusNavigator} can also provide information assistance services for shoppers. For example, the product description, sale information, or other necessary information can be displayed on the screen when the phone is under a certain LED light. Another example is information assistance services in museum~\cite{Grobe2013,Kim2019}. This can help to reduce the number of close physical contacts in these places.
	
	\paragraph{Autonomous robot} 
	Similar to the information assistance systems for reducing close physical contacts, autonomous robots using the VLC technology for communication and localization can also be deployed to assist people in certain circumstances, for example, elderly-assistant robots, walking-assistant robots, shopping-assistant robots, etc.,~\cite{Guan2020,Zhuang2019}. 
	Moreover, visible light signals do not cause any interference to RF signals, and thus they can be effectively deployed in diverse indoor environments such as hospitals, schools, and workplaces.
	
	\subsubsection{Traffic Control}
	In the context of social distancing, high demand traffic can cause a large concentration of people in a certain area (e.g., city center). By adopting smart traffic light systems in~\cite{Jin2017,Wei2018}, we can deploy an intelligent traffic controlling system using the VLC technology to control large traffic flows as illustrated in Fig.~\ref{Fig:VLC}(c). That can help to reduce vehicle density in public areas. The VLC technology provides a communication method between vehicles and the light infrastructure (e.g., traffic lights, street lights). First, vehicles can send their information (e.g., their IDs) to the light infrastructure by using its headlights as transmitters, thus the system can detect and monitor the traffic flow. However, in this case, it is required that the light infrastructure must be equipped with VLC receivers (e.g., traffic cameras or photodiodes). Second, based on the awareness of the traffic, the system can control the vehicles by sending instructions to guide them. In this case, the system uses traffic lights, or street lights as transmitters to send information and the vehicles use dash cameras to receive the information. For example, the system will notify them about hot zones that have a high density of vehicles and do not allow them to enter, so that they can avoid these zones.		
	
	\textit{Summary:} The availability of smart retail systems is proof of the superior performance and convenience of VLC technology compared to other RF technologies in high precise indoor localization and navigation. By leveraging such commercial approaches, we can deploy the cost-effective crowd monitoring system on a large scale, not only in shopping malls or hypermarkets but also in other public places, such as airports, train stations, and hospitals, based on the existing illuminating infrastructures. Building/facilities managers can immediately alert or notify the users if they are in the middle of a crowd (e.g., varying the color temperature of the lights in the high-density zones). People can also take the initiative in planning their move to the desired locations without encountering the crowds. On the other hand, assistance systems help to reduce the number of staff/volunteers, nurses inside public buildings; or limit the close contacts between them and customers, patients. 
	Moreover, the combination with other RF technologies such as Bluetooth and Infrared also ensures the location-based services are not interrupted when the smartphone is not being actively used by the user (e.g., the phone is in the pocket). 
	Last but not least, the VLC technology can be a potential communication method between the intelligent traffic controlling system and vehicles in the outdoor environment.
	However, the main disadvantage of the VLC technology is that interference from ambient and sun lights have significant impacts on the visible light communication channels~\cite{Komine2004,Pathak2015}. It results in poor performance of the RSS-based positioning approaches and outdoor communications.

	\subsection{Thermal}

	Thermal based positioning systems can be classified into two main categories which are infrared positioning (IRP) systems and thermal imaging camera (THC). Typical IRP systems such as~\cite{IR_local_posit_1,Active_Badge,Optotrak} are low-cost, short-range (up to 10 meters) systems that use infrared (IR) signals to determine the position of targets via AOA or TOA measurement method. On the other hand, the THC, which constructs images from the object's heat emission, can operate at a larger range (up to a few kilometers)~\cite{Thermal_camera}. Because of this difference, IRP and THC can be applied in different social distancing scenarios as discussed below.

	\subsubsection{Keeping Distance}
	In keeping distance scenarios, IRP systems such as Active Badge~\cite{Active_Badge}, Firefly~\cite{Firefly_1}, and OPTOTRAK~\cite{Optotrak} can be utilized. In the Active Badge, badges that periodically emit unique IR signals are attached to the targets. Based on the distances from the fixed infrared sensors to the badges, the target's position can be calculated. As a result, this application can be useful to determine the distance between two people as well as to identify crowds in indoor environments. The main advantages of this solution are low cost and easy implementation. However, it requires users to wear tag devices to track their locations.
	
	To achieve a higher positioning accuracy, the Firefly~\cite{Firefly_1} and OPTOTRAK~\cite{Optotrak} systems can be implemented. These systems contain infrared camera arrays and infrared transmitter called markers. Due to the difference in setups (one target is attached with one tag in Firefly and multiple tags in OPTOTRAK), the Firefly system can accurately determine the target's 3D position, whereas the OPTOTRAK system can capture the target's movement. The main disadvantage of these systems is that they are prone to interference from other radiation sources such as sunlight and light bulbs. Combined with their short-range, IRP is mostly applicable in small rooms with poor-light conditions. 
	
	\subsubsection{Physical Contact Monitoring}
\begin{figure}[!]
		\centerline{\includegraphics[width=.8\linewidth]{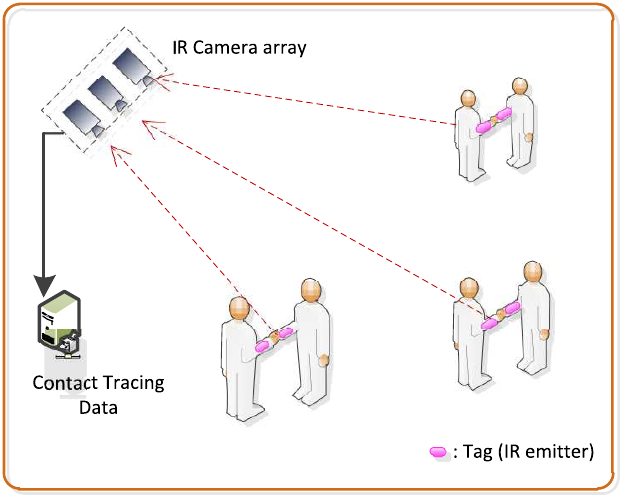}}
		\caption{Physical contact monitoring by infrared system~\cite{Firefly_1}.}
		\label{fig:IR_contac_trace}
	\end{figure}

	Since the Firefly and OPTOTRAK systems can accurately capture movements, they can be useful for contact tracing scenarios in social distancing. For example, markers can be attached to the target's body parts which are usually used in physical contacts, e.g., hands for handshakes and body for hugs. The movement of these body parts can then be captured by the IR camera as illustrated in Fig.~\ref{fig:IR_contac_trace}, and the recorded data can be analyzed later to determine if there are close contacts between the target and other people. Based on this information, the contacts that the target made can be traced later if necessary.
	\begin{figure*}[!]
		\centering{\includegraphics[width=.8\linewidth]{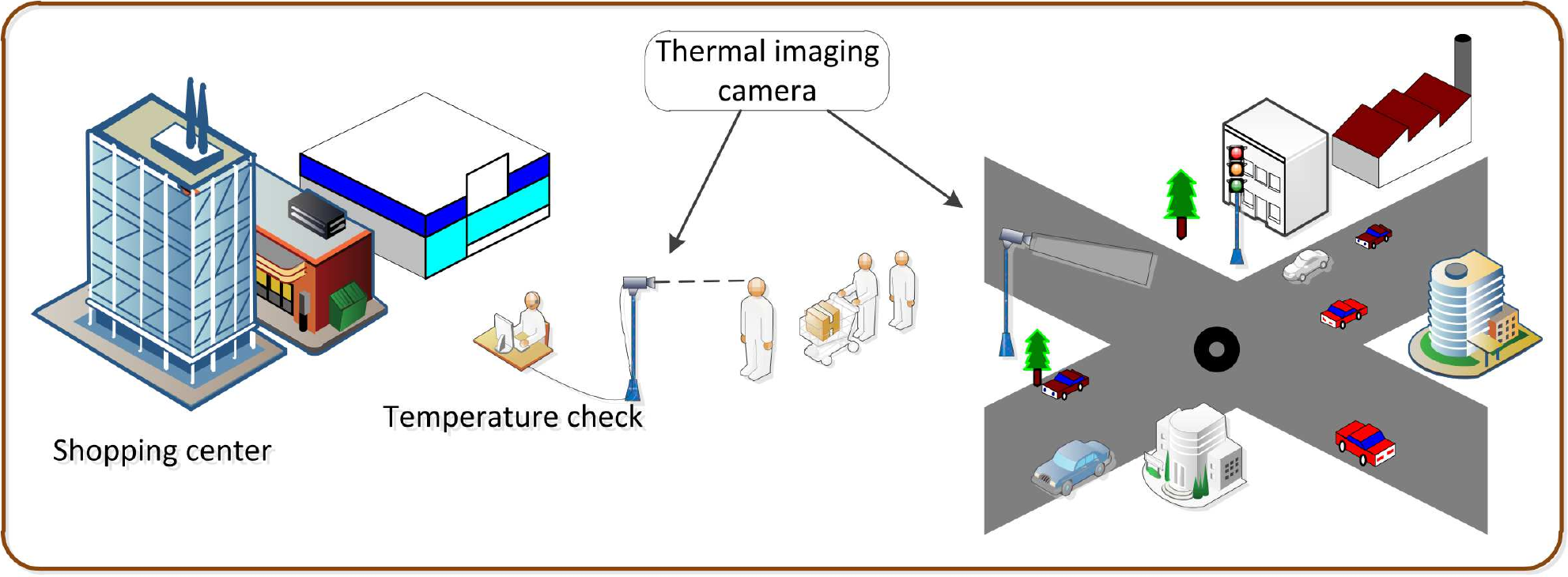}}
		\caption{Thermal camera used in susceptible group detection and traffic monitoring.
		}
		\label{fig:Thermal_SD}
	\end{figure*}
	\subsubsection{Real-time Monitoring}    
	For traffic monitoring in social distancing contexts, both IRP and THC can be utilized, especially in poor-light conditions. 
	The authors in~\cite{IR_Trafic} propose a robust vehicle detector based on the IRP under the condition to quantify traffic level and flow. The collected data can be sent to assist the authorities in social distancing monitoring. However, since IRP has a short range, THC systems such as~\cite{Thermal_Trac2} can be a better choice in a larger area with high vehicle density.
	
	Due to its very high observation range (a few kilometers)~\cite{Thermal_Trac,Thermal_Trac2,Thermal_TracHuman}, THC is particularly effective for real-time monitoring scenarios, such as public building monitoring, detecting closure violation, and non-essential travel detection, which does not require high positioning accuracy. THC systems such as those proposed in~\cite{IR_local_posit_1,IR_led_phase} are efficient in these scenarios since they are light-weight and can cover a wide area with medium accuracy. 
	
	\subsubsection{Symptom Dectection and Monitoring and Susceptible Group Detection}    	
	Another application of thermal technology is to detect susceptible groups. Since the THCs measure heat emitted from people or other objects, they can be used for checking people's temperature quickly from a far distance~\cite{Thermal_temp,Thermal_temp2}. Further, the THC system has the ability to detect slight temperature differences with a resolution of 0.01 degree~\cite{Thermal_temp3}. Thus, it can be a good means to check health conditions and sickness trends of patients. Moreover, the system can be deployed in shopping centers to measure customers' temperature remotely. This can help to detect infection symptoms early and also the prevent disease spread.
	
	
	\textit{Summary:} Thermal based positioning systems are helpful in some social distancing scenarios, especially in poor-light conditions. For short-range communication applications, the IRP is cost-effective and can be used for positioning and tracing purposes. Whereas, some light-weight THC systems can be leveraged for real-time monitoring over long distances due to their high range. However, the high cost of THC should be considered when implementing in practice.

	\subsection{Artificial Intelligence}

\begin{figure*}[!]
	\centering
	\includegraphics[scale=0.28]{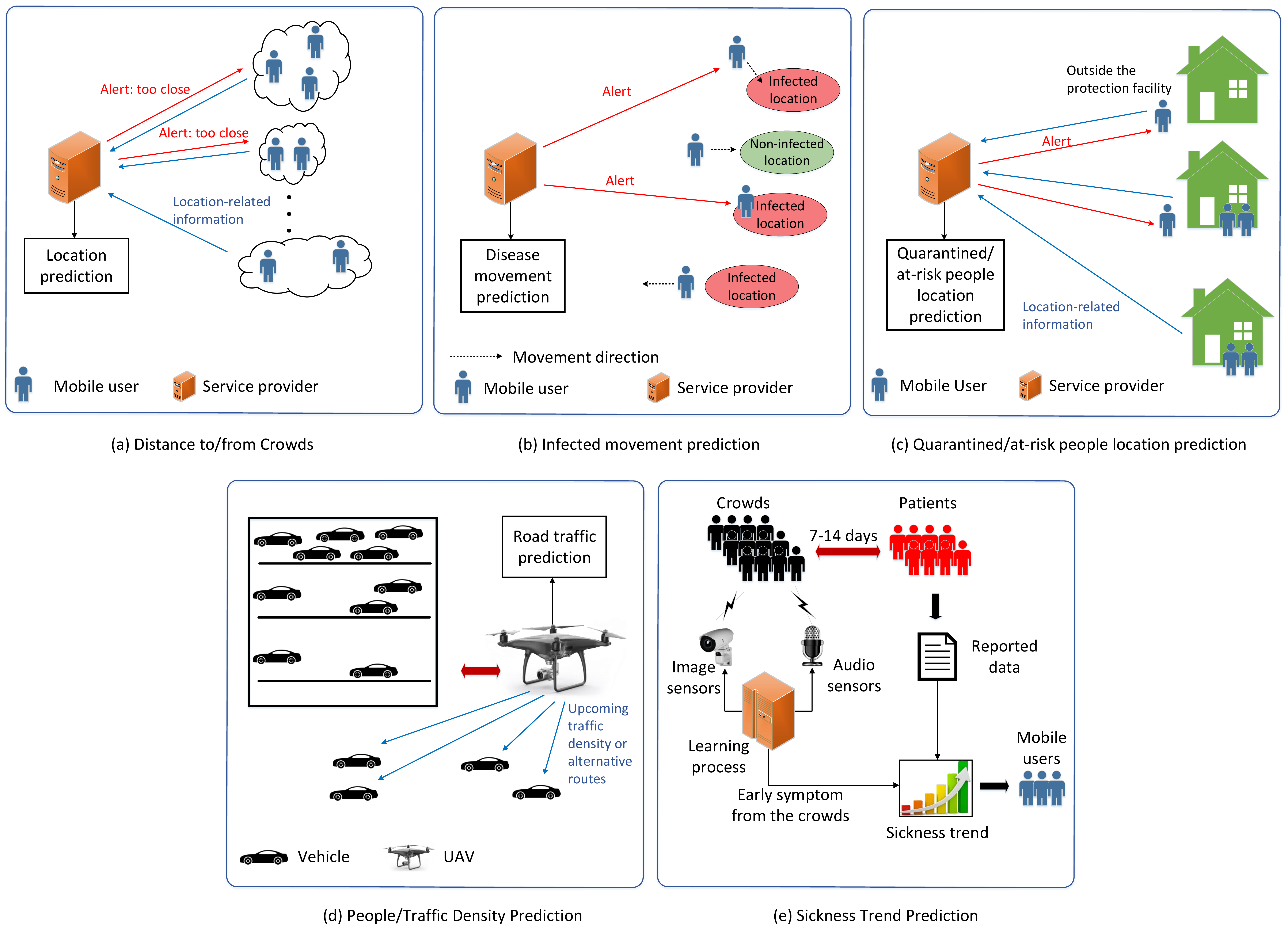}
	\caption{Application of artificial intelligence to social distancing.
	}
	\label{fig:AI1}
\end{figure*}

Over the last 10 years, we have witnessed numerous applications of AI in many aspects of our lives such as healthcare, automotive, economics, and computer networks~\cite{AIbook1}. The outstanding features of AI technologies are the ability to automatically ``learn'' useful information from the obtained data. This leads to more intelligent automation, operating cost reduction as well as the great compatibility to adapt to changing environments. For that, AI (and its underlying machine learning algorithms) can also play a key role in social distancing, especially in modern lives, with many practical applications, as discussed below.
\subsubsection{Distance to/from Crowds and Contact Tracing}
Applications of machine learning to users' location data allow us to effectively monitor the distance between people and trace the close contacts of infected people. In~\cite{Jia:2016}, the authors analyze the accuracy of a user's location prediction based on his/her friends' location datasets. In this case, a temporal-spatial Bayesian model is developed to select influential friends considering their influence levels to the user. Thus, the service provider can predict the exact location of a mobile user by using the temporal-spatial Bayesian model. Then, when the user is too close to other mobile users/people at crowded public places or his/her friends when they go in a group as illustrated in Fig.~\ref{fig:AI1}(a), his/her smartphone can alert to keep a safe distance. In addition, using the list of influential friends based on their ranks, the service provider can utilize it for the contact tracing purpose when the mobile user or one of his/her influential friends in the list gets infected.

\subsubsection{Infected Movement Prediction}
Another application of machine learning is to predict infected people movement from one location to another one and hence can potentially predict the geographic movement of the disease. The prediction is particularly crucial as infected people may travel to various places and can accidentally infect others before know that they carry the disease. In~\cite{Cho:2016}, the authors introduce a smartphone-based location recognition and prediction model to detect current location and predict the destination of mobile users. In particular, the location recognition is implemented using the combination of k-nearest neighbor and decision tree learning algorithms, and the destination prediction is realized using hidden Markov models. Given the history of infected people movement, we can adopt the above model to recognize and predict the potential geographic movement of the disease. Using the information, people can be advised to stay away from the possible infected locations through alerts from their smartphones as illustrated in Fig.~\ref{fig:AI1}(b). 

\subsubsection{Quarantined/At-Risk People Location Prediction}
The current location prediction of quarantined people, e.g., infected people, and at-risk people, e.g., sick and old people, is very important to monitor whether they currently stay at the self-quarantined and self-protection areas, e.g., their homes, or not. To this end, machine-learning-based location prediction approach can help to detect the current position of those people in a certain area. In~\cite{Brighente:2019}, the authors apply the auto-encoder neural networks and one-class support vector machines to verify whether a user is within a specific area or not. Considering various channel models, i.e., path-loss, shadowing, and fading, the proposed solutions can achieve Neyman-Pearson optimal performance by observing the probability of miss-detections and false-alarms. The authors in~\cite{Yin:2020} propose a novel localization system leveraging the federated learning to allow mobile users to collaboratively provide accurate location services without revealing mobile users' private location. As such, the authors utilize deep neural networks with the Gaussian process to accurately predict the desired location of the mobile users. As a result, we can apply the proposed solutions to detect if infected people or at-risk people currently move away from their homes as illustrated in Fig.~\ref{fig:AI1}(c). Moreover, we can utilize the proposed solutions to determine the movement frequency of the self-isolated people outside the protection facility. Using the movement frequency history, the authorities can enforce them to stay at the protection facility for further infection prevention. 

\subsubsection{People/Traffic Density Prediction}
Predicting the density of people or the number of people in public places allows us to efficiently schedule or guide people to stay away or refrain from coming to soon-to-be over-crowded places. For example, when the predicted number of people in a certain place almost reaches a pre-defined threshold (e.g., according to the social distancing requirement), the service provider can broadcast a local notification to incoming people via cellular networks, aiming at encouraging them to move to another area. In~\cite{Polese:2019}, the authors adopt advanced machine-learning-based approaches for edge networks to predict the number of mobile users within base stations' coverages. Particularly, the framework first groups the base stations into clusters according to their network data and deployment locations. Then, using various machine learning algorithms, e.g., the Bayesian ridge regressor, the Gaussian process regressor, and the random forest regressor, we can predict the number of mobile users within their network coverages. From the preceding architecture, one can utilize Wi-Fi hotspots and cluster them based on their locations. By doing so, we can predict the number of people within each cluster's coverage. Using the same architecture, we can extend the application to predict the traffic level on the roads. Specifically, upon predicting the number of vehicular users on the roads, we guide the drivers to choose particular routes to satisfy the social distancing requirements, e.g., suggest alternative routes to avoid crowded areas. In~\cite{Challita:2019}, the authors introduce a UAV-enabled intelligent transportation system to predict road traffic conditions using the combination of convolutional and recurrent neural networks. In particular, sensor cameras on the UAVs are utilized to capture the current road traffic. By using this information, the UAVs can then predict the road traffic conditions using the aforementioned deep learning methods. Thus, from the traffic prediction, the UAVs can work as mobile road side units to orchestrate road traffic for over-crowding avoidance through informing the upcoming road traffic conditions to vehicular users via cellular networks accordingly (Fig.~\ref{fig:AI1}(d)).

\subsubsection{Sickness Trend Prediction}

Machine-learning-based location prediction method is also of importance to predict the sickness trend in specific areas. This sickness trend prediction can be used to inform people to stay safe from possible infected places. For example, the work in~\cite{Hossain:2020} designs a contactless surveillance framework, i.e., FluSense, to predict the influenza-like disease 7-14 days before the real disease occurs in the hospital waiting areas. In particular, a set of real-time sensors including a microphone array to detect normal speech/cough sounds and a thermal camera to detect crowd density are embedded into an edge computing platform. Considering millions of non-speech audio samples and hundred thousands of thermal images for audio and image recognition models, the proposed framework can accurately predict the number of daily influenza-like patients with Pearson correlation coefficient of 0.95. The prediction model from this work can be correlated/combined with the localized medical/health information (e.g., from local hospitals/clinics) to further improve the prediction accuracy as shown in Fig.~\ref{fig:AI1}(e). We then can inform the local mobile users about the sickness trend prediction to avoid the potential areas where many influenza-like patients exist. 

\subsubsection{Symptom Detection and Monitoring}	
Coughing is one of the most common and detectable symptoms of influenza pandemics. In the presence of a pandemic, the early detection of such symptoms can play a key role in limiting the disease spread from the infected to the susceptible population. For example, if a coughing person can be detected and identified in public places, that person and the people in close proximity can be tested for the disease. 

In several studies, such as~\cite{cough1,cough2,cough3,cough4}, AI technologies are leveraged to identify the cough patterns in audio recordings collected from microphones or acoustic sensors. In~\cite{cough1}, audio signals are analyzed using recurrent and convolutional neural networks to detect coughs with a high accuracy (up to 92\%). Similarly, a hidden Markov model is proposed in~\cite{cough2} to detect cough from continuous audio recordings. In addition to audio signals, signals from motion sensors are also analyzed in~\cite{cough3} by a novel classification algorithm. However, a common limitation of these approaches is that they require the sensors to be attached to the person, which is not always possible in social distancing scenarios. To address this problem, a cough detection system is proposed in~\cite{cough4}. This system utilizes a wireless acoustic sensors network connected to a central server for both cough detection and localization. In particular, when a sound is detected, the sensors first localize the sound source by the AOA technique. Then, the sensors send the measured sound signals to the central server for cough identification using a novel classification algorithm. In the social distancing context, this system can be applied directly to monitor and detect coughing people in public places. Nevertheless, a limitation of this system is that the localization and measurement errors increase significantly when the sound source is too far from the sensors.

\textit{Summary:} Various AI technologies can be leveraged to facilitate social distancing implementations, especially in the scenarios that require modeling and prediction. In particular, AI technology can help to predict people's locations, traffic density, and sickness trends. Moreover, AI-based classifications algorithms can be utilized to detect symptoms such as coughs in public areas.

Table~\ref{tab:Summary} provides a summary of the technologies presented in this Section. Generally, each technology has a special characteristic that makes it a very effective solution for a specific scenario. For example, Computer Vision is the only technology capable of identifying a person without any attached device. As a result, this technology is particularly effective for scenarios such as quarantined people detection and monitoring. Moreover, ultrasound signals are confined by walls, which enables low-cost ultrasonic positioning system to efficiently monitor people in a small room. Furthermore, since inertial sensors are built-in in most smartphones, they can be quickly utilized for keeping distance smartphone applications. In addition, visible light technology can be leveraged for building information assistance systems which help to reduce human presence. Finally, thermal camera is the only technology that can detect people over a large distance (a few kilometers) without the need for attached devices, which makes it an ideal solution to detect violation of quarantines or closures.    
\begin{table*}[]
	\caption{Summary of Other Emerging Technologies}
	\begin{tabular}{|>{\raggedright\arraybackslash}m{1.5cm}|>{\raggedright\arraybackslash}m{1.5cm}|>{\raggedright\arraybackslash}m{2cm}|>{\raggedright\arraybackslash}m{1.2cm}|>{\raggedright\arraybackslash}m{1.2cm}|>{\raggedright\arraybackslash}m{1.2cm}|>{\raggedright\arraybackslash}m{1.8cm}|>{\raggedright\arraybackslash}m{4.0cm}|}
		\hline 
		\multicolumn{1}{|>{\centering\arraybackslash}m{1.5cm}|}{\multirow{1}{*}{\textbf{Technology}}} & \multicolumn{1}{|>{\centering\arraybackslash}m{1.5cm}|}{\multirow{1}{*}{\textbf{Range}}}&
		\multicolumn{1}{>{\centering\arraybackslash}m{2cm}|}{\multirow{1}{*}{\textbf{Accuracy}}} & \multicolumn{1}{>{\centering\arraybackslash}m{1.2cm}|}{\multirow{1}{*}{\textbf{Cost}}} & \multicolumn{1}{>{\centering\arraybackslash}m{1.2cm}|}{\multirow{1}{*}{\textbf{Privacy}}}&
		\multicolumn{1}{>{\centering\arraybackslash}m{1.2cm}|}{\multirow{1}{*}{\textbf{Readiness}}} & \multicolumn{1}{>{\centering\arraybackslash}m{1.8cm}|}{\multirow{1}{*}{\textbf{Indoor/Outdoor}}} & \multicolumn{1}{>{\centering\arraybackslash}m{4.0cm}|}{\multirow{1}{*}{\textbf{Integrate with existing systems}}}\\
		\hline 
		\hline 
		Computer Vision \cite{vision:humanbehavior2,vision:2dpose,vision:3dpose,vision:mask_rcnn,vision:yolov3}      &  Depends on the cameras  & Low to High         &   Medium to High             &   Low      &   High        &       Both          &   Public and private camera systems                             \\ \hline

		Ultrasound \cite{Activebat_1,ultrasound_Principle,UAV_Ul,Medical_robot}  & Short, confined by walls    &  Less than 14cm~\cite{Activebat_1}        & Low to High           &   Low~\cite{Activebat_1} to
		
		High~\cite{Cricket}
		&  High         & Indoor                &   None                              \\ \hline
		
		Inertial sensors \cite{INS1,INS2,INS3,INS4,INS9,INS14}   &  Not applicable &  Less than 1m~\cite{INS1}        & Low              &  High
		&  High         & Both                &   Smartphone                              \\ \hline
		Visible light \cite{Sahin2015,pureLifi,Qualcom2016,AtriusNavigator,Philips,Guan2020,Zhuang2019,Kim2019}& Short &$\le$ 1cm~\cite{Guan2020},  $\le$ 10cm \cite{Sahin2015,AtriusNavigator,Zhuang2019}, $\le$~20cm~\cite{Eroglu2015}        & Low  & High & High & Both & Smartphone, Smart Retail  \\ \hline
		
		Thermal \cite{Firefly_1,Thermal_Trac,Thermal_Trac2,Thermal_camera,Thermal_temp}      & IRP~10m, THC~a few km  & Medium                   & Low to High  & Low to High    &    High     &  Both         &                                 None               \\ \hline
		
	\end{tabular}
	\label{tab:Summary}
\end{table*}

	\section{Open Issues and Future Research Directions}
	\label{open}
	In this section, we discuss the open issues of social distancing implementation such as security and privacy concerns, social distancing encouragement, work-from-home, and the increased demands in healthcare appointments, home healthcare services, and online services. To addressed these issues, potential solutions are also presented.
	
	\subsection{Security and Privacy-Preserving in Social Distancing}
	
	\begin{figure*}[!]
		\centering
		\includegraphics[scale=0.28]{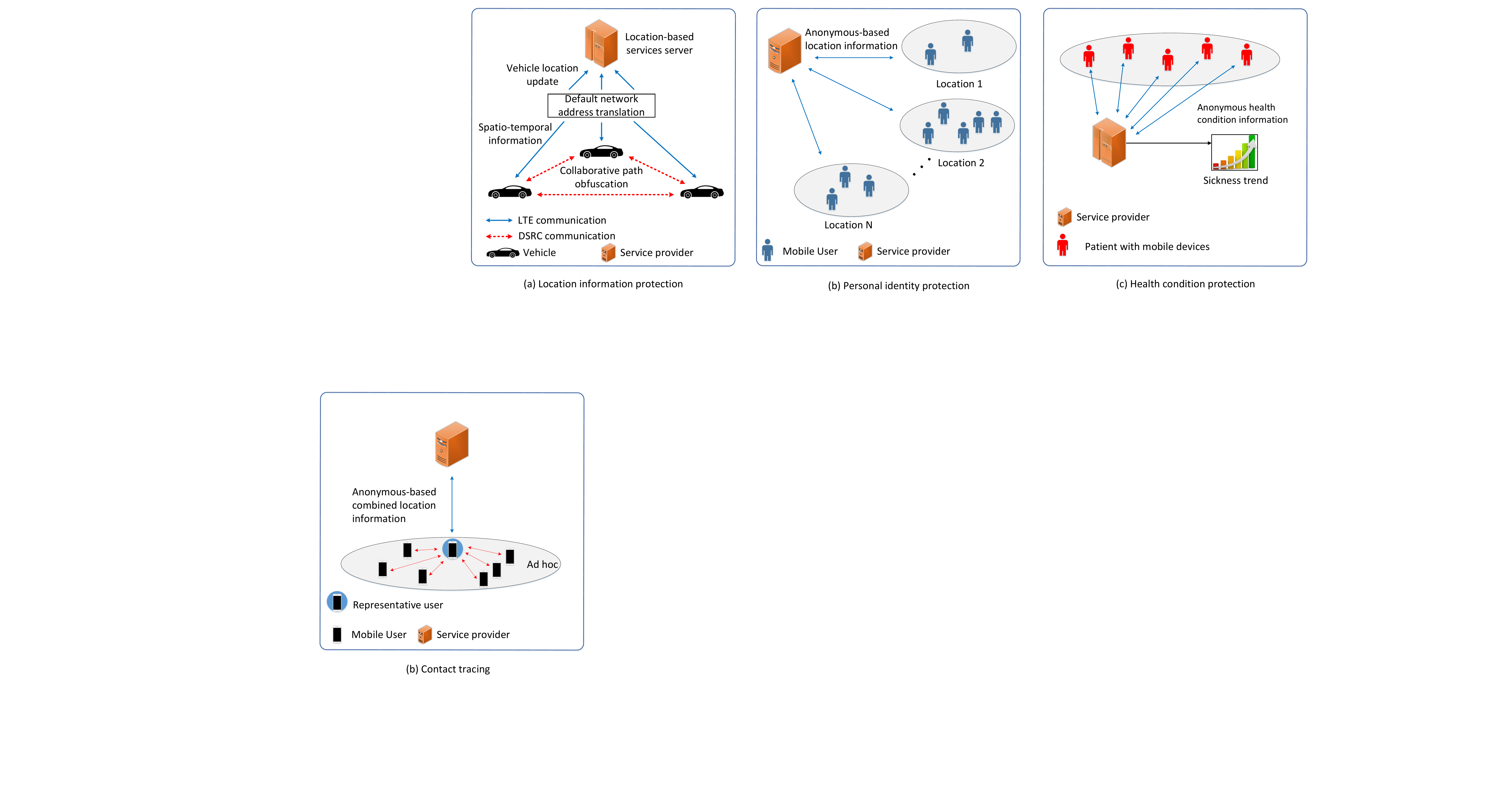}
		\caption{Location-based privacy preserving for social distancing scenarios.
		}
		\label{fig:privacy1}
	\end{figure*}
	
	Most aforementioned social distancing scenarios (see Table~\ref{tab:Summary} for more details) call for people's private information, to a different extent, ranging from their face/appearance to location, travel records, or health condition/data. These data, if not protected properly, attract cyber attackers and can turn users into victims of financial, criminal frauds, and privacy violation~\cite{Menzies:2015}. Users' data like health conditions can also adversely impact people's employment opportunities or insurance policy. Given that, to enable technology-based social distancing, it is critical to develop privacy-preserving and cybersecurity solutions to ensure that users' private data are properly used and protected. 
	
	The general principle of users' privacy-preserving is to keep each individual user's sensitive information private when the available data are being publicly accessed. To do so, data privacy-preserving mechanisms including data anonymization, randomization, and aggregation can be utilized~\cite{Tran:2019}. For example, Apple, Google, and Facebook have developed people mobility trend reports while preserving users' privacy during the COVID-19 outbreak. In particular, Apple utilizes random and rotating identifiers to preserve mobile users' movements privacy~\cite{Apple:2020}. Meanwhile, Google aggregates and uses anonymized datasets from mobile users who turn on their location history settings in their Android smartphones. In this case, a differential privacy approach is applied by adding random noise to the location dataset with the aim to mask individual identification of a mobile user~\cite{Google:2020}. Similarly, Facebook utilizes aggregated and anonymized user mobility datasets and maps to determine the mobility trend in certain areas including the social connectedness intensity among nearby locations~\cite{Facebook:2020}. In addition to the Apple's, Google's, and Facebook's latest privacy-preserving implementation, in the following, we will thoroughly discuss how the latest advances in security and privacy-preserving techniques can help to facilitate social distancing without compromising users' interest/privacy.

	\subsubsection{Location Information Protection}
	
	To protect the exact location/trajectory information of participating mobile users in social distancing, some advanced location-based privacy protection methods can be adopted. Specifically, we can anonymize/randomize/obfuscate/perturb the exact location of each mobile user to avoid malicious attacks from the attackers using the following mechanisms. For example, the authors in~\cite{Shaham:2020} develop a privacy-preserving location-based framework to anonymize spatio-temporal trajectory datasets utilizing machine-learning-based anonymization (MLA). In this case, the framework applies the $K$-means machine learning algorithm to cluster the trajectories from real-world GPS datasets and ensure the $K$-anonymity for high-sensitive datasets. Using the $K$-anonymity~\cite{Wu:2016, Zhao:2018}, the framework can collect location information from $K$ mobile users within a cloaking region, i.e., the region where the mobile users' exact locations are hidden~\cite{Martinez-Balleste:2013, Koh:2017}. In~\cite{Corser:2016}, the use of $K$-anonymity is extended into a continuous network location privacy anonymity, i.e., $KDT$-anonymity, which not only considers the average anonymity size $K$, but also takes the average distance deviation $D$ and the anonymity duration $T$ into account. Leveraging those three metrics, the mobile users under realistic vehicle mobility conditions can control the changes of anonymity and distance deviation magnitudes over time.
	
	The authors of~\cite{Lim:2017} propose a mutually obfuscating paths method which allows the vehicles to securely update accurate real-time location to a location-based service server in the vehicular network. In this case, the vehicles first hide their IP addresses due to the default network address translation operated by mobile Internet service providers. Then, they generate fake path segments that separate from the vehicles' actual paths to prevent the location-based service server from tracking the vehicles. Exploiting dedicated short-range communications (DSRC) among vehicles and road navigation information from the GPS, the vehicles can mutually generate made-up location updates with each other when they communicate with the location-based service server (to obtain spatio-temporal-related information). In~\cite{Cui:2018}, vehicles which use location-based services can dynamically update virtual locations in real-time with respect to the relative locations of current nearby vehicles. This aims to provide deceptive information about the driving routes to attackers, thereby enhancing location privacy protection. 
	
	In addition to the anonymization and obfuscating methods, randomization and perturbation are the methods that can be employed to protect user's location privacy in social distancing scenarios. In~\cite{Cao:2019}, a location privacy-preserving method leveraging spatio-temporal events of mobile users in continuous location-based services, e.g., office visitation, is investigated. Specifically, an $\epsilon$-differential privacy is designed to protect spatio-temporal events against attackers through adding random noise to the event data~\cite{Xu:2018, Yin:2018, Wei:2019}. In~\cite{Natgunanathan:2019}, the authors present a location privacy protection mechanism using data perturbation for smart health systems in hospitals. In particular, instead of reporting the patient's real locations directly, a processing unit attached to a patient's body can adaptively produce perturbed locations, i.e., the relative change between different locations of the patient. In this case, the system considers the patient's travel directions and computes the distance between the patient's current locations and the patient's sensitive locations (i.e., patient's pre-defined locations which he/she does not want to reveal to anyone, e.g., patient's treating room). Using this dynamic location perturbation, the need for a trusted third party to store real locations can be removed. Leveraging the aforementioned methods, we can also prevent the service provider to access mobile users' and vehicles' exact locations/trajectories/paths when they implement social distancing for crowd/traffic density and movement detection. Specifically, a platoon of mobile users/vehicles in a certain area can collaborate together to mix their real locations/trajectories/paths anonymously (Fig.~\ref{fig:privacy1}(a)). In this way, the service provider will only obtain the aggregated location/trajectory/path information of the platoon instead of each individual's exact location/trajectory/path for its location privacy.
	
	\subsubsection{Personal Identity Protection}
	
	In addition to protecting mobile users' location-related information, preserving their personal identities is of importance to improve users' acceptance of the latest technologies to social distancing. Specifically, we can exchange or anonymize personal identities among trusted mobile users to avoid the attackers identifying the actual identity of each individual user. In~\cite{Sun3:2017}, the authors develop a pseudo-identity exchanging protocol to swap/exchange identity information among mobile users when they are at the same sensitive locations, e.g., hospital and residential areas. In particular, when a mobile user receives another trusted user's identity and private key, the mobile user will verify if the encryption of another user's identity hash function and public key is equal to the encryption of the received private key. If that condition holds, the mobile user will change his/her identity with that user's identity and vice versa.
	
	Another method to protect personal identity in social distancing scenarios is individual information privacy protection through indirect- or proxy-request as proposed in~\cite{Han:2018}. In particular, instead of directly submitting a request to the server, a mobile user can have his/her social friends through the available social network resources, i.e., trusted social media, to distribute his/her request anonymously to the server. The request result can be returned to his/her social friends and then forwarded to the requested mobile user, thereby preserving the requested mobile user's identity. In fact, there may exist some malicious friends which expose the identity of the mobile user. Therefore, the authors in~\cite{Sun4:2017} investigate a user-defined privacy-sharing framework on social network to choose his/her particular friends who are trusted to obtain the mobile user's identity information. In this case, the mobile user only shares his/her identity information with the particular friends whose pseudonyms match the mobile user's identity through the authorized access control. Using the same approaches from the above works, we can use local wireless connections, e.g., Bluetooth and Wi-Fi Direct, to anonymously exchange actual location information in a mobile user group, i.e., between a mobile user and his/her trusted nearby mobile users, in an ad hoc way. As shown in Fig.~\ref{fig:privacy1}(b), when the service provider requires to collect location-related information for the current crowd density detection, a representative mobile user from the group can send the group's anonymous location information to the service provider, aiming at preserving the personal identity of each mobile user in the group.
	
	Moreover, Apple and Google have recently introduced a key schedule for contact tracing to ensure the privacy of users~\cite{News3}. Specifically, there are three types of key: (i) tracing key, (ii) daily tracing key, and (iii) rolling proximity identifier. The tracing key is a 32-byte string that is generated by using a cryptographic random number generator when the app is enabled on the device. The tracing key is securely stored on the device. The daily tracing key is generated for every 24-hour window by using the SHA-256 hash function with the tracing key. The rolling proximity identifier is a privacy-preserving identifier which is sent in Bluetooth advertisements. This identifier is generated by using the SHA-256 hash function with the daily tracing key. Each time the Bluetooth MAC address is changed, the app can derive a new identifier. When a positive case is diagnosed, its daily tracing keys are uploaded to a server. This server then distributes them to the clients who use the app. Based on this information, each of the clients will be able to derive the sequence of the rolling proximity identifiers that were broadcasted from the user who tested positive. In this way, the privacy of the users can be protected because, without the daily tracing key, one cannot obtain the user's rolling proximity identifier. In addition, the server operator also cannot track the user's location or which users have been in proximity.
	
	Similarly, several solutions have been proposed in~\cite{PACT},~\cite{Pan}. The key idea of these solutions is generating a unique identifier and broadcasting it to nearby devices. In particular, PACT~\cite{PACT} regularly (every few seconds) emits a data string, called chirps, generated by cryptographic techniques based on the current time and the current seed of the user to ensure the privacy. Similarly, in~\cite{Pan}, the identifier $EphID$ (called ephemeral ID) is created as follows:
	\begin{equation}
	EphID = PRG\big(PRF(SK_t, broadcast \mbox{ } key)\big),
	\end{equation}
	where $PRF$ is a pseudo-random function (e.g., SHA-256), $broadcast \mbox{ } key$ is a fixed and public string, and PRG is a stream cipher (e.g., AES in counter mode). $SK_t$ is the secret key of each user during day $t$ which is computed as follows:
	\begin{equation}
	SK_t = H(SK_{t-1}),
	\end{equation}
	where $H$ is a cryptographic hash function. Upon receiving the identifier, other nearby devices will keep it as a log. If a user is diagnosed with the disease, other users who may have encountered the infected person will receive a warning of a potential contact.
	
	With outstanding performance in data integrity, decentralization, and privacy-preserving, blockchain technology can be an effective solution to preserve privacy to enable technology-based social distancing scenarios. A blockchain is a distributed database shared among users in a decentralized network. This decentralized nature of blockchain ensures its immutability property, i.e., the data stored within cannot be altered without the consensus of the majority of network users~\cite{PoS}. Another advantage of blockchain technology is that the users' anonymity is ensured thanks to the public-private keys pair mechanism~\cite{blockchain}. As a result, blockchain technology can effectively address the personal identity issue in social distancing scenarios where people have to share their movement and location information but not their exact identities. For example, in the infected movement data scenario, we only need to know the movement path of a person, and whether or not that person is infected. In this case, the person anonymity can be ensured with the public-private keys pair mechanism, since there is no way to link the public key to that person true identity. 
	
	\subsubsection{Health-Related Information Protection}
	
	To monitor the sickness trend in a certain place, e.g., the hospital, for the social distancing purpose (i.e., to inform the upcoming mobile users not to enter a high-risk area/building), the health-related condition information of visiting mobile users has to be shared to provide reliable learning dataset. To protect this highly sensitive information, the authors in~\cite{Lin:2016} propose a differential privacy-based protection approach to preserve the electrocardiogram big data by utilizing body sensor networks. In particular, non-static noises are applied to produce sufficient interference along with the electrocardiogram data, thereby preventing the malicious attackers to point out the real electrocardiogram data.
	
	To provide secure health-related information access for authenticated users, a dynamic privacy-preserving approach leveraging the biometric authentication process is introduced in~\cite{Zhang5:2018}. Specifically, when a user wants to access the medical server containing his/her health condition, a secure biometric identification at the server for the user's validity is employed where the exact value of his/her biometric template remains unknown to the server. In this way, the personal identity of the authenticated user can be preserved. To further enhance the anonymity of his/her medical information, the random number that is used to protect the biometric template is updated after every successful login. Then, the authors in~\cite{Vijayakumar:2020} propose a secure anonymous authentication model for wireless body area networks (WBANs). Specifically, this framework enables both patients and authorized medical professionals to securely and anonymously examine their legitimacies prior to exchanging biomedical information in the WBAN systems. Motivated from the above works, we can utilize mobile devices, secure service provider, and the aforementioned privacy-preserving approaches to anonymously collect people's health condition information for illness monitoring in the hospital/medical center (Fig.~\ref{fig:privacy1}(c)). In this way, the social distancing through monitoring the sickness trend can be implemented efficiently while preserving the sensitive information of the people in the illness areas.

	\subsection{Real-time Scheduling and Optimization}
\begin{figure}[!]
		\centering
		\includegraphics[width=.5\textwidth]{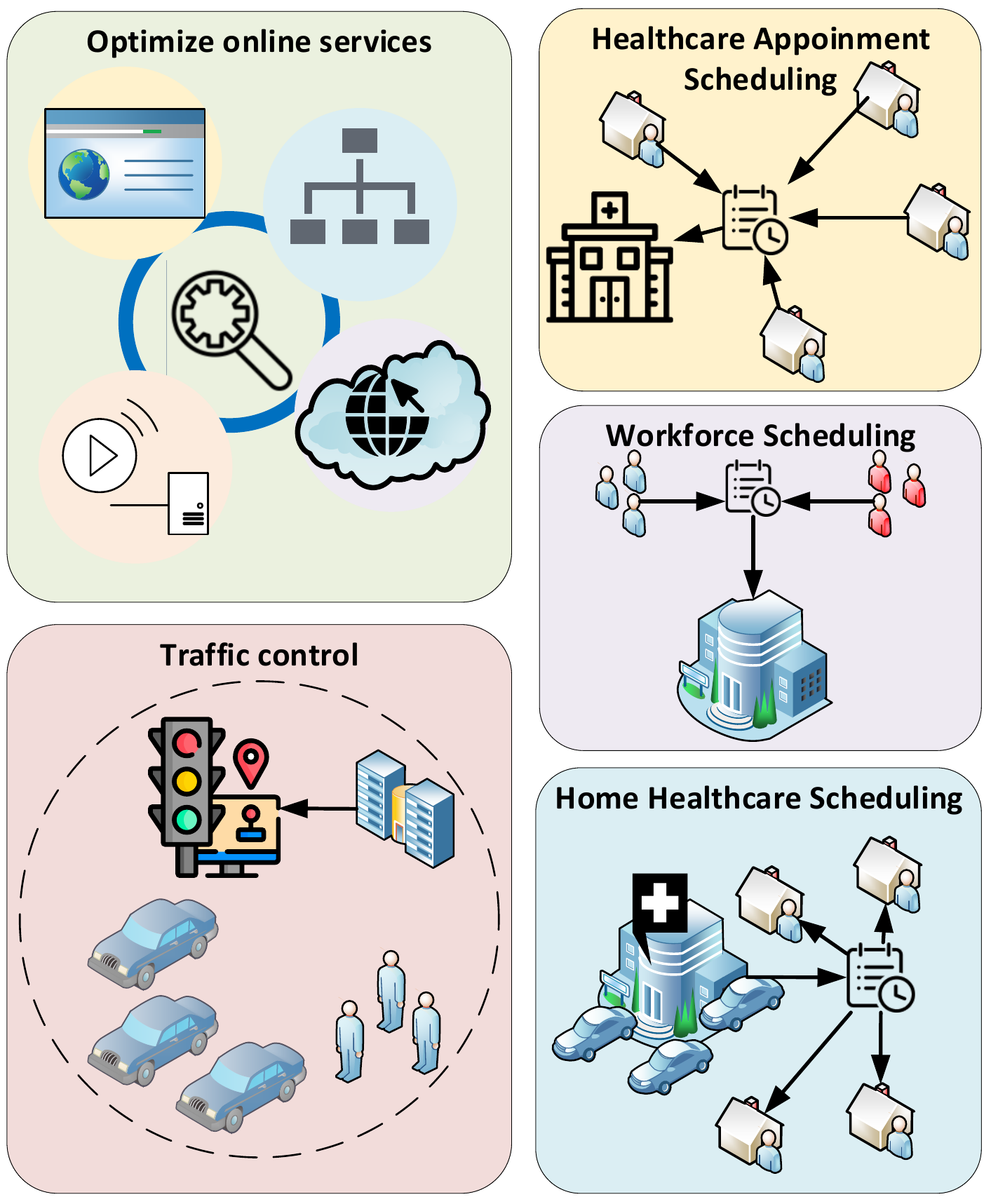}
		\caption{Scheduling and optimization for several social distancing scenarios.}
		\label{fig:sched}
	\end{figure}

	In the context of social distancing, real-time scheduling and optimization techniques can play a key role in preventing an excessive number of people at the a given place (e.g., supermarkets, hospitals) while maintaining a reasonable Quality-of-Service level. Fig.~\ref{fig:sched} illustrates several social distancing scenarios where scheduling and optimization techniques can be applied. In particular, proper scheduling can help reduce the number of necessary employees at the workplace and the number of patients coming to the hospital, thereby minimizing the physical contacts among people. Moreover, traffic scheduling can help to reduce the peak number of vehicles and pedestrians, and network resource optimization (e.g., network/resource slicing) can meet surging demands on the online services while more people are working remotely from home. 
	
	\subsubsection{Workforce Scheduling}
	Workforce scheduling can help to limit the number of people at the workplaces while ensuring the necessary work is done. While working from home is encouraged in social distancing, some essential work requires people to be present at the workplace for important tasks (e.g., health, transportation and manufacturing). Moreover, different types of tasks impose various constraints such as due date (time constraints), dependence among tasks (precedence constraints), skill requirements (skill constraints), and limited resources usage (resource constraints) which further complicate the scheduling problem. For such scenarios, workforce scheduling techniques can be utilized to optimally align and reduce the number of required employees to practice social distancing. In~\cite{sched1}, a novel three-phase algorithm is proposed for workforce scheduling to optimize the operational cost and service level simultaneously. Another Genetic-Algorithm-based hybrid approach is presented in~\cite{sched2}, which optimizes the schedules of the workforce according to multiple objectives including urgency, skill considerations, and workload balance. Similarly, in~\cite{sched3}, a Mixed-Integer-Programming-based approach is developed to minimize the operational cost with consideration of skill constraints. It is worth noting that the main objective of these approaches is to minimize cost, which is not the highest priority in the context of social distancing. In~\cite{sched4},~\cite{sched5}, and~\cite{sched6}, several methods are proposed to optimize the workforce schedules with consideration of rotating shifts, which indirectly reduce the number of employees to a certain extent. Nevertheless, the main objective of these approaches is reducing costs. Therefore, developing techniques to reduce the physical contacts or distance among employees at the workplace is critical for workforce scheduling in social distancing scenarios.
	
	\subsubsection{Medical/Health Appointment Scheduling}Besides workforce planning, scheduling techniques can also help to optimize healthcare services, especially healthcare appointments and home healthcare services, thereby decreasing unnecessary traffic and the number of patients coming to hospitals. Several approaches have been proposed to effectively schedule appointments. In particular, a local search algorithm is proposed in~\cite{sched7} to minimize patient waiting times, doctor idle times, and tardiness (lateness). Moreover, a two-stage bounding approach and a heuristic are presented in~\cite{sched8} and~\cite{sched10}, respectively. However, a common limitation of these techniques is that they do not take into account the uncertainties in the duration of the appointments and the possibility that the patient will not come to the scheduled appointment. To address that, the uncertainty in the processing times (e.g., of surgeries) is considered by a conic optimization approach in~\cite{sched9}. Similarly, a multistage stochastic linear program is developed in~\cite{sched11} to minimize patient waiting times and overtime, which takes into account the unpredictable appointment duration and unplanned cancellations. Although there are many effective approaches to optimize appointment scheduling, the open issue is to develop techniques that specifically minimize or control the number of patients simultaneously coming to the hospitals to maintain a suitable level of social distancing, similar to that of the workforce scheduling scenario. 
	
	\subsubsection{Home Healthcare Scheduling}Similar to appointment scheduling, home healthcare services (HHS) can help to reduce the pressure on hospitals and traffic in the social distancing context. In~\cite{sched12}, a multi-heuristics approach is proposed for HHS scheduling to minimize the total traveling times of HHS staff. An extended problem is presented in~\cite{sched13}, where the objective also includes minimizing the tardiness and additional skills and time constraints are considered. For this problem, local search-based heuristics are proposed in the paper. Another local search-based heuristic is proposed in~\cite{sched14} for HHS scheduling with the objective to minimize traveling times and optimize Quality-of-Service while considering workload and time constraints. In~\cite{sched15}, a Genetic-Algorithm-based hybrid approach is proposed for HHS scheduling with uncertainty in patient's demands to minimize transportation costs. Also addressing uncertainties, a branch-and-price algorithm is proposed in~\cite{sched16} to minimize the traveling costs and delay of services while considering stochastic service times. Unlike in workforce planning and appointment scheduling, HHS scheduling techniques can be more effectively applied to social distancing scenarios because they can minimize the traveling distances while ensuring Quality-of-Service.
	
	\subsubsection{Traffic Control}Scheduling techniques have also been applied for traffic control. In social distancing scenarios, scheduling techniques can help to regulate the traffic level, especially the number of pedestrians. In~\cite{sched17}, a novel scheduling algorithm is developed for traffic control, considering both vehicles and pedestrians, to minimize the delays. Similarly, a macroscopic model and a scheduling algorithm are proposed for traffic control, which jointly minimize both the pedestrians and vehicle delays in~\cite{sched18}. Another scheduling approach is proposed in~\cite{sched19} that considers both pedestrians and vehicles. Different from the previously mentioned approaches, this approach only focuses on minimizing pedestrian delay. Although there is a vast literature on traffic scheduling techniques, the social distancing implications have not been taken into account. For example, to maintain social distancing, a more meaningful objective would be to reduce/constrain the peak number of pedestrians on the street at the same time.
	\subsubsection{Online Services Optimization}
	When social distancing measures are implemented, more people will be staying at home e.g., working from home. Physical meetings/gatherings will move to virtual platforms, e.g., webinars. That results in much higher Internet traffic and corresponding virtual service demands (e.g., video streaming, broadcasting, and contents delivery). Therefore, optimizing online services delivery is a challenging issue in the social distancing context. Fortunately, online services optimization is a well-studied topic with a substantial body of supporting literature. 
	
	For example, in~\cite{opti1}, a novel algorithm is proposed to optimize the contents delivery process in a \textit{CDN semi-federation} system. In particular, the algorithm optimally allocates the content provider's demand to multiple Content Delivery Networks (CDNs) in the federation. The results show that the latency can be reduced by 20\% during peak hours. 
	Another technique to reduce the delay and network congestion is edge-caching, which brings the contents closer to the network users. In~\cite{opti2}, the performance of two edge-caching strategies, i.e., coded and uncoded caching, are analyzed. Moreover, two optimization algorithms are developed to minimize the content delivery times for the two caching strategies. 
	
	Besides the contents delivery, the demands on video streaming traffic are also much higher during social distancing implementation because there are many people who work from home. In that context, emerging networking technologies can be an effective solution. For example, an architecture utilizing HTTP adaptive streaming~\cite{opti3} and software-defined networking technology is proposed to enable video streaming over HTTP. Moreover, a novel algorithm is developed to optimally allocates users into groups, thereby reducing communication overhead and leveraging network resources. The results show that the proposed framework can increase video stability, Quality-of-Service, and resource utilization.
	
	Scheduling and optimization are well-studied topics with a vast literature available, which can be utilized for different social distancing scenarios such as workforce, healthcare appointment, home healthcare, and traffic scheduling, and optimization of online services delivery. Nevertheless, except for the home healthcare service scenario, the existing techniques' objectives do not align with the objectives of social distancing. Moreover, scheduling algorithms are often developed such that they are only efficient for specific problems. Therefore, developing novel optimization/scheduling algorithms in operations research and adopting social distancing as a new performance metric or design parameter is very much desirable. Furthermore, the optimization of Internet-based services such as content delivery can help to encourage people to stay at home during social distancing periods by ensuring the service levels.

	\subsection{Incentive Mechanism to Encourage Social Distancing}
	
	Due to the people's self-interested/selfish nature characteristics in their daily life~\cite{Zhang:2017} (especially during the pandemic outbreak), incentive mechanisms can be very helpful in encouraging people to accept or share relevant information to enable new social distancing methods. These mechanisms have been thoroughly discussed in crowdsourcing as implemented in~\cite{Ma:2018, Zhang2:2017, Zhang:2012, Liu2:2018, Xu:2015}. Therein, the service providers can provide incentives to a large number of people to attract their contributions in data collection for crowdsourcing process. For example, the contract theory-based incentive mechanism for crowdsourcing is discussed in~\cite{Ma:2018, Zhang2:2017}. In particular, this approach is considered as an efficient mechanism to leverage common agreements between the participating entities, e.g., a service provider and its mobile users, in a certain area under complete and incomplete information from the participants~\cite{Bolton:2005}. The use of a game theory-based incentive mechanism to encourage a set of mobile users to form a crowdsourcing community network is investigated in~\cite{Zhang:2012, Liu2:2018}. Then, in~\cite{Xu:2015}, the authors utilize an auction theory-based approach incentive mechanism to stimulate mobile users participation in crowdsourcing tasks such as traffic monitoring. In the following, we also highlight the existing incentive mechanisms and how they can be further adopted to encourage social distancing applications.
	
	\subsubsection{Distance Between any Two People and Distance to/from Crowds}
	
	\begin{figure*}[!]
		\centering
		\includegraphics[scale=0.215]{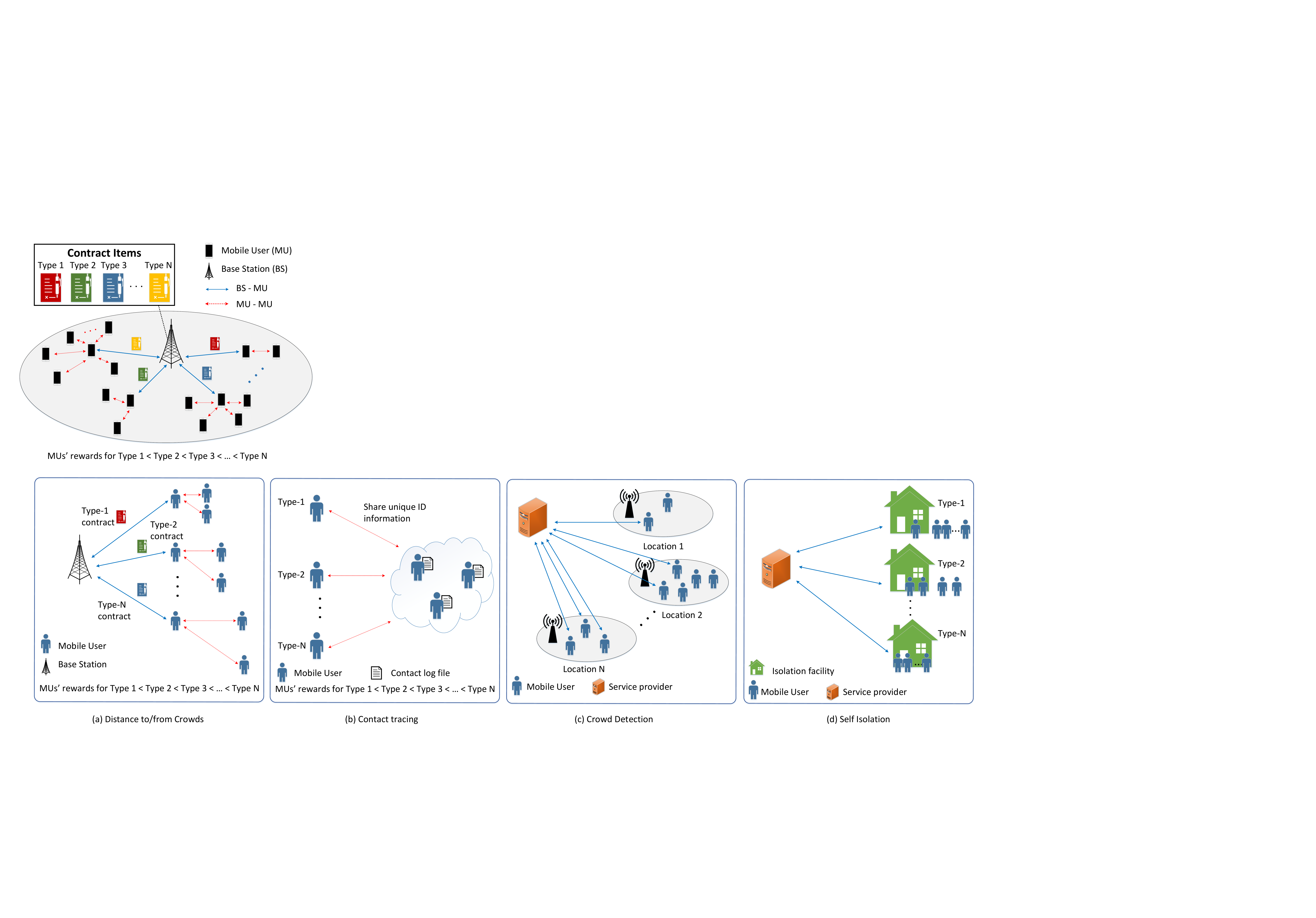}
		\caption{Contract-based incentive design scenarios to encourage social distancing.
		}
		\label{fig:incentive1}
	\end{figure*}
	
	To motivate people to keep \textit{safe} distances from themselves to others, contract theory-based incentive models via D2D communications, e.g., Bluetooth, Wi-Fi Direct, can be employed. In~\cite{Zhang:2015}, the authors propose a contract theory-based mechanism to provide a higher reward for D2D-capable mobile users if they send the information to a requesting mobile user with a higher transmission data rate. Taking into account the number of potential nearby mobile users in proximity, the authors in~\cite{Chen:2017} introduce the same mechanism such that a mobile user will receive a higher payment if they can share the information to more nearby users. Likewise, the same approach considering a higher reward for a mobile user who has shorter distances in sharing its information to nearby D2D pairs is presented in~\cite{Chen:2018}. Inspired by the aforementioned works, we can consider the contract theory-based method along with D2D communications to encourage people to keep distances from other people/crowds. Specifically, mobile service providers can be subsidized/funded or requested by the government to provide incentives to their users to keep a distance from others when they are in public. Specifically, a service provider can offer contracts to mobile users, as illustrated in Fig.~\ref{fig:incentive1}(a). Considering the current distances from the nearby mobile users and capability to inform them through D2D communications, those mobile users can obtain more rewards when they successfully keep a sufficient distance (e.g., at least 1.5 meters) from other people/users. A violation (e.g., getting closer than 1.5 meters to someone) can lead to a ``penalty'' (e.g., losing part of the previous rewards).

	\subsubsection{Contact Tracing}
	In a pandemic outbreak, contact tracing is considered as one of the most important actions to contain the spread of the disease. To trigger each mobile user for information sharing, e.g., mobile user's public identity, the network operator requires to offer incentives to those who contribute such information (besides privacy-preserving solutions). In~\cite{Ma:2018}, the authors introduce a contract theory-based incentive mechanism in a crowdsourced wireless community network. In particular, the network operator offers contracts to network-sharing mobile users containing a Wi-Fi access price (for their nearby mobile users accessing the network sharing) and a subscription fee (for the network-sharing mobile users). Motivated from this work, we can also develop a contact-tracing framework which allows a mobile user to broadcast his/her public identity to the nearby mobile users as long as their distances are within 1.5 meters. Then, the nearby mobile users can store this public identity in their close-contact log files including the time and location when they receive that public identity as shown in Fig.~\ref{fig:incentive1}(b). Mobile users who store such log files will pay the sharing mobile user to compensate for the information sharing. In this way, when at least one of the mobile users in the log files is infected by the contagious disease, the mobile service provider can alert the mobile users with the log files to implement social distancing.
	
	\subsubsection{Crowd Detection}
	A high density of people in specific areas can make contagious diseases to spread the infection more quickly due to people's close proximity. To support social distancing, an incentive mechanism approach can also be applied to detect the people density in public areas or the number of people in a building. In~\cite{Zhang2:2017}, the authors present a tournament model-based incentive mechanism to encourage mobile users (with various performance ranks) connected to the local wireless networks, e.g., Wi-Fi hotspots, to send the location and unique identifier of the networks to the service provider (Fig.~\ref{fig:incentive1}(c)). From the hotspots' location information, the service provider can then determine the people density in each hotspot area or the number of people in a building (which may have several hotspot areas). Using the above method, we can also encourage mobile users to avoid non-essential public places, e.g., restaurants and shopping malls. In this case, the reward can be adapted according to the locations and essential level of the services (e.g., cinemas, restaurants, grocery stores, schools, and hospitals).
	
	In addition to the people density detection, we can adopt incentive mechanisms to monitor the density of vehicles on the city roads for traffic crowd avoidance purposes. In fact, the contagious diseases, e.g., coronavirus, can remain on the surfaces for fours hours up to several days~\cite{Doremalen:2020}. Thus, avoiding traffic jams on the roads can reduce the possibility of disease infection. In~\cite{Duan:2012}, the authors propose a reward-based smartphone collaboration method to support data acquisition for location-based services. Specifically, a client will attract surrounding smartphone users, e.g., vehicular users on a highway, to collaborate together with the aim to build a big database containing location information as implemented in Google's Android smartphones and Apple's iPhone~\cite{News3}. The joining smartphone users then receive shared rewards from the client considering their collaboration costs. Based on this database, the client can determine the traffic levels according to the vehicles' density on the roads dynamically and sell this information to the authorities or service provider. Such information can be useful for several social distancing scenarios such as crowd detection, traffic/movement monitoring, and traffic control. 
	\subsubsection{Location/Movement Sharing Stay-at-home Encouragement}
	To further drive people away from high-density public places, one can also consider incentive mechanisms for better social distancing efficiency (especially for the people with their mobile devices). In particular, the authors in~\cite{Tian:2017} study the uneven distribution of the crowdsourcing participants when maximizing the social welfare of the network. To address this problem, a movement-based incentive mechanism to stimulate the participants to move from popular areas to unpopular ones was introduced. This approach guarantees that the participants will announce their actual costs for further reward processes. Likewise, an incentive mechanism in spatial crowdsourcing considering budget constraints to reduce imbalanced data collection is discussed in~\cite{Liu:2017}. Particularly, the service provider will provide a higher reward when the mobile users are willing to participate in remote locations instead of nearby locations where they belong to (based on their daily routines). A similar work utilizing a redistribution algorithm to incentivize crowdsourced service providers from oversupplied areas to undersupplied ones is also investigated in~\cite{Neiat:2018}. The above works are then extended in~\cite{Wang:2019}. Instead of encouraging mobile users to completely move to faraway locations, the service provider will offer a task-bundling containing the nearby and remote tasks for each participating mobile user. All of these works show that the proposed incentive mechanisms can efficiently balance the various location popularity such that we can encourage people to move to low-density places. 
	
	In a narrow-down scenario, we can also utilize an incentive mechanism to encourage family-isolation/group-isolation for the possible vulnerable/at-risk people, e.g., sick people and older people. For example, the authors in~\cite{Rahman:2017} propose a spatio-temporal-based incentive mechanism using both smartphone and human intelligence in an ad hoc social network. This framework allows a very large crowd to work together in providing information sharing, i.e., geo-tagged multimedia resources, while receiving incentives from the system. Based on this method, we can also engage the vulnerable/at-risk groups to isolate themselves and deliver incentives for them at a certain location during a particular period (Fig.~\ref{fig:incentive1}(d)). The larger number of vulnerable/at-risk members in a group, the higher incentives will be given. Furthermore, we can design a real-time incentive mechanism to encourage people to implement self-isolation by providing more rewards for those who spend more time at a given location, e.g., at home. In this case, the reward can be negative, i.e., penalty, to discourage people from going to crowded places.
	
	\subsection{Pandemic Mode for Social Distancing Implementation}
	
	\begin{figure*}[!]
		\centering
		\includegraphics[scale=0.2]{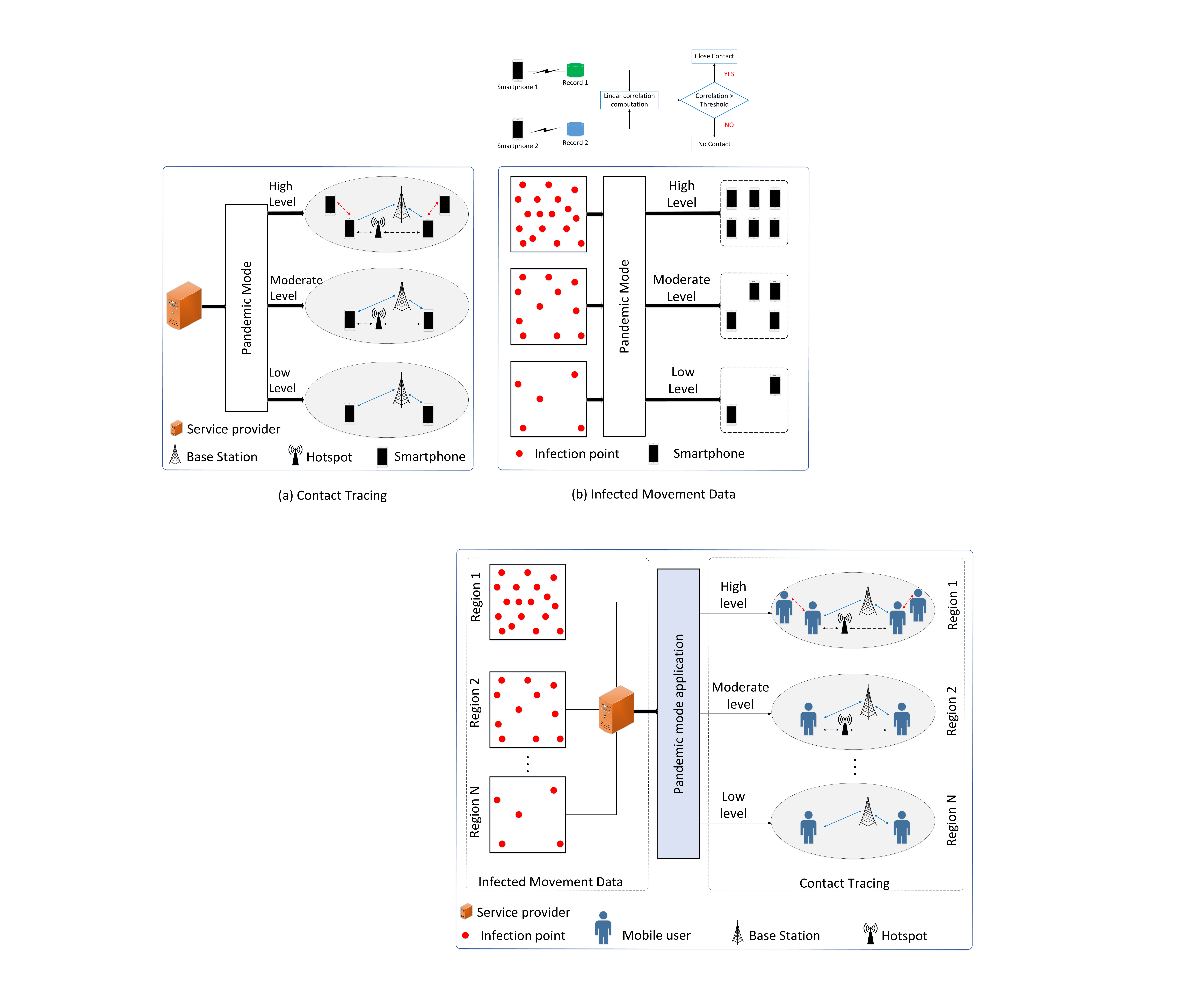}
		\caption{Pandemic mode in future infrastructures to support social distancing.}
		\label{fig:pandemic1}
	\end{figure*}
	
	An occasional pandemic outbreak in a particular period can drive the mobile service providers, e.g., Google and Apple, to build up a pandemic mode application for current users' mobile devices, e.g. smartphones. This application represents a comprehensive framework utilizing the current pandemic situation, i.e., infected movement data, to help the mobile users stay aware of the contagious diseases and perform cautious actions to slow down the spread of the diseases through implementing social distancing. To this end, the use of users' smartphones is very crucial to realize this pandemic mode application as similarly implemented for smartphone-based disaster mode application in~\cite{Hossmann:2011,Neumann:2012,Fujiwara:2014,Hossain:2016,Lu:2017,Kamruzzaman:2017,Thomas:2018}. When a contagious disease outbreak is imminent, the government can first broadcast an urgent notification for mobile users to install/deploy the official pandemic mode application in their smartphones. Then, based on the current infected movement data, e.g., the current reported number of infected people and currently infected areas, from the government officials, the service providers can determine the risk levels of the pandemic and activate a certain level in the smartphones. Considering the risk level, the smartphones can leverage the existing sensors and wireless connections to perform effective contact tracing activity for contagious disease containment.
	
	
	\subsubsection{Infected Movement Data}
	
	To determine the risk levels of pandemic mode, the authorities first need to monitor the current infected movement information, i.e., infected areas and the number of infected people. Based on this observation, the authorities then can orchestrate the pandemic mode risk levels and notify mobile users such that they can avoid the areas where the highly-likely infection exists according to the current risk level. In~\cite{Wang:2016}, the authors introduce an identification framework to observe the spatial infection spread based on the arrival records of infectious cases in subpopulation areas. Considering susceptible and infectious people movement in metapopulation networks, the framework first splits the whole infection spread into disjoint subpopulation areas. Then, a maximum likelihood estimation is applied to predict the most likely invasion pathways at each subpopulation area. Using a dynamic programming-based algorithm, the framework can finally reconstruct the whole spread by iteratively assembling the invasion pathways for each subpopulation to produce the final invasion pathways. Then, the authors in~\cite{Zhou:2018} present a spatial-temporal technique to locate real-time influenza epidemics utilizing heterogeneous data from the Internet. In particular, the technique constructs a multivariate hidden Markov model through aggregating influenza morbidity data, influenza-related data from Google, and international air transportation data. This aims to identify the spatial-temporal relationship of influenza transmission which will be used for surveillance application. Through experimental results, the technique can predict an influenza epidemic ahead of the actual event with high accuracy. Recently, Google and Apple also create a framework to demonstrate the community mobility trend with respect to the COVID-19 outbreak~\cite{Google:2020, Apple:2020}. In particular, this framework is generated based on the regions of mobile users and changes in visits monitoring at various public places, e.g., groceries, pharmacies, parks, transit stations, workplaces, and residential areas.
	
	Motivated by the above works, the authorities can first collect the spatio-temporal infectious disease-related information from the Internet and official reports. Using the aforementioned methods, the authorities can then extract meaningful information about the spread locations/pathways and time of the infectious diseases, which leads to various spatio-temporal disease spread levels. Based on these disease spread levels, the authorities can customize the pandemic mode risk level for different regions, e.g., states, cities, and provinces, at different times. For example, if the disease spread level, e.g., the density of infected people, at a particular city is high, the authorities can set the pandemic mode into a high-risk level for a week (as shown in Fig.~\ref{fig:pandemic1}). Otherwise, the pandemic mode level can be set at a low-risk level. 
	
	\subsubsection{Contact Tracing}
	
	After determining the risk levels of pandemic mode based on the infected movement data, the authorities can broadcast the risk level notification through smartphones' pandemic mode application. Afterward, the smartphones can perform contact tracing to help quickly discovering infected people for efficient outbreak containment~\cite{Tsui:2013,Chen:2019}. Based on the risk level of the pandemic mode, the smartphones can automatically trace contacts using certain sensors and wireless connections. For example, Google and Apple currently collaborate together to develop a contact tracing application utilizing Bluetooth technology, aiming to quickly detect past contacts among mobile users in close proximity~\cite{News3}. In this case, the Bluetooth is used to exchange beacon signals containing unique keys between two smartphones prior to storing these keys to the cloud server for infected people notification. Similarly, the work in~\cite{Sun:2016} develops a wireless sensor system to exchange beacon signals between a mobile device with other nearby mobile devices as its contact information. In another work, an epidemiological data collection scheme utilizing users' smartphones is described in~\cite{Hashemian:2012}. Specifically, a user's smartphone can be used as a sensor platform to collect high accurate information including the user's location, activity level, and contact history between the user and certain locations. Then, a smartphone-based contact detection system leveraging the smartphone's magnetometer history is investigated in~\cite{Jeong:2019}. To determine the close contact, the system measures the linear correlation between two smartphones' magnetometer records. 
	
	Inspired by the aforementioned works, the smartphones can be utilized as crucial tools to implement contact tracing considering the current risk level of the pandemic mode activated by the authorities (as illustrated in Fig.~\ref{fig:pandemic1}). In particular, if the authorities activate low-risk levels, i.e., the current number of infected people and areas are small, the smartphones can trace close contacts using cellular networks only. In this case, the pandemic mode application will disable certain sensors, Bluetooth, and Wi-Fi by default. However, if high-risk level pandemic mode, i.e., the current number of infected people and areas are large, is activated, the pandemic mode application will enable all of the wireless connections including Bluetooth, Wi-Fi, and cellular network, as well as relevant sensors automatically to trace contacts faster.

	\section{Conclusion}
	\label{conclusion}
	
	Social distancing has been considered to be a crucial measure to prevent the spread of contagious diseases such as COVID-19. In this article, we have presented a comprehensive survey on how technologies can enable, encourage, and enforce social distancing. Firstly, we have provided an overview of the social distancing, discussed its effectiveness, and introduced various practical social distancing scenarios where the technologies can be leveraged. We have then presented and reviewed various technologies to encourage and facilitate social distancing measures. For each technology, we have provided an overview, examined the state-of-the-art, and discussed how it can be utilized in different social distancing scenarios. Finally, we have discussed open issues in social distancing implementations and potential solutions to address these issues.
	

\end{document}